\documentclass[aps, onecolumn, superscriptaddress, 12pt]{revtex4}

\usepackage[english]{babel}
\usepackage{hyperref}
\usepackage[pdftex]{graphicx}
\usepackage{color}
\usepackage[toc,page]{appendix}

\usepackage{amsmath, bbold}
\usepackage{amssymb}
\usepackage{mathtools}
\usepackage{braket}

\usepackage[sort&compress]{natbib}

\newcommand{\I}{\mathrm{i}}

\begin{document}

\title{Spectral and transport properties of a $\mathcal{PT}$-symmetric
tight-binding chain with gain and loss}

\author{Adrian Ortega}
\email{adrian.ortegar@alumnos.udg.mx }
\affiliation{Departamento de F\'isica, Universidad de Guadalajara, Blvd. Gral.
Marcelino Garc\'ia Barrag\'an 1421, C.P. 44430, Guadalajara, Jalisco, M\'exico}

\author{Thomas Stegmann}
\email{stegmann@icf.unam.mx}
\affiliation{Instituto de Ciencias F\'isicas, Universidad Nacional Aut\'onoma de M\'exico,
Av. Universidad s/n, Col. Chamilpa, C.P. 62210 Cuernavaca, Morelos, M\'exico}

\author{Luis Benet}
\email{benet@icf.unam.mx}
\affiliation{Instituto de Ciencias F\'isicas, Universidad Nacional Aut\'onoma de M\'exico,
Av. Universidad s/n, Col. Chamilpa, C.P. 62210 Cuernavaca, Morelos, M\'exico}

\author{Hern\'an Larralde}
\email{hernan@icf.unam.mx}
\affiliation{Instituto de Ciencias F\'isicas, Universidad Nacional Aut\'onoma de M\'exico,
Av. Universidad s/n, Col. Chamilpa, C.P. 62210 Cuernavaca, Morelos, M\'exico}

\date{\today}


\begin{abstract}
  We derive a continuity equation to study transport properties in a
  $\mathcal{PT}$-symmetric tight-binding chain with gain and loss in 
  symmetric configurations. This allows us to identify the density 
  fluxes in the system, and to define a transport coefficient to 
  characterize the efficiency of transport of each state. These quantities 
  are studied explicitly using analytical expressions for the eigenvalues 
  and eigenvectors of the system. We find that in states with broken 
  $\mathcal{PT}$-symmetry, transport is inefficient, in the sense that 
  either inflow exceeds outflow and density accumulates within the 
  system, or outflow exceeds inflow, and the system becomes depleted. 
  We also report the appearance of two subsets of interesting eigenstates 
  whose eigenvalues are independent on the strength of the coupling to 
  gain and loss. We call these opaque and transparent states. Opaque states 
  are decoupled from the contacts and there is no transport; transparent 
  states exhibit always efficient transport. Interestingly, the appearance 
  of such eigenstates is connected with the divisors of the length of the 
  system plus one and the position of the contacts. Thus the number of 
  opaque and transparent states varies very irregularly.
\end{abstract}

\maketitle

\section{Introduction and motivation}
\label{sec:Introduction}

Non-Hermitian quantum mechanics has been extremely useful in the description of
a great deal of physical systems, from scattering, resonance phenomena and
ionization, to effective models of open systems \cite{Gamow1928, Siegert1939,
Peierls1959, Hatano1996, Nelson1998, Moibook2011, SchomerusPTRS2013}.
In particular, non-Hermitian systems described by $\mathcal{PT}$-symmetric 
Hamiltonians~\cite{BenderPRL1998, Benderbook2018} have found many 
applications \cite{HeissJPA2012,BenderJP2015, FengNature2017, El-GanainyNature2018, Miri2019}. 
$\mathcal{PT}$-symmetric Hamiltonians may have
real or complex eigenvalues, corresponding to unbroken or broken
$\mathcal{PT}$-symmetry phases. The transition between these phases occurs at
the so called exceptional points, at which two (or more)
eigenvalues and eigenfunctions coalesce and the Hamiltonian becomes defective
\cite{Katobook1995}. Several remarkable phenomena have been reported recently
in the vicinity of these points, for example, topological
states \cite{Lee2016, MartinezEPJST2018, FoaPRB2018, Ni2018, Yuce2018, Lin2019, Caspel2019},
chirality \cite{DembowskiPRL2001, Mailybaev2005, Peng2016}, unidirectional
invisibility \cite{Regensburger2012, Lin2011, Feng2012}, unidirectional zero sonic
reflection \cite{Merkel2018}, enhanced sensing \cite{Chen2017} and the possibility
to stop light \cite{GoldzakPRL2018}.

In this broad context, simple models are valuable as they permit a thorough
understanding of the phenomena taking place in the system. Indeed, analysis of even the 
simplest $2\times 2$ $\mathcal{PT}$-symmetric matrices has led to important insights
\cite{BenderJPA2004, WangPTRS2013}, though certain
aspects cannot be captured with such simple model, such as the simultaneous coalescence of
more than two eigenvalues~\cite{FoaPRB2018,GraefeJPA2012}.

In this paper we study a simple
one-dimensional tight-binding chain, with gain and loss at arbitrary
($\mathcal{PT}$-symmetric) positions along the chain.
This system, and extensions of it, have already been extensively
studied \cite{KottosPRL2009, JinPRA2009, YogeshPRA2010, YogeshPRA2011, LonghiOL2014, ElenewskiPRB2014,
  GarmonPRA2015, ZhangPRA2017, HarterNature2018}. We focus here in the
$\mathcal{PT}$-symmetric tight-binding chain from the perspective of quantum 
transport. The outline and main results of our work are as follows: First we 
provide the necessary definitions
in Sec.~\ref{sec:ModelSolution}. In Sec.~\ref{sec:LocEigstates} we outline the
derivation of a continuity equation for the density on the 
$\mathcal{PT}$-symmetric tight-binding chain. We also propose a parameter, the 
transport coefficient $\xi$, that measures the efficiency
of transport through the system. 
The explicit analytical expressions of the eigenvalues and eigenvectors of the system, which
will be used to study transport through the chain, are presented in 
Sec.~\ref{sec:egvalsegvecs}, while a detailed derivation can be found in the  
Appendix \ref{sec:egvalsol}. It is worth noting that this derivation does not 
make use of the Bethe ansatz as in previous works \cite{JinPRA2009, 
YogeshPRA2010}, but uses straightforward algebra in the ring of 
semi-infinite sequences. This is a simpler, or at least alternative,
solution of the problem. From the explicit results for the eigenvalues and eigenvectors of the system, we show that under certain circumstances, the system may have a set of eigenstates characterized by having eigenvalues that 
are independent of the strength of the gain and loss in the system. Some of these states do not couple to the gain and loss, and thus, are non-conducting or ``opaque'', whereas another subset ---that we call ``transparent''--- 
always conduct efficiently. We show that the 
condition for the appearance of such states depends on the divisors of the length of the 
system plus one, and of the positions of the loss and gain. This implies that the number of both opaque and transparent states 
varies very irregularly dependent on the size of the system, and the precise position of the leads, even for
large sizes. Next, some specific cases of how the eigenvalues behave 
in the $\mathcal{PT}$-unbroken
(-broken) phases are analyzed thoroughly. We also develop a simple
perturbation scheme to obtain the eigenvalues and eigenvectors
around an exceptional point. With the full solution to the problem, in
Sec. \ref{sec:transportPT} we analyze analytically and numerically the
transport in a $\mathcal{PT}$-symmetric tight-binding chain as a function of 
the chain length and position of the gain and loss.
We analyze some eigenfunctions along the parameter space and show how
the $\mathcal{PT}$-unbroken (-broken) phase affect their behaviour.
We give our conclusions in Sec. \ref{sec:Conclusions} and provide an outlook
based on our results.

\section{The PT-symmetric tight-binding chain}
\label{sec:ModelSolution}

A system is $\mathcal{PT}$-symmetric if the Hamiltonian commutes 
with the operator $\mathcal{PT}$, where $\mathcal{P}$ and 
$\mathcal{T}$ are the parity and the time reversal operators,
respectively. This is commonly referred
to as space-time reflection symmetry (see e.g. 
\cite{BenderCP2005, GraefeJPA2008}). Following
Bender~\cite{Bender2007}, for a $\mathcal{PT}$-symmetric operator 
we say that $\mathcal{PT}$ symmetry is unbroken
if all the eigenfunctions of the Hamiltonian are also eigenfunctions
of $\mathcal{PT}$; otherwise, we say that $\mathcal{PT}$ symmetry is broken. 
In this work we consider spinless particles for which the effect 
of the time reversal operator $\mathcal{T}$ can be defined simply 
as complex conjugation \cite{WangPTRS2013}:
\begin{equation}
  \begin{matrix}
    \mathcal{T} \equiv {}^*, & \mathcal{T}^2=\mathbb{1},
  \end{matrix}
\end{equation}
where $\mathbb{1}$ is the identity. For a matrix $M$, the action of
$\mathcal{T}$ is $\mathcal{T}M\mathcal{T} = M^*$. The
parity operator $\mathcal{P}$ is defined by the properties
\begin{equation}
  \begin{matrix}
    \mathcal{P} = \mathcal{P}^* & \text{and} &  \mathcal{P}^2 = \mathbb{1}.
  \end{matrix}
\end{equation}
Fixing a basis in a Hilbert space, we choose $\mathcal{P}$ as the
matrix $J$ with components
\begin{equation}
  J_{ij} = \delta_{i,N-j+1},
  \label{eq:J}
\end{equation}
which is commonly known as the exchange matrix in the mathematical
literature \cite{CantoniLAA1976}, sometimes called sip matrix
\cite{GraefeJPA2008}.

\begin{figure}
  \centering
  \hspace*{-6mm}
  \includegraphics[width=0.5\linewidth]{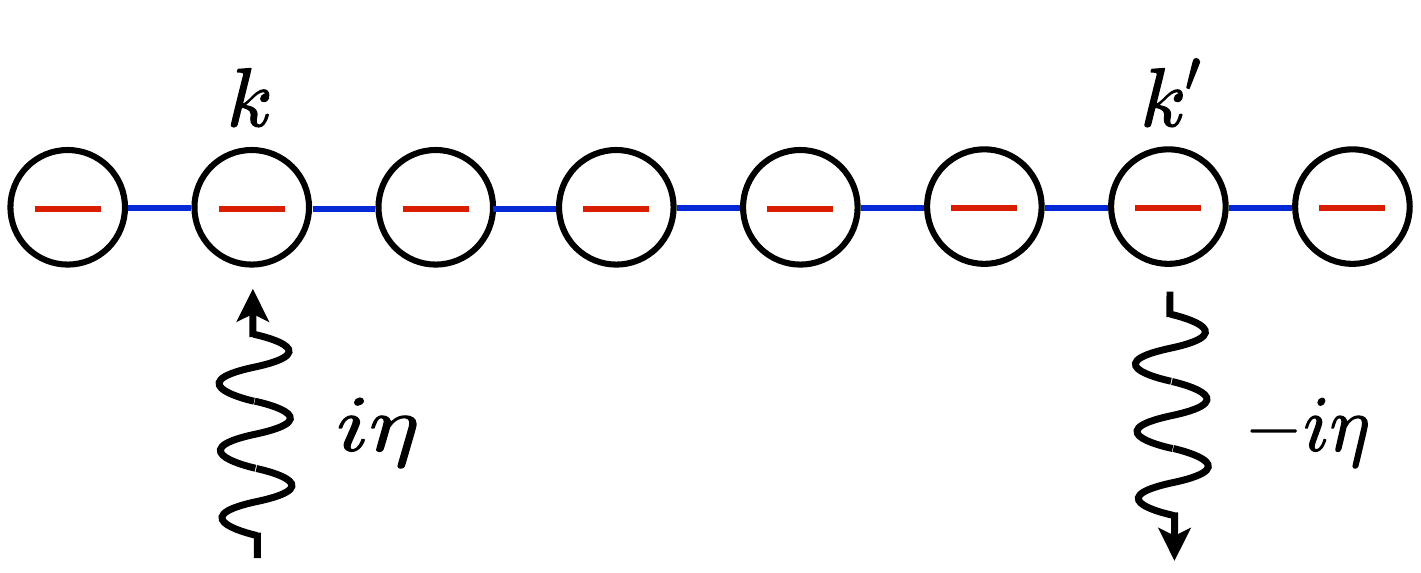}
  \caption{One-dimensional tight-binding chain, Eq. (\ref{eq:1}), with
  gain and loss in a $\mathcal{PT}$-symmetric configuration. We 
  illustrate the case $N=8$ with the contacts at sites $k=2$ and 
  $k'=N-k+1=7$; the strength of the coupling of the contacts is 
  characterized by $\eta$.}
  \label{fig:modeldiagram}
\end{figure}

In the site basis, a tight-binding chain in one dimension with gain and 
loss in a symmetric configuration (see Fig. \ref{fig:modeldiagram})
is described by the $\mathcal{PT}$-symmetric Hamiltonian

\begin{equation}
  H =  t \sum_{i=1}^{N-1} \Big(\ket{i}\bra{i+1} + \ket{i+1}\bra{i}\Big)
  + \I \eta \Big(\ket{k}\bra{k}  - \ket{N-k+1}\bra{N-k+1}\Big),
  \label{eq:1}
\end{equation}

where $t$ is the nearest-neighbor coupling, $N$ is the length of 
the chain and corresponds to the dimension of the Hilbert space of 
the system, $k$ is the position of the gain, $k'=N-k+1$ the position
of the loss, and $\eta$ is a real number that describes the strength 
of the gain and loss. Without loss of generality we fix $t=1$.

In what follows, we will denote $\ket{E_\theta}=\sum_{j=1}^N u_j(\theta)
\ket{j}$ the eigenstates of the Hamiltonian Eq. (\ref{eq:1}), corresponding
to the energy $E_\theta$, where $\theta$ is the pseudo momentum that
characterizes each state.

\section{Transport properties}
\label{sec:LocEigstates}

\subsection{The continuity equation}
\label{ssec:Continuity}

In this section we derive the continuity equation for the density in
the chain described by the effective Hamiltonian Eq.~(\ref{eq:1}). We illustrate the derivation for the case 
where the contacts are in the end-to-end configuration, i.e., at 
$k=1$ and $k'=N$, and write its generalization to other configurations.

Let $\ket{\Psi(t)}$ be a solution of the Schr{\"o}dinger equation. 
Then we have
\begin{equation}
    \frac{\partial\ket{\Psi(t)}}{\partial t} = \frac{H}{\I}\ket{\Psi(t)},
\end{equation}
and its adjoint
\begin{equation}
    \frac{\partial\bra{\Psi(t)}}{\partial t} = -\frac{1}{\I}\bra{\Psi(t)}H^\dagger,
  \label{eq:cs}
\end{equation}
where we have set $\hbar=1$. Using the site basis we write
\begin{equation}
  \ket{\Psi(t)} = \sum_{j=1}^N c_j (t) \ket{j},
  \label{eq:Psi}
\end{equation}
where the expansion coefficients are given by
\begin{equation}
  \begin{split}
    c_n(t) &= \langle n|\Psi(t)\rangle,\\
    c_n^*(t) &= \langle n|\Psi(t)\rangle^* = \langle\Psi(t)| n\rangle.
  \end{split}
\end{equation}
In order to derive the continuity equation, we consider the time derivative of the diagonal elements of the 
density matrix $\rho_{nn}(t) = \langle n|\Psi(t)\rangle \langle \Psi(t)|n\rangle$, obtaining
\begin{equation}
  \begin{split}
    \frac{\partial\rho_{nn}}{\partial t} &= \frac{1}{\I}\big( \langle n|H|\Psi(t)\rangle
    \langle\Psi(t)|n\rangle  - \langle n|\Psi(t)\rangle \langle \Psi(t)|H^\dagger|n\rangle\big) \\
    & = \frac{1}{\I}\sum_m \big( H_{nm} c_m(t) c_n^*(t) - H_{mn}^* c_n(t) c_m^*(t) \big).
  \end{split}
  \label{eq:Hs}
\end{equation}
By using the explicit form
of the Hamiltonian, Eq.~(\ref{eq:1}), assuming that the gain and loss are at the end points of the chain, we find
\begin{equation}
\frac{\partial\rho_{nn}}{\partial t}=\frac{1}{\I}\begin{cases}
                  c_{2}(t)c_1^*(t)- c_1(t) c_{2}^*(t)+ 2\I|c_1(t)|^2\eta, & n=1, \\
                  c_{N-1}(t)c_N^*(t) - c_N(t) c_{N-1}^*(t) - 2\I |c_N(t)|^2\eta, & n=N,\\
                  c_{n+1}(t)c_n^*(t) + c_{n-1}(t)c_n^*(t) - c_n(t) c_{n+1}^*(t) - c_n(t)
                    c_{n-1}^*(t), & \textrm{otherwise}
              \end{cases}
   \label{eq:continuity1}
\end{equation}
Equation~(\ref{eq:continuity1}) can succinctly be written as
\begin{equation}
\frac{\partial\rho_{nn}}{\partial t}+J_{n+1}-J_n= 2\eta
|c_1(t)|^2\delta_{n,1} - 2\eta|c_N(t)|^2\delta_{n,N},
\label{eq:continuity2}
\end{equation}
where we have introduced the local fluxes
\begin{equation}
J_{n}(t)  \equiv \I \big(c_n(t) c_{n-1}^*(t) - c_{n-1}(t) c_n^*(t)) 
= -2 \operatorname{Im}(c_{n}(t)c_{n-1}^*(t)), 
    \label{flux}
\end{equation}

which represent the density flux from site $n-1$
to site $n$ ($n\neq 1, N+1$), with the boundary conditions
$J_1= J_{N+1}= 0$.
Equation~(\ref{eq:continuity2}) is a continuity equation with
source and sink terms representing the 
inflow and outflow due to the presence of gain and loss in the 
chain. Its generalization to other gain and loss configurations
is straight forward and reads
\begin{equation}
\frac{\partial\rho_{nn}}{\partial t}+J_{n+1}-J_n= 2\eta
|c_k(t)|^2\delta_{n,k} - 2\eta|c_{N+1-k}(t)|^2\delta_{n,N+1-k}.
\label{eq:continuity_gen}
\end{equation}

It should be stressed that the inflow and outflow terms are,
as expected, proportional to $\eta$, which are the components
that make the Hamiltonian non-Hermitian.

\subsection{The transport coefficient}
\label{ssec:TransportCoeff}

Now consider $|\Psi(t)\rangle$ to be a time-dependent eigenstate
which we write in the site basis
\begin{equation}
\ket{E_\theta(t)}\equiv e^{-\I E_\theta t}\ket{E_\theta}=e^{-\I E_\theta t}\sum_{j=1}^N u_j(\theta) \ket{j}.
\end{equation}
For these states, $c_n(t)=e^{-\I E_\theta t}u_n(\theta)$.
When all eigenvalues are real, products of the form
$c_n(t)c^*_m(t)=u_n(\theta)u^*_m(\theta)$, as those appearing in 
the definition for the flux in Eq.~(\ref{flux}), are independent 
of time. 

In the broken $\mathcal{PT}$-symmetric phase the 
corresponding eigenvalues are complex, and terms of the form
$c_n(t)c^*_m(t)=e^{-2\operatorname{Im}(E_\theta) t} u_n(\theta)u^*_m(\theta)$
increase or decrease exponentially in time.
In view of this, we define
\begin{equation}
    \xi_{E_\theta}=\left\vert\frac{c_{k'}(t)}{c_k(t)}\right\vert^2,
    \label{transp_coeff}
\end{equation}
which corresponds to the
ratio of the outflow to the inflow in Eq.~(\ref{eq:continuity_gen}) with $k'=N-k+1$.
Evaluated in the states corresponding to the eigenfunctions
$\ket{E_\theta(t)}$, the transport coefficient is independent of 
time and of the normalization. If $\xi > 1$,
the inflow at $k'$ is larger than the outflow at $k$ and there is a
buildup of density within the chain. Conversely, if $\xi<1$ the outflow is
larger than the inflow and the system becomes depleted. When $\xi=1$, the
gain and loss are equally coupled, the inflow and outflow are the same,
and, in this sense, transport is efficient.

Now, given that $\ket{E_\theta}$ 
are eigenvectors of a $\mathcal{PT}$-invariant Hamiltonian, then
\begin{equation}
H\mathcal{PT}\ket{E_\theta} =E^*_\theta\,\mathcal{PT}\ket{E_\theta}.\label{PTket}
\end{equation}
Thus, since in the unbroken
$\mathcal{PT}$-symmetry phase the eigenvalues are real, 
the eigenfunctions fulfill
\begin{equation}
  \mathcal{PT}\ket{E_\theta} \propto \ket{E_\theta},
  \label{sym1}
\end{equation}
while in the broken $\mathcal{PT}$ symmetry phase,
some eigenvalues $E_\theta$  are
complex and come in conjugate pairs. The corresponding 
eigenstates satisfy
\begin{equation}
  \mathcal{PT}\ket{E_\theta} \propto \ket{E^*_\theta}.
  \label{eq:sym2}
\end{equation}

Consequently, in the unbroken $\mathcal{PT}$-symmetry
phase we have $\xi_{E_\theta}=1$. This indicates that
transport in the eigenstates with real eigenvalues 
is efficient if the gain and loss couple with such states.
On the other hand, for states in the 
$\mathcal{PT}$-broken symmetry
phase with complex eigenvalues, their eigenstates
localize around the gain and decouple from the loss
or viceversa; in this case $\xi_{E_\theta}$ is no 
longer equal to one, indicating that transport between
loss and gain is deficient. Yet, in view of
Eq.~(\ref{eq:sym2}), it is straight forward to see that
in this phase $\xi_{E_\theta}\xi_{E_\theta^*}= 1$.
As we shall see below, even if there are some states
with complex eigenvalues, others may still have real 
eigenvalues, and therefore efficient transport is still
possible. 
Further, under certain circumstances 
there may be states with real eigenvalues that are independent 
of $\eta$. These can be divided depending on whether (or not) 
their amplitudes vanish at the gain and loss positions.  If the 
amplitude vanishes at the position of the leads, the state does 
not couple to the gain and loss and there is no transport through 
this state in the system ($\xi_{E_\theta}$ is undefined). We refer 
to these as opaque states. If, on the other hand, the amplitude 
does not vanish at the leads, then $\xi_{E_\theta}=1$ for all 
values of $\eta$. We call these transparent states.

\section{Eigenvectors and eigenvalues}
\label{sec:egvalsegvecs}
To investigate the transport properties of the $\mathcal{PT}$ 
symmetric tight binding chain we require the spectra and eigenvectors
of the system. For this particular system, the eigenvalues have been
obtained using the Bethe ansatz \cite{YogeshPRA2010, YogeshPRA2011}.
Here, however, we obtain the eigenvalues and eigenvectors 
of the  Hamiltonian using symbolic calculus
\cite{Losonczi1992, Yueh2005, Chengbook2003};
see Appendix \ref{sec:egvalsol} for the full derivation. 
The eigenvalues are given by
\begin{equation}
  E_\theta = 2\cos{\theta},
  \label{eq:2}
\end{equation}
where the values of the pseudo momentum $\theta$ are those non-trivial
solutions ($\theta \neq m \pi$, with $m\in \mathbb{Z}$) that fulfill
the equation
\begin{equation}
\sin{(N+1)\theta} + 
\frac{\eta^2}{\sin^2\theta}\sin[(N-2k+1)\theta]\sin^2(k\theta) = 0,
  \label{eq:3}
\end{equation}
where the gain and loss are located at sites $k$ and $k'=N-k+1$,
respectively.

Writing the eigenvectors in the site basis
$\ket{E_\theta} = \sum_{j=1}^N u_j(\theta) \ket{j}$, as required 
above to calculate the fluxes and transport coefficients in the system, 
the $u_j(\theta)$ component of the eigenvector is given by
\begin{align}
  \begin{split}
    u_j(\theta) &= \langle j \ket{E_\theta} =
    \frac{u_1(\theta)}{\sin\theta}\Big[ \sin j\theta
     - \I\eta \Theta(j-k-1) \frac{\sin k\theta\,\sin(j-k)\theta}{\sin\theta} \\
    & + \Theta(j-N+k-2) \frac{\sin(j-N+k-1)\theta}{\sin\theta} \\
    & \qquad \times \Big( \I\eta \sin(N-k+1)\theta
      - \eta^2 \frac{\sin(N-2k+1)\theta\,\sin k\theta}{\sin\theta} \Big)
    \Big],
  \end{split}
  \label{eq:4}
\end{align}
where $\Theta(x)$ is the unit step function defined by $\Theta(x)=1$ 
if $x\geq0$
and $\Theta(x)=0$ otherwise. In Eq.~(\ref{eq:4}), $u_1(\theta)$ is the
first component of each eigenvector, which can be used to fix
the normalization.

Before presenting results for specific configurations of 
the system, we discuss some general properties that follow
directly from expressions (\ref{eq:3}) and (\ref{eq:4}).

First of all, clearly, in the limit $\eta\to0$, the
Hamiltonian becomes a symmetric (real Hermitian) 
matrix, actually a {\it centrosymmetric} matrix, and the 
results in \cite{CantoniLAA1976} hold. From Eq. (\ref{eq:3}) 
we obtain $\theta=\frac{r\pi}{N+1}$ ($r=1,2, \dots,N$), 
which using Eq.~(\ref{eq:2}) yields the well-known 
solution for the eigenvalues 
$E_{\theta, \eta=0}$~\cite{Losonczi1992, Yueh2005}. From the
centrosymmetry of $H$ it follows that the eigenvectors are 
symmetric or skew-symmetric with
respect to the exchange matrix $J$, i.e., they fulfill
$J \ket{E_{\theta,\eta=0}} = \pm  \ket{E_{\theta,\eta=0}}$.
It has been shown~\cite{OrtegaAdP2015, OrtegaPRE2016, OrtegaPRE2018} 
that centrosymmetry is relevant in achieving good transport properties
in disordered systems.

In the opposite limit, when $\eta\to\infty$, we expect the system 
to be divided into several subsystems depending on the positions 
$k$ and $k'$ of the gain and loss: two of them correspond to 
the uncoupled gain and loss, and the remaining ones to disjoint
tight-binding chains. From Eq.~(\ref{eq:3}), the real parts of
$\theta$ for the disjoint tight-binding chains are given by
$\theta=r\pi/(N-2k+1)$ for $r=1, 2, \dots N-2k$, and the 
double-roots $\theta = r\pi/k$ for $r = 1, \dots k-1$. All 
these $N-2$ eigenvalues have, in the limit $\eta\rightarrow\infty$, 
an imaginary part which is or tends asymptotically to zero. 
The asymptotic behavior of the two remaining eigenvalues, which 
are purely imaginary, can be obtained by writing $\theta=\I\phi$. 
Thus, Eq. (\ref{eq:3}) is transformed to
\begin{equation}
\sinh(N+1)\phi\,\sinh^2\phi + \eta^2\sinh(N-2k+1)\phi\,\sinh^2 k\phi = 0,
\end{equation}
which in the limit $\operatorname{Re}(\phi)\gg1$, reduces to
$e^{2\phi} \sim -\eta^2$. Taking the logarithm we obtain
$\phi \sim \log \eta  + \frac{\I\pi}{2} + \I\pi m$, $m\in\mathbb{Z}$
which, using Eq. (\ref{eq:2}), yields
\begin{equation}
  E_{\theta,\eta\rightarrow\infty} =
  E_{\I\phi,\eta\rightarrow\infty} \sim \pm \I\Big(\eta - \frac{1}{\eta}\Big).
\label{eq:5}
\end{equation}
We note that this equation holds independently of $k$, i.e., for 
any symmetric configuration of the gain and loss. 

We now discuss the conditions for the presence of opaque and transparent states, and the criteria to distinguish between them. Consider the pseudo momentum $\theta^\textrm{Op}_r = r \pi / M$, where both $r$ and $M$ are 
integers, $r=1,\dots,M-1$, and $M>1$ is a divisor 
of $N+1$ and $k$ simultaneously. It follows that $M$
divides $k$ and $k'=N-k+1$ as well. 
In this case, it is clear that $\theta^\textrm{Op}_r$ are 
solutions of Eq.~(\ref{eq:3}), {\it independently} of the 
value of $\eta$, and the corresponding eigenvalues are 
real. Also, using Eq.~(\ref{eq:4}), it is straight
forward to verify that the corresponding eigenvectors satisfy 
$u_k(\theta^\textrm{Op}_r) = u_{N-k+1}(\theta^\textrm{Op}_r)=0$.
Thus, the gain and loss are not coupled to these states and, 
as mentioned previously, these states are opaque. Notice, for example, that for
$k=1$ and $k'=N$, the end-to-end configuration, there 
is no such $M$ and there will be no opaque states. Whereas in configurations in which $k$ and $N+1$ are not relative primes, there will exist one or more integers $M$ that divide both $k$ 
and $N+1$, giving rise to opaque states in the system.
Similarly, for the transparent states we define  $\theta^\textrm{T}_r=r\pi/A$
for $r=1,\dots, A-1$, as those solutions of Eq.~(\ref{eq:3}) 
such that $A$ simultaneously divides $N+1$ and $N-2k+1=k'-k$
but does not divide $k$ (which then would fulfill the
definition of an opaque state). If $A$ divides simultaneously 
$N+1$ and $k'-k$, it also divides $2k$ and $2k'$. Arguing as above, 
the solutions $\theta^\textrm{T}_r$ are also independent from $\eta$ and real. Now the corresponding states are coupled to the gain and loss, transport is efficient through these states, and their eigenvalues are insensitive to the strength of the coupling. 

\subsection{Spectra and exceptional points for $N=10$ and $N=23$}
\label{sec:spectra}

In the following, we discuss the spectra for two specific chains of length 
$N=10$ and $N=23$, varying (symmetrically) the position of the contacts 
$k$ and $k'$. The choice of these values of $N$ is to
illustrate the case in which $N+1$ is a prime number ($N=10$), 
and no solutions $\theta^\textrm{Op}_r$ or $\theta^\textrm{T}_r$ 
exist for any position
of the contacts. In turn, when $N=23$, $N+1=24$ is a highly composite
number (i.e., it has more divisors than any smaller integer),
and we encounter the opposite situation.

\subsubsection{Results for $N=10$}
\label{sec:N10}

\begin{figure}
  \centering
  \includegraphics[scale=0.3]{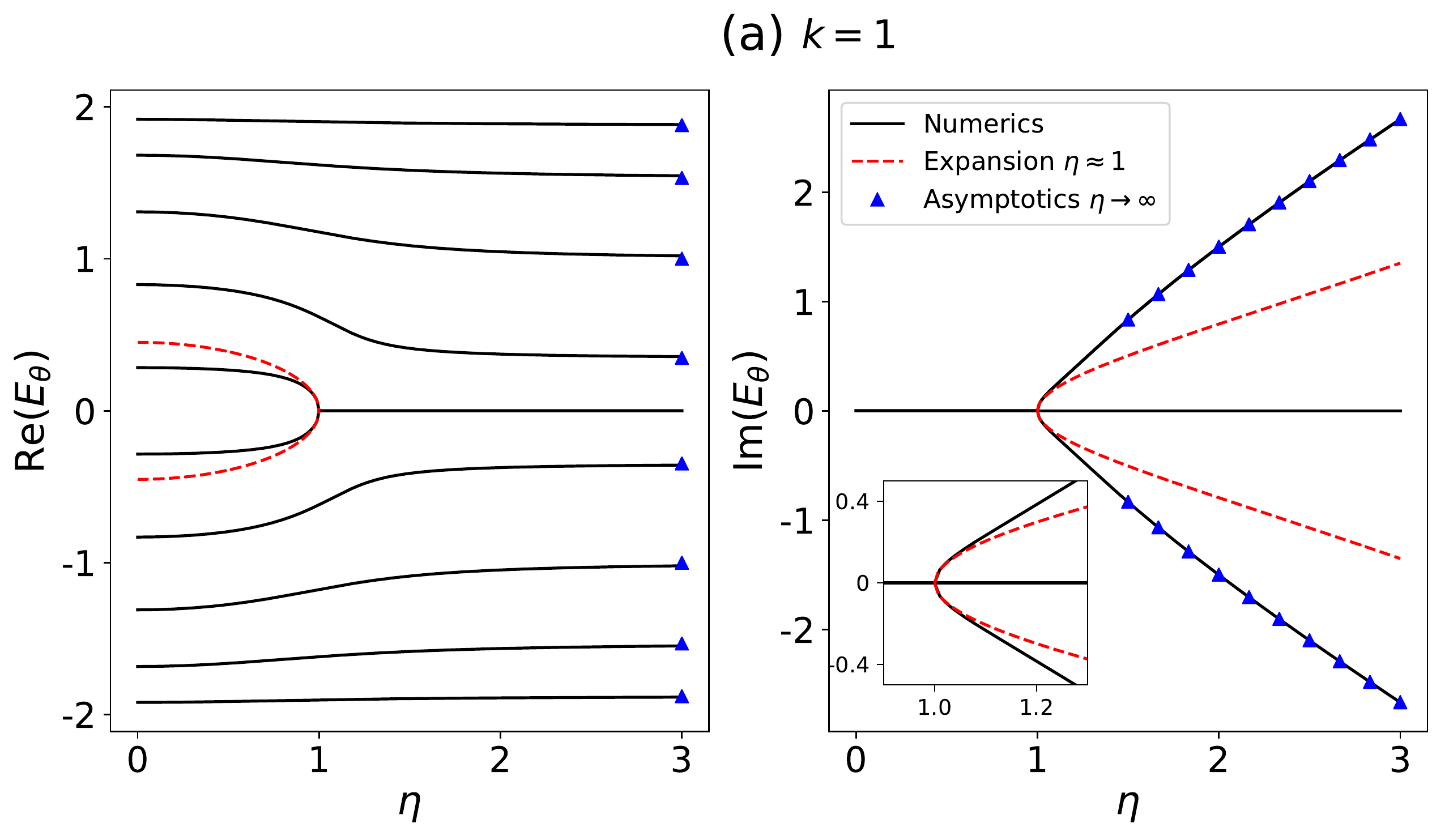}
  \includegraphics[scale=0.3]{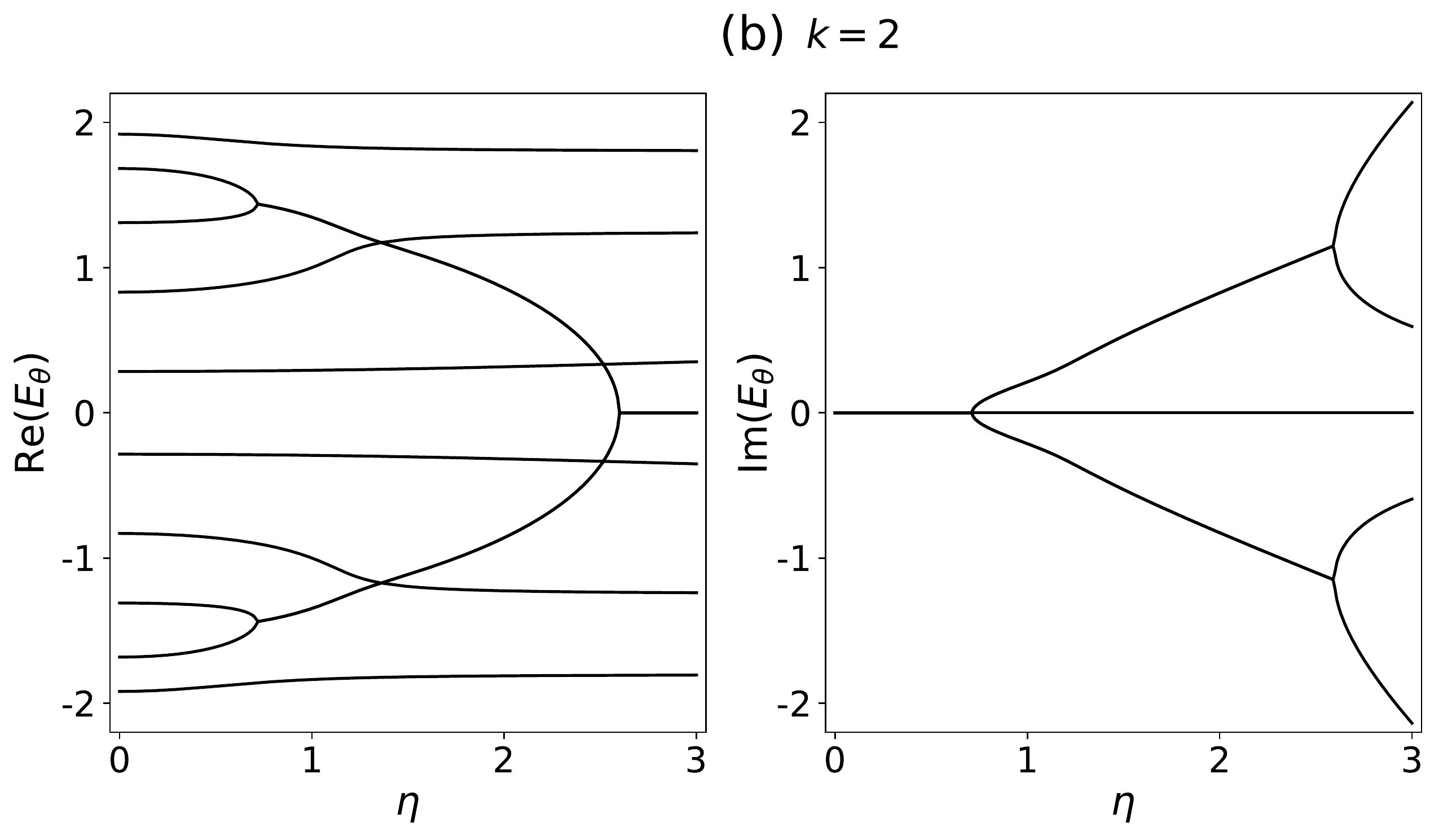}\\
  \includegraphics[scale=0.3]{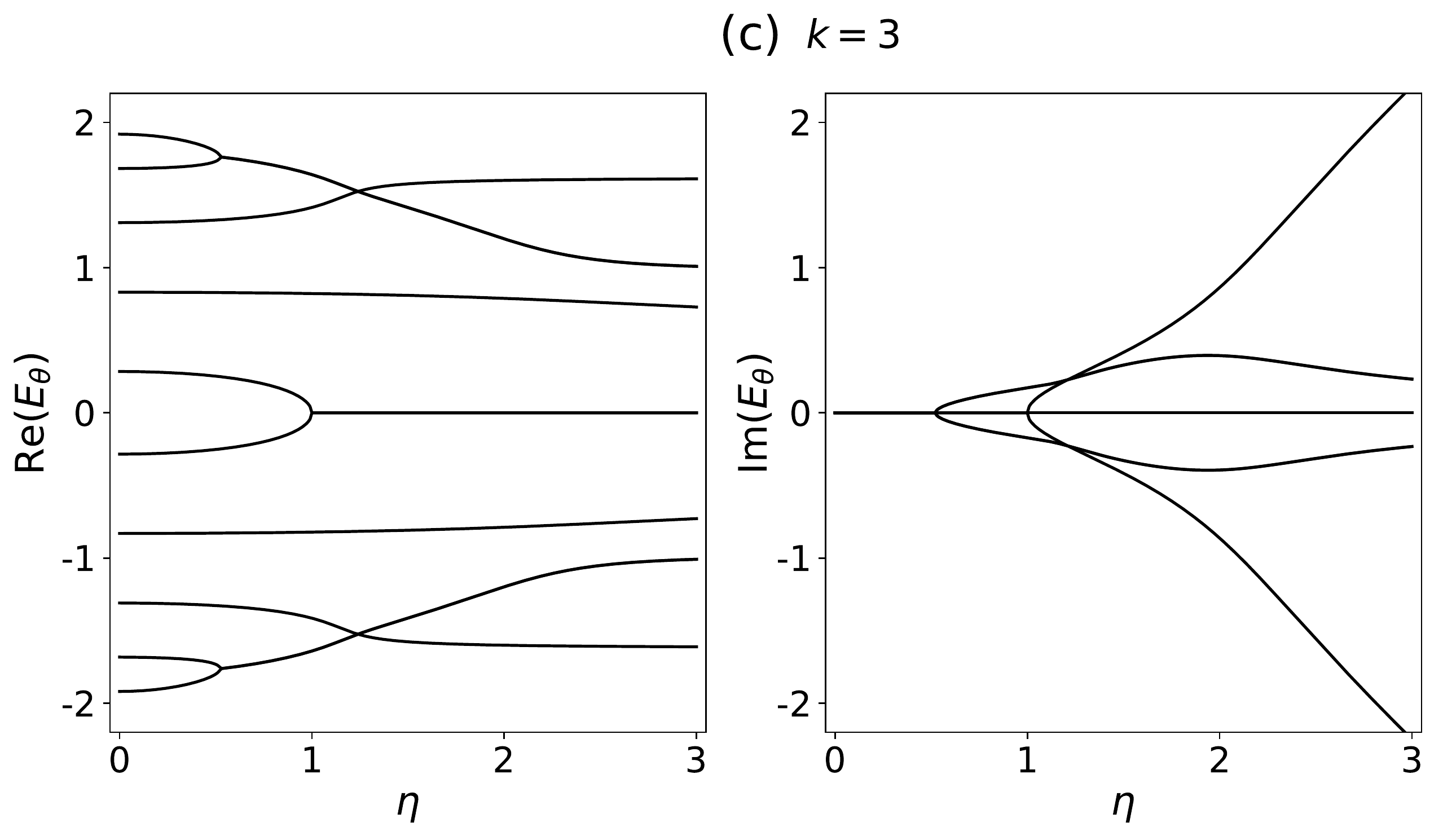}
  \includegraphics[scale=0.3]{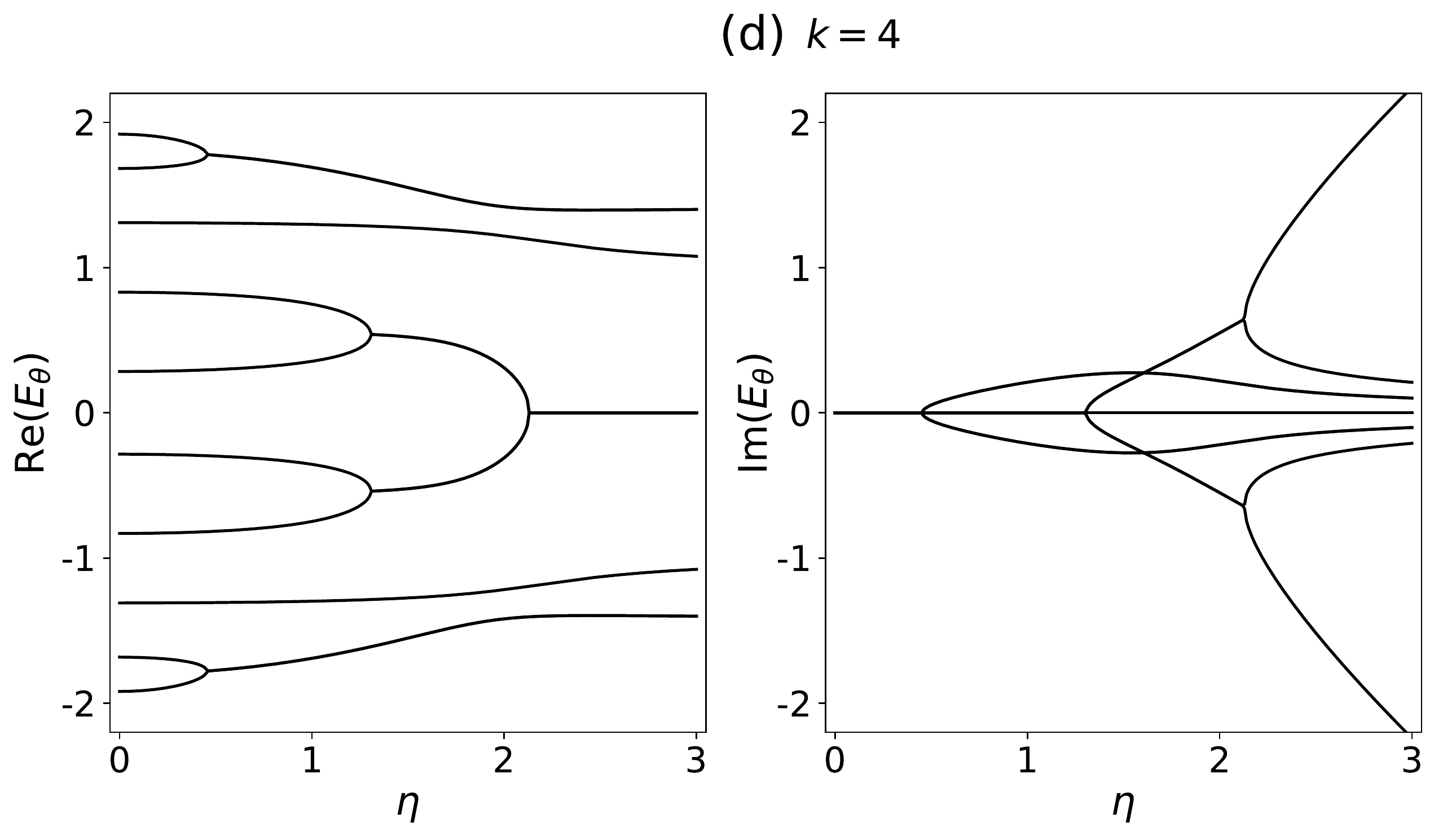}\\
  \includegraphics[scale=0.3]{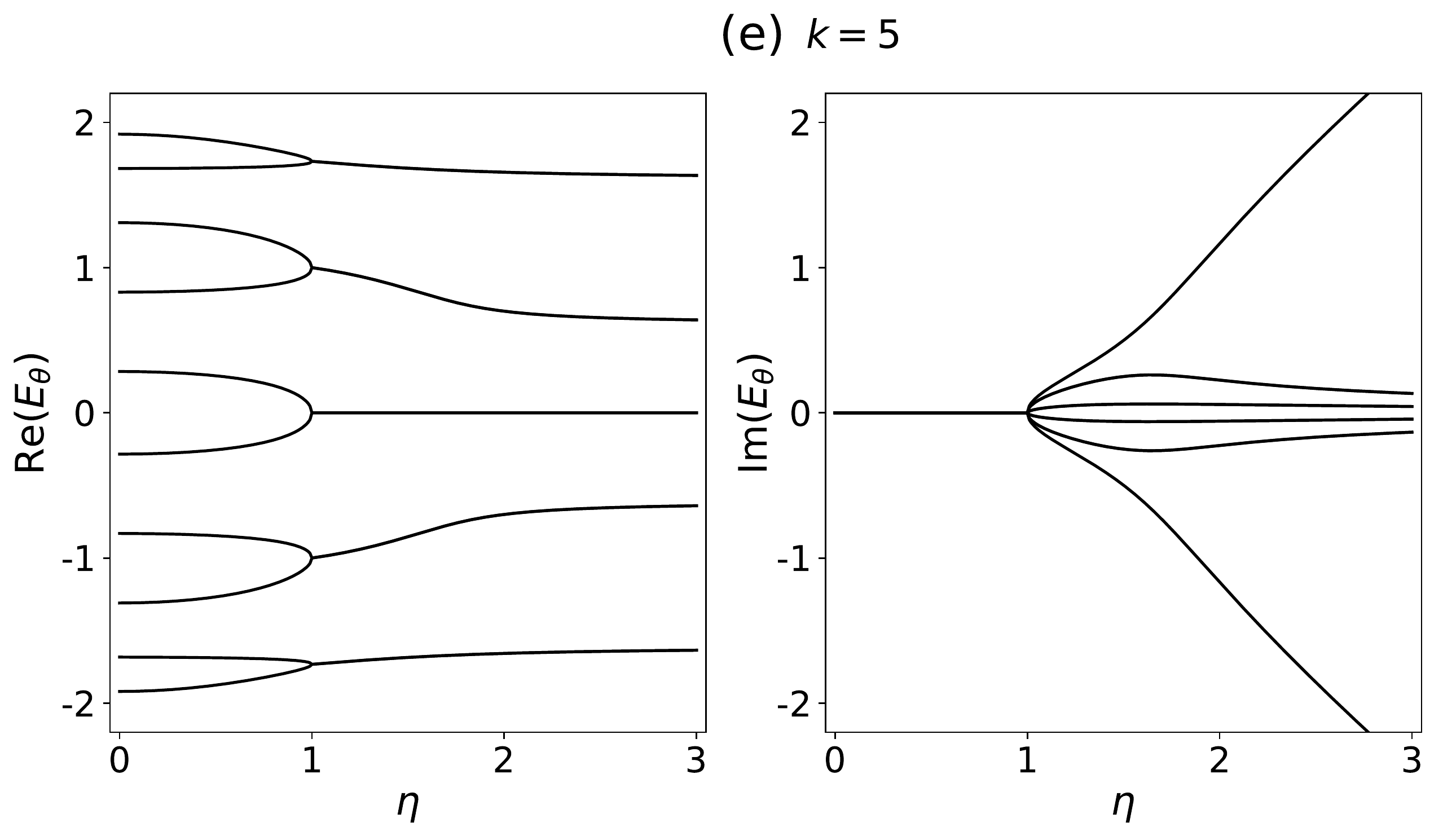}
  \caption{Real and imaginary parts of the 
    eigenvalues $E_\theta$ as a function of the 
    coupling parameter $\eta$ for $N=10$. The values of $k$ which
    define the specific configuration of the contacts, 
    are indicated in the figures.
    The dashed curves included in the end-to-end 
    configuration (a) show the approximation given by
    Eq.~(\ref{eq:Eplusmin}) 
    around $\eta\approx 1$, which is enlarged in the inset of
    the right panel. The triangles illustrate
    the asymptotic result for $\eta\to\infty$, Eq.~(\ref{eq:5}).}
  \label{fig:2}
\end{figure}

We begin with the spectra for $N=10$. 
Figure~\ref{fig:2} shows the real and imaginary part
of the spectra for all values of the contact positions $k$ and $k'$. As stated above, 
for this value of $N$ all eigenvalues depend on $\eta$, and
there are no opaque states. 

In Fig.~\ref{fig:2}(a) we show
the case $k=1$ and $k'=10$, where the contacts are in 
the configuration. In this case there
is only one exceptional point, located at $\eta_0=1$; 
the behavior
of the eigenvalues close to the exceptional point is detailed in 
Sect.~\ref{sec:perturb}. The blue triangles in this figure
correspond to the asymptotic results for $\eta\to\infty$; c.f.
Eq.~(\ref{eq:5}). Figure~\ref{fig:2}(b) displays the spectrum
for $k=2$ and $k'=9$; we note that there are two 
exceptional points for a value
of $\eta < 1$, signaled by the coalescence of two pairs of
eigenvalues and the appearance of two (doubly degenerate)
imaginary parts that branch out. In addition, there are
two (real) eigenvalue crossings
that do not correspond to exceptional points. The real 
part of the four
complex eigenvalues that emanated from the exceptional points
coalesce again at a value $\eta > 2$ and $\operatorname{Re} 
(E_\theta)=0$ thereafter. After this coalescence, two
of the imaginary parts tend to zero as $\eta$ increases, while the other
two tend to infinity according to Eq.~(\ref{eq:5}).

As we move inwards the position of $k$ and $k'$, richer
behavior of the eigenvalues is observed, with more 
occurrences of exceptional
points, some of them again involving coalescences of complex
eigenvalues as well as some crossing of eigenvalues 
which do not represent exceptional points. Interestingly,
for $k=5$ and $k'=6$, when the contacts are at the center of
the chain, there are 5 distinct coalescences leading to
exceptional points, all appearing at the same value $\eta=1$.

\subsubsection{Results for $N=23$}
\label{sec:N23}

\begin{figure}
  \centering
  \includegraphics[scale=0.25]{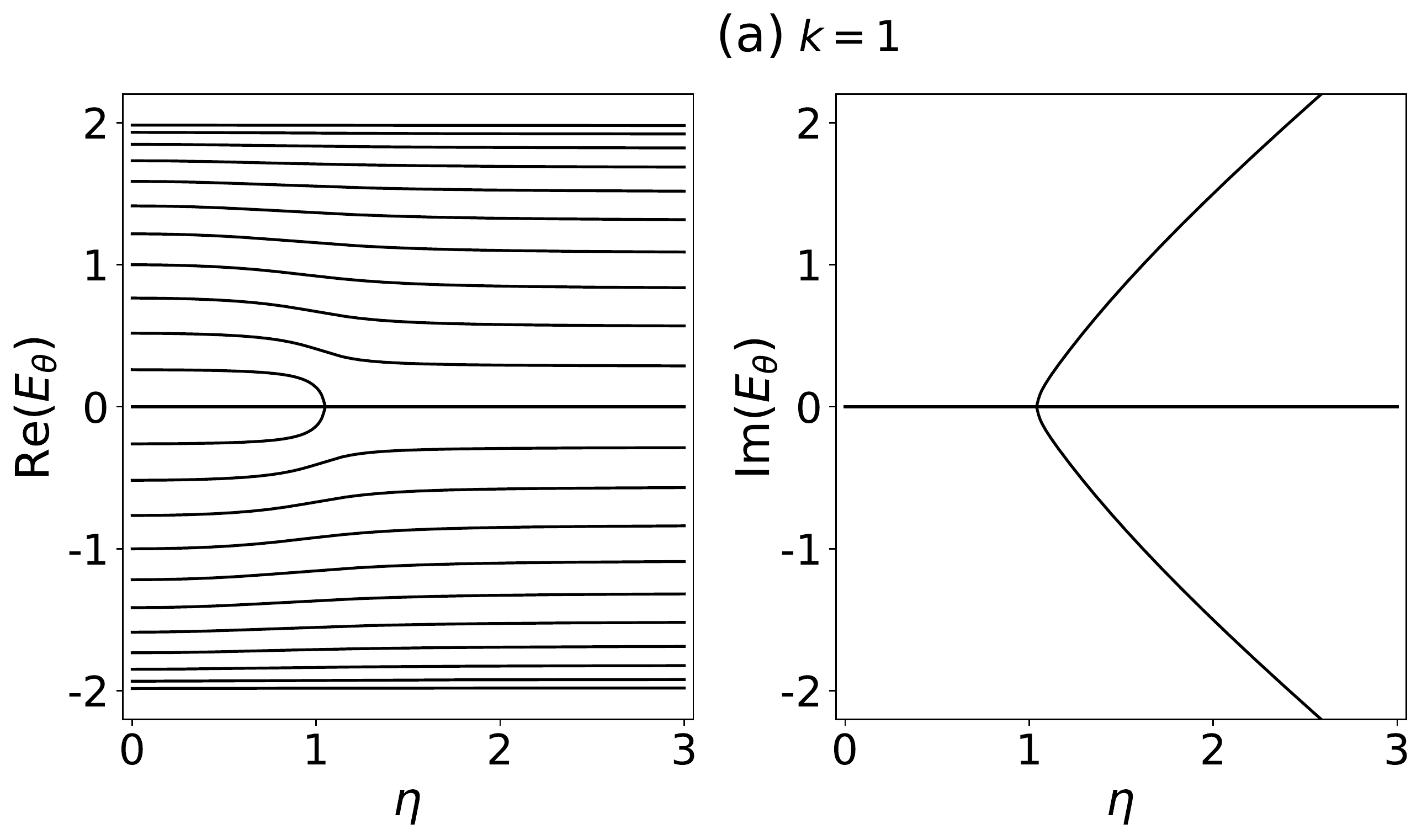}
  \includegraphics[scale=0.25]{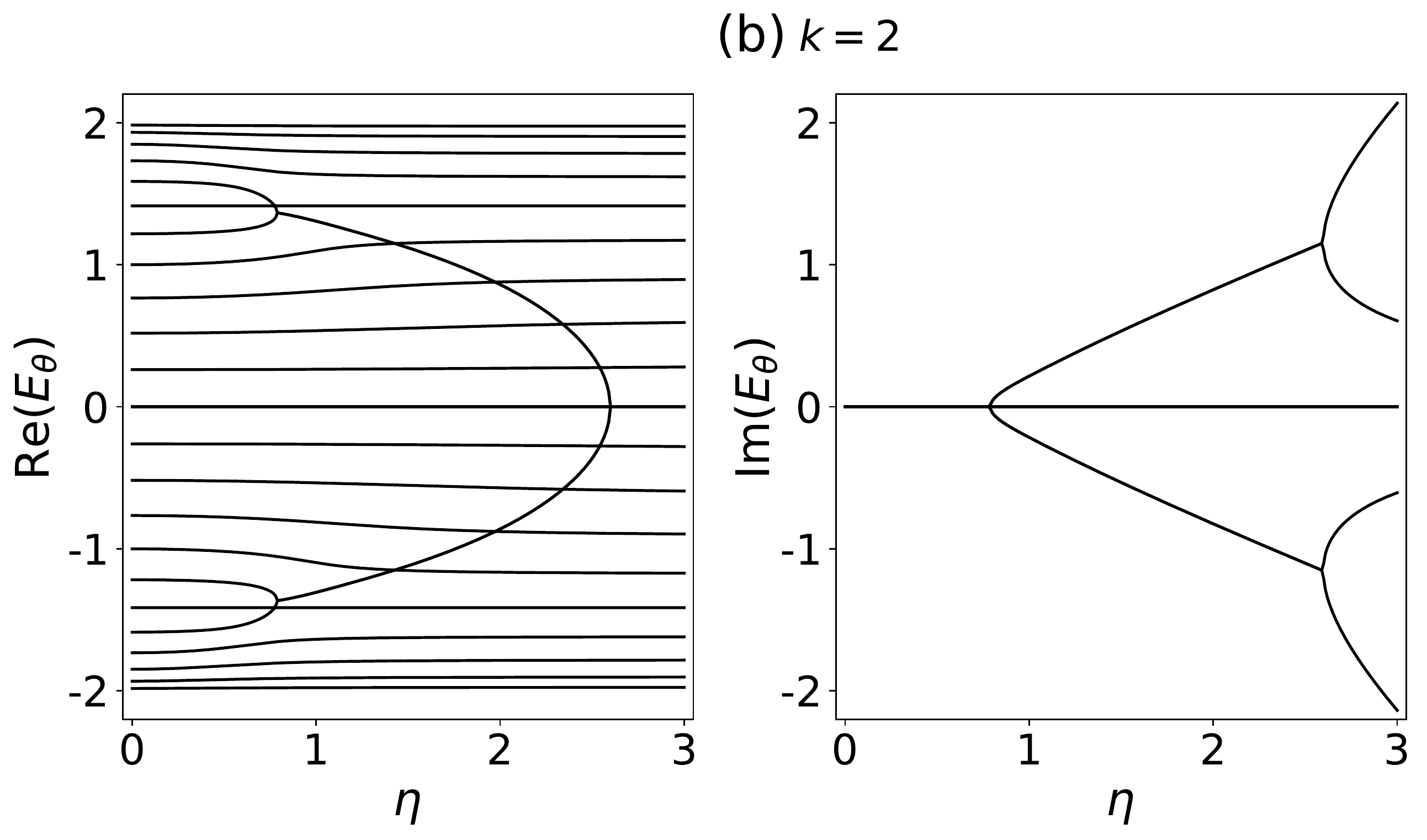}\\
  \includegraphics[scale=0.25]{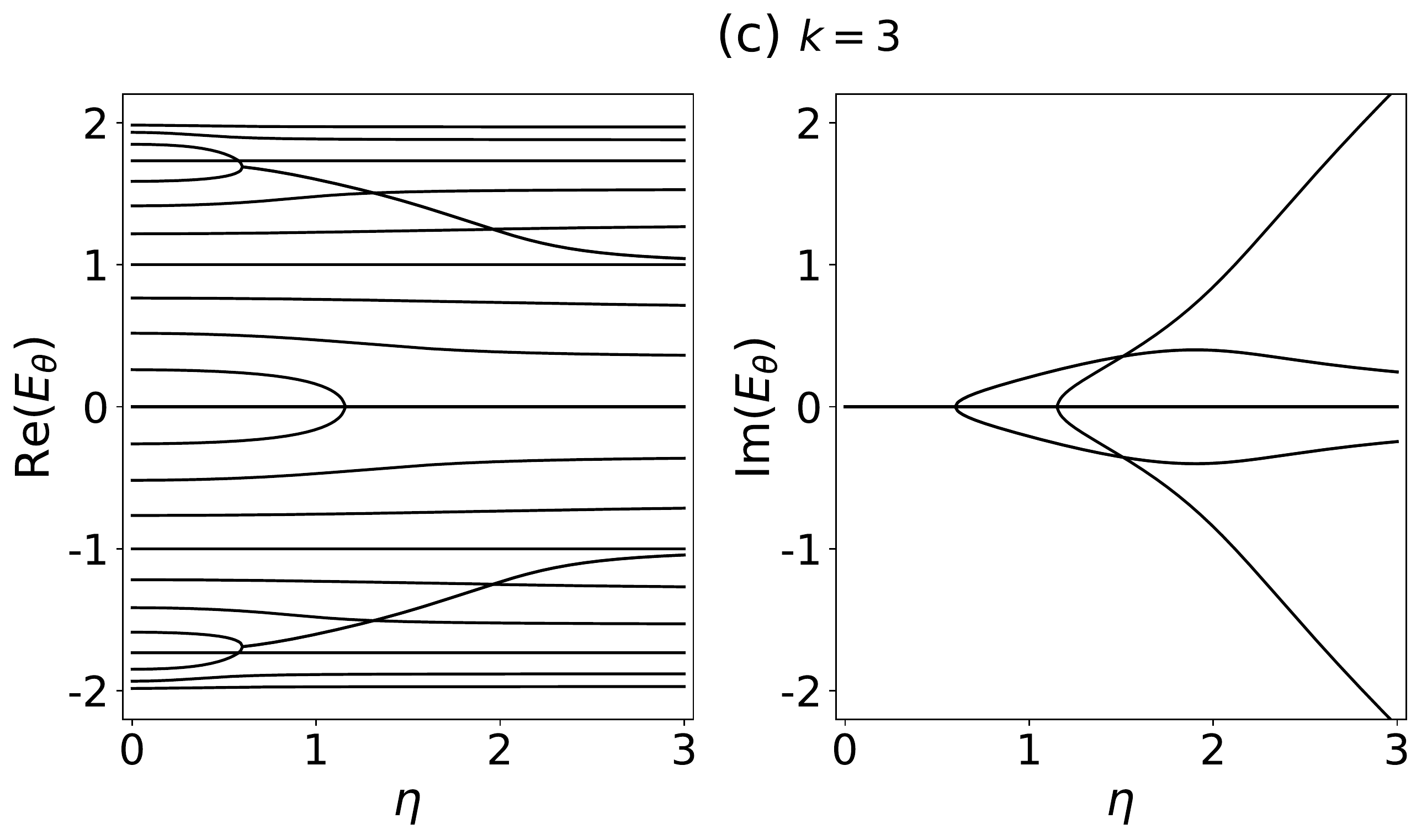}
  \includegraphics[scale=0.25]{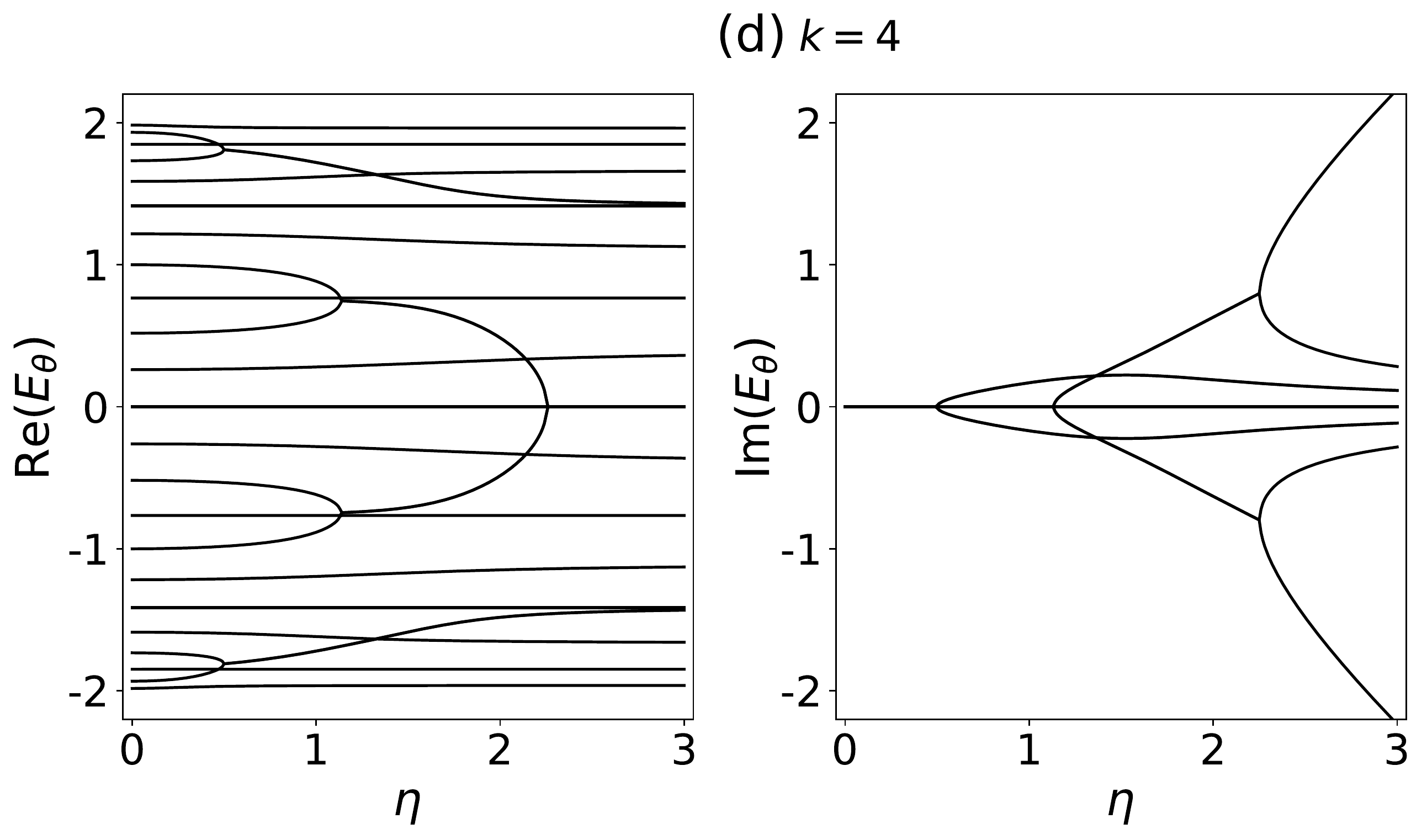}\\
  \includegraphics[scale=0.25]{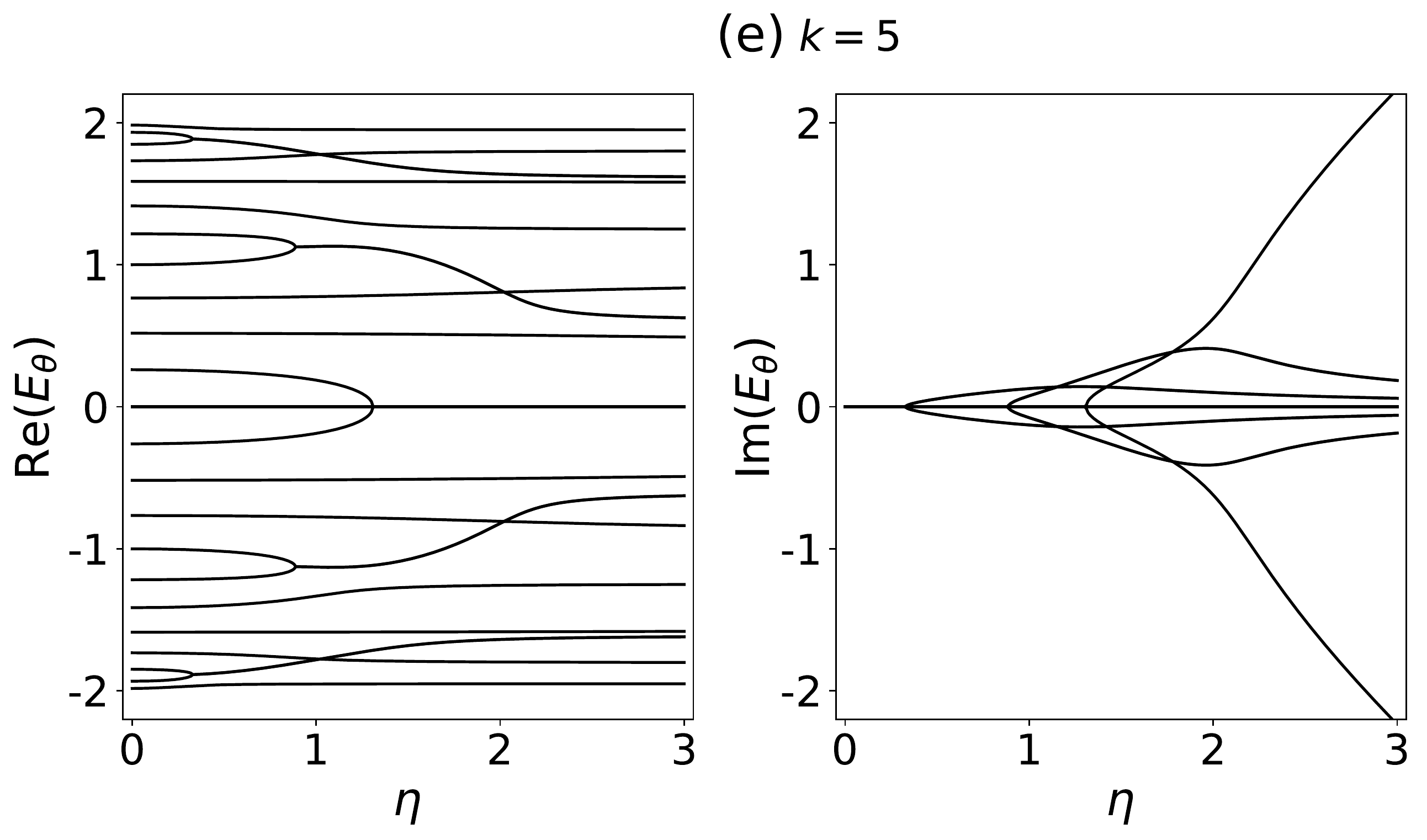}
  \includegraphics[scale=0.25]{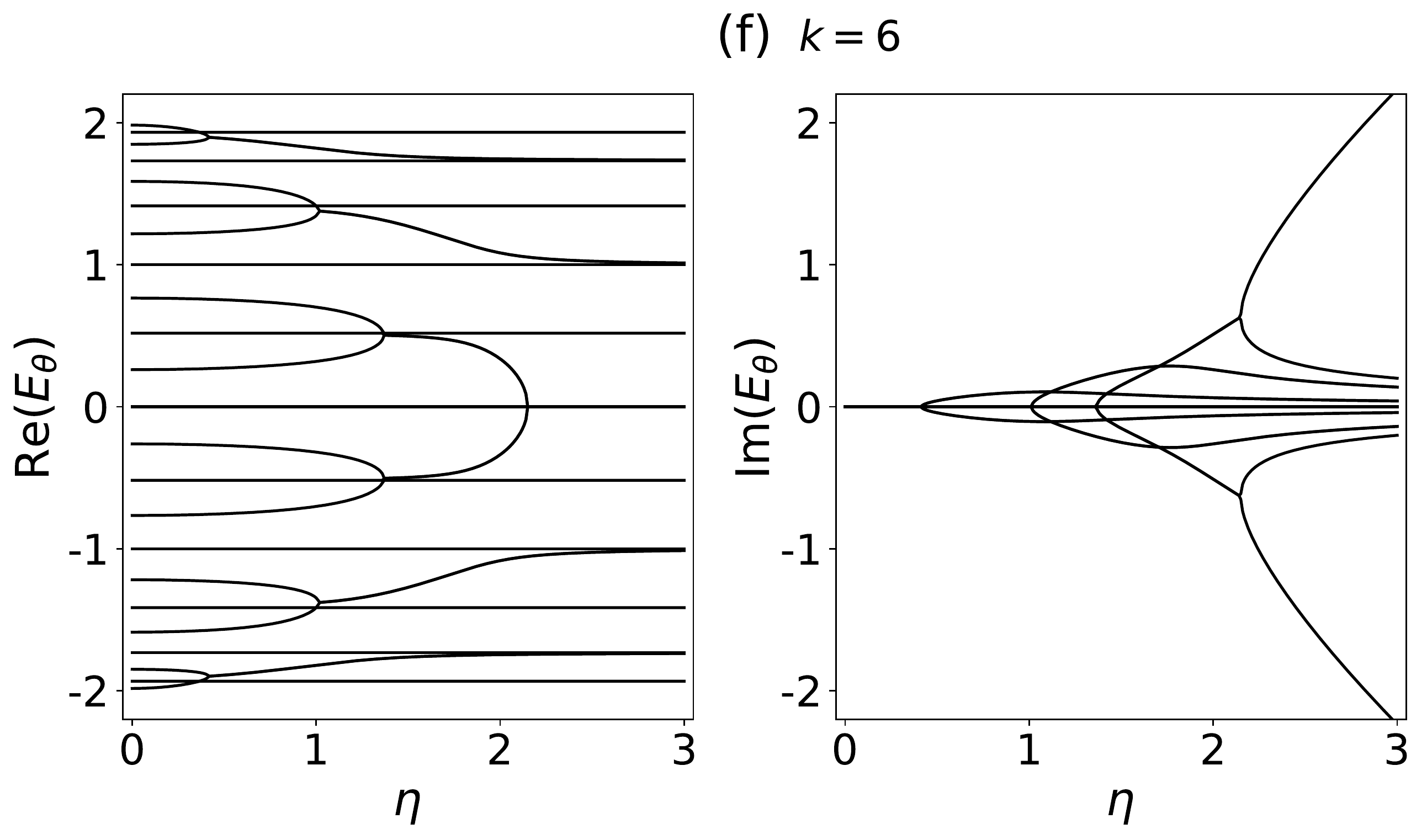}\\
  \includegraphics[scale=0.25]{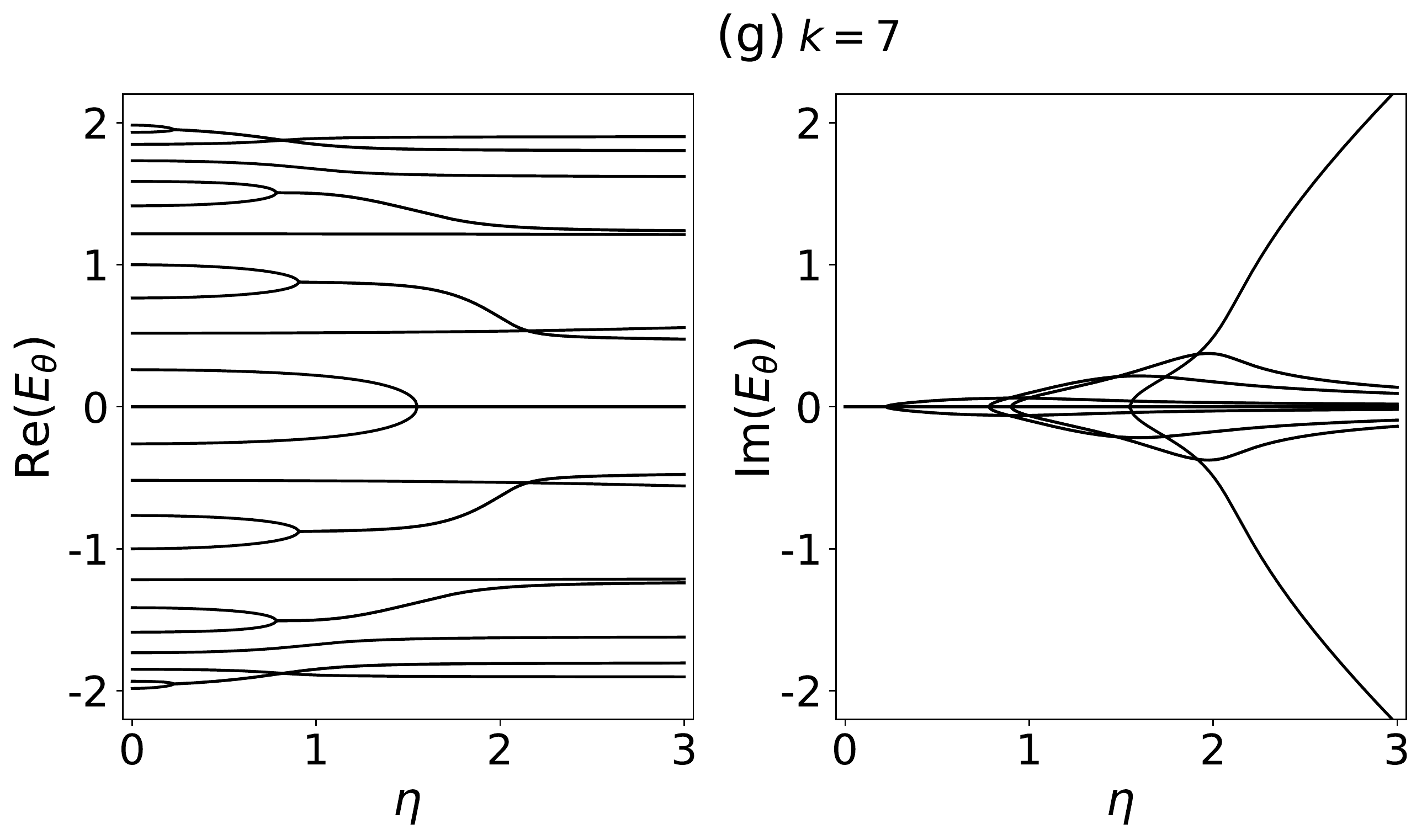}
  \includegraphics[scale=0.25]{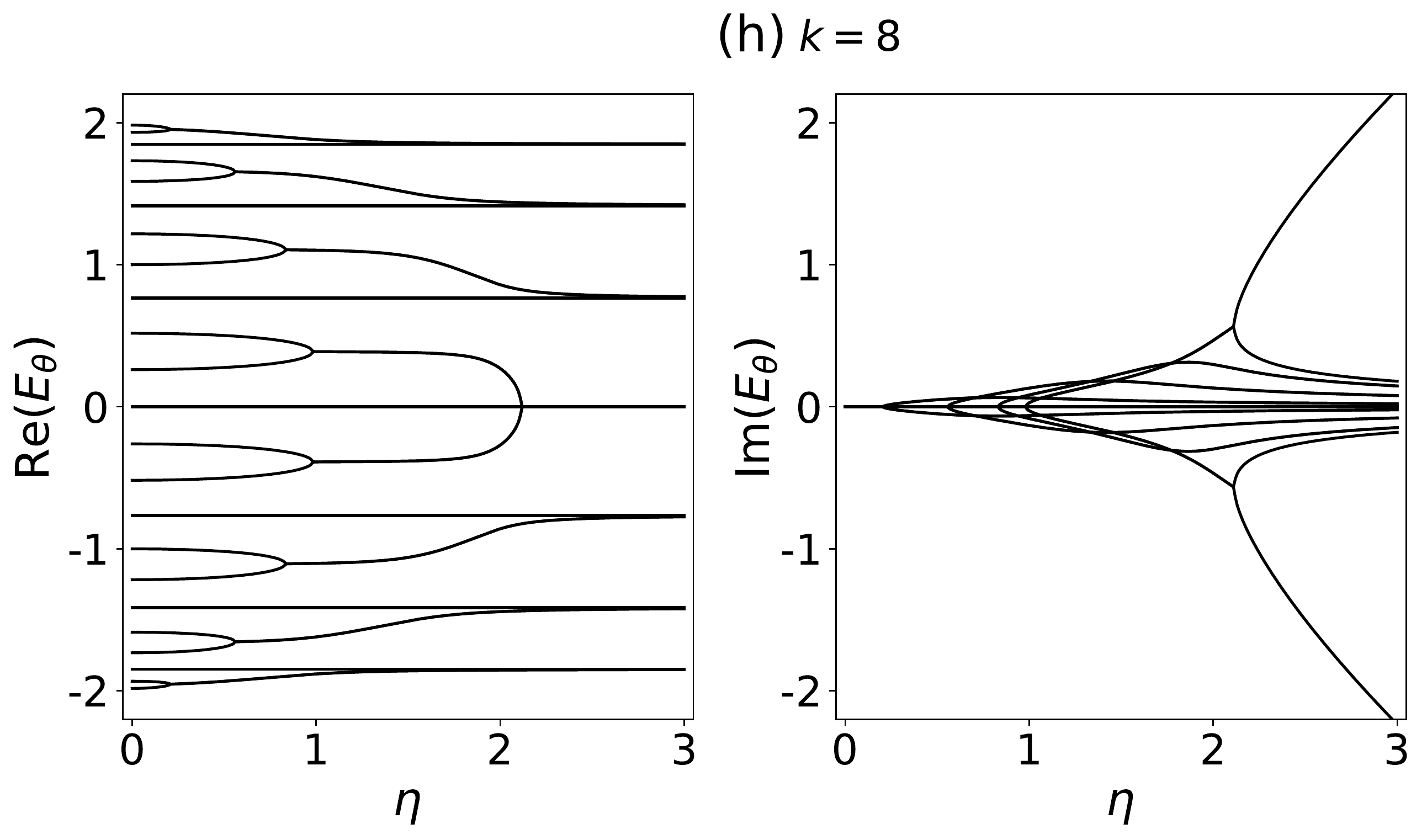}\\
  \includegraphics[scale=0.25]{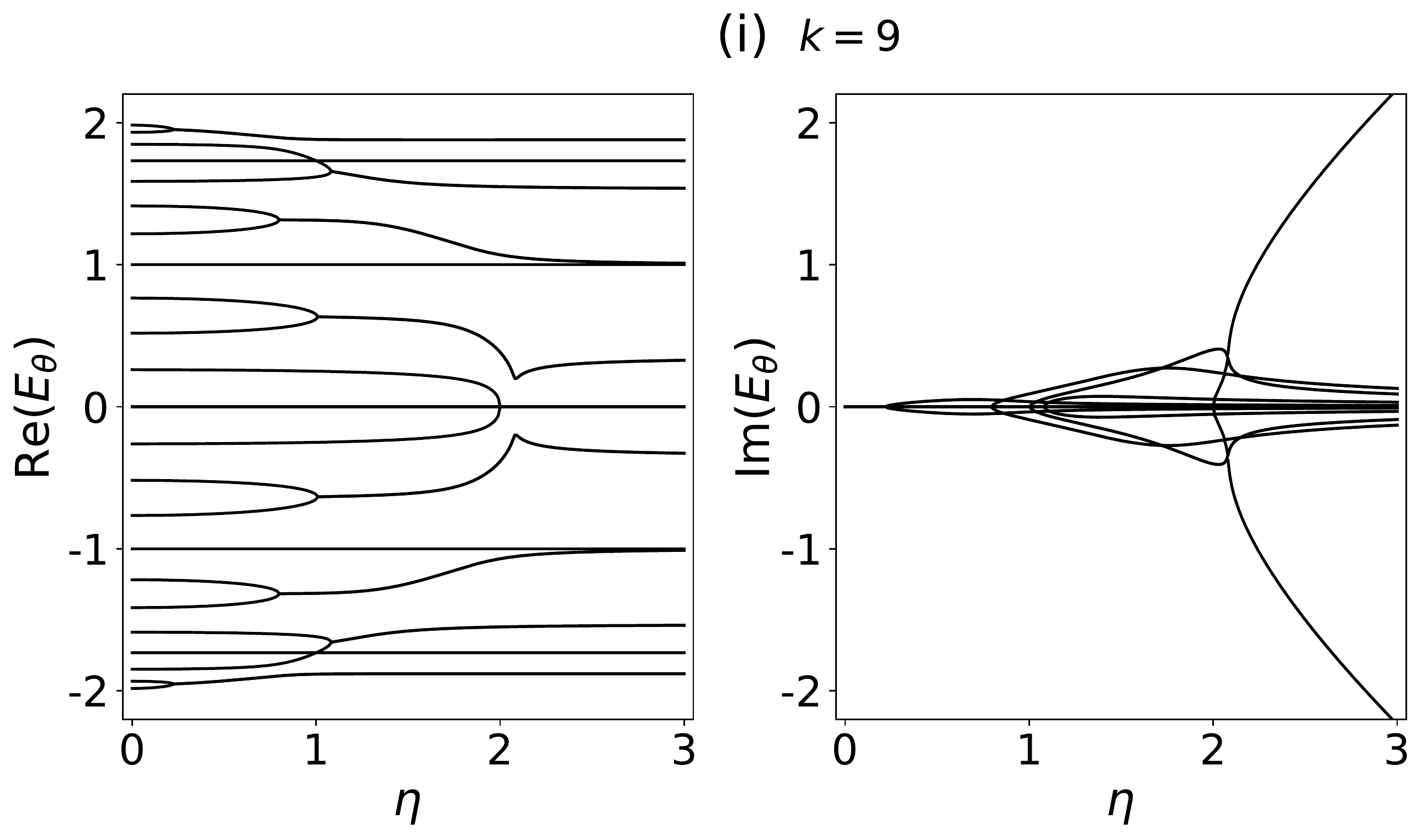}
  \includegraphics[scale=0.25]{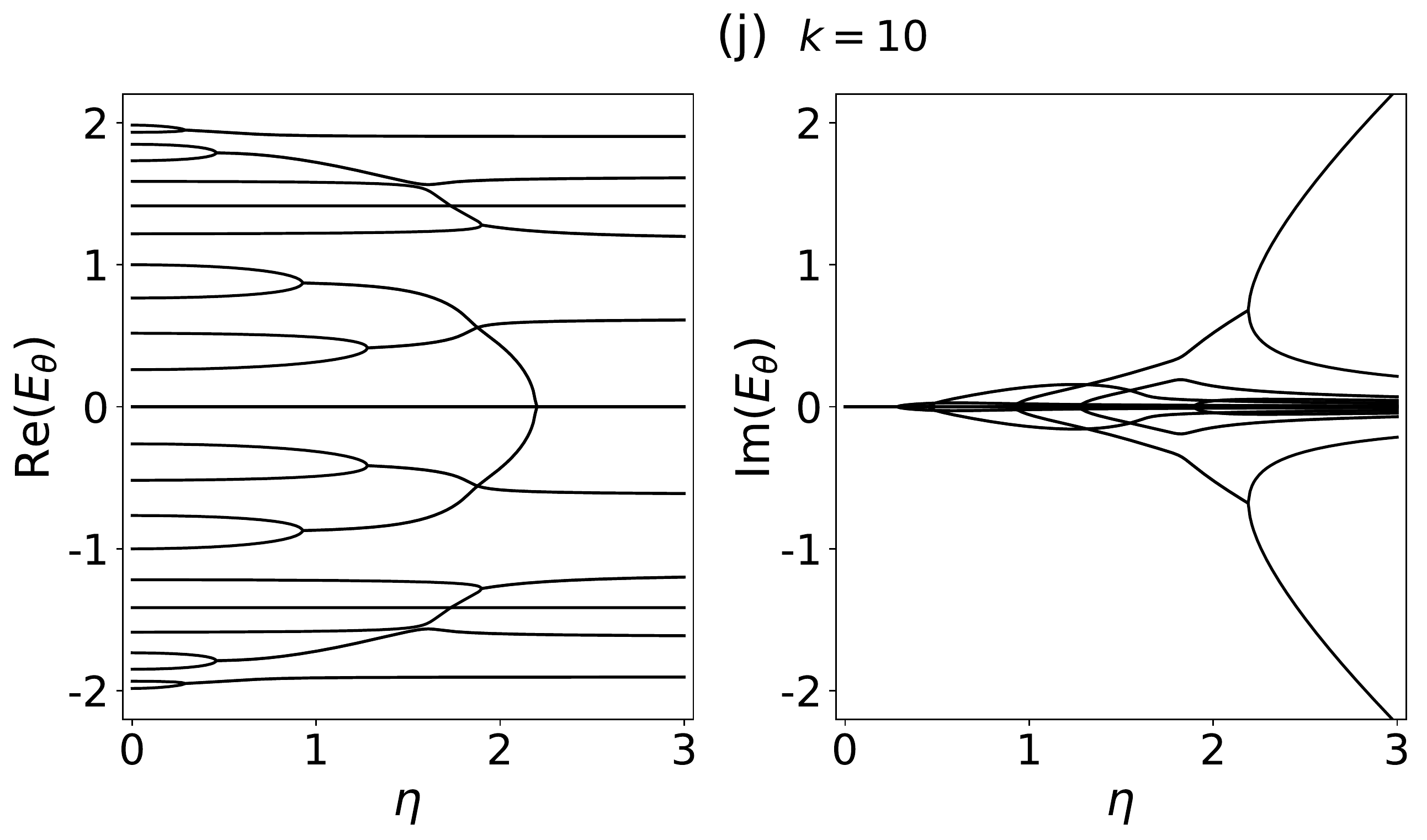}\\
  \includegraphics[scale=0.25]{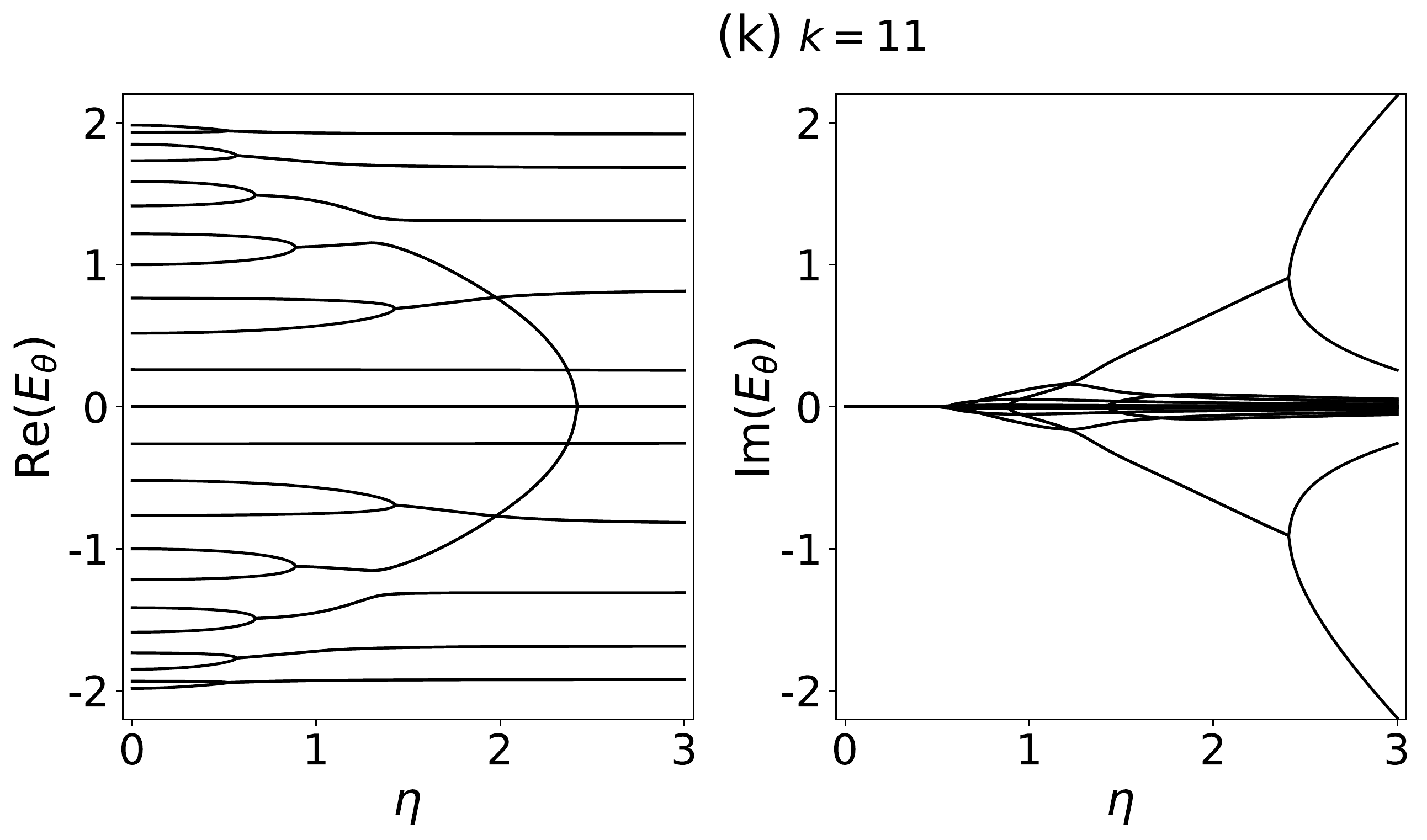}
  \caption{Real and imaginary parts of the 
    eigenvalues $E_\theta$ as a function of the 
    coupling parameter $\eta$ for all 
    configurations of $N=23$.}
  \label{fig:3}
\end{figure}

We now consider the case $N=23$. As mentioned above, $N+1=24$ can be 
divided by more integers than any smaller integer; in this case, it can 
be divided by $M,A=2, 3, 4, 6, 8, 12$. The spectra for this value 
of $N$ are shown in Fig.~\ref{fig:3}. In contrast to
Fig.~\ref{fig:2}, we observe that $E_{\theta_0}=0$
is an eigenvalue of the Hamiltonian, independently of the
location of the contacts and the value of $\eta$. This is a 
consequence of the fact that $\theta_0=\pi/2$ always 
satisfies Eq.~(\ref{eq:3}) for odd $N$. As we shall
see later, this state is either an opaque or a transparent state,
depending if $k$ is even or odd, respectively.

Figure~\ref{fig:3}(a) shows that the end-to-end configuration 
($k=1$ and $k'=23$) exhibits one exceptional point 
at $\eta=1$. The remaining eigenvalues have 
a smooth dependence on $\eta$. In this configuration, 
$E_{\theta_0}=0$, being independent of $\eta$, is the
only transparent state, because $k$ is odd and $A=2$
divides $N+1=24$ and $2k$ simultaneously. It is easy
to see from Eq.~(\ref{eq:4}) that $u_1(\pi/2)$ and 
$u_N(\pi/2)$ are both non zero.

Figure~\ref{fig:3}(b) shows the case $k=2$ and $k'=22$. In this
case there are various eigenvalue crossings, and two values
of $\eta$ where the coalescence corresponds to (pairs of)
exceptional 
points; note that the second one involves a coalescence
of the real part of four complex eigenvalues. In this case, 
the only integer 
$M>1$ that divides simultaneously $k=2$ and $N+1=24$ is $M=2$.
Therefore, $\theta^\textrm{Op}_{r=1} = \pi/2$ satisfies 
Eq.~(\ref{eq:3}) and
$u_2(\theta^\textrm{Op}_{r=1}) = u_{22}(\theta^\textrm{Op}_{r=1})=0$ 
corresponding to the only opaque state of this case. Similarly,
$A=2,4$ divides simultaneously $N+1$ and $2k$; the states with $A=2$
are opaque states and therefore we have two transparent states
$\theta^\textrm{T}_{r=1}=\pi/4$ and $\theta^\textrm{T}_{r=3}=3\pi/4$.

The different panels in Fig.~\ref{fig:3} illustrate how
the spectra become more complex in terms of
eigenvalue crossings, exceptional points, 
and opaque or transparent states, as the contacts are moved.
In some cases, one can also observe
avoided level crossings; see Fig.~\ref{fig:3}(j) or~(k).

\begin{figure}[t]
   \hspace*{-12mm}
  \includegraphics[scale=0.7]{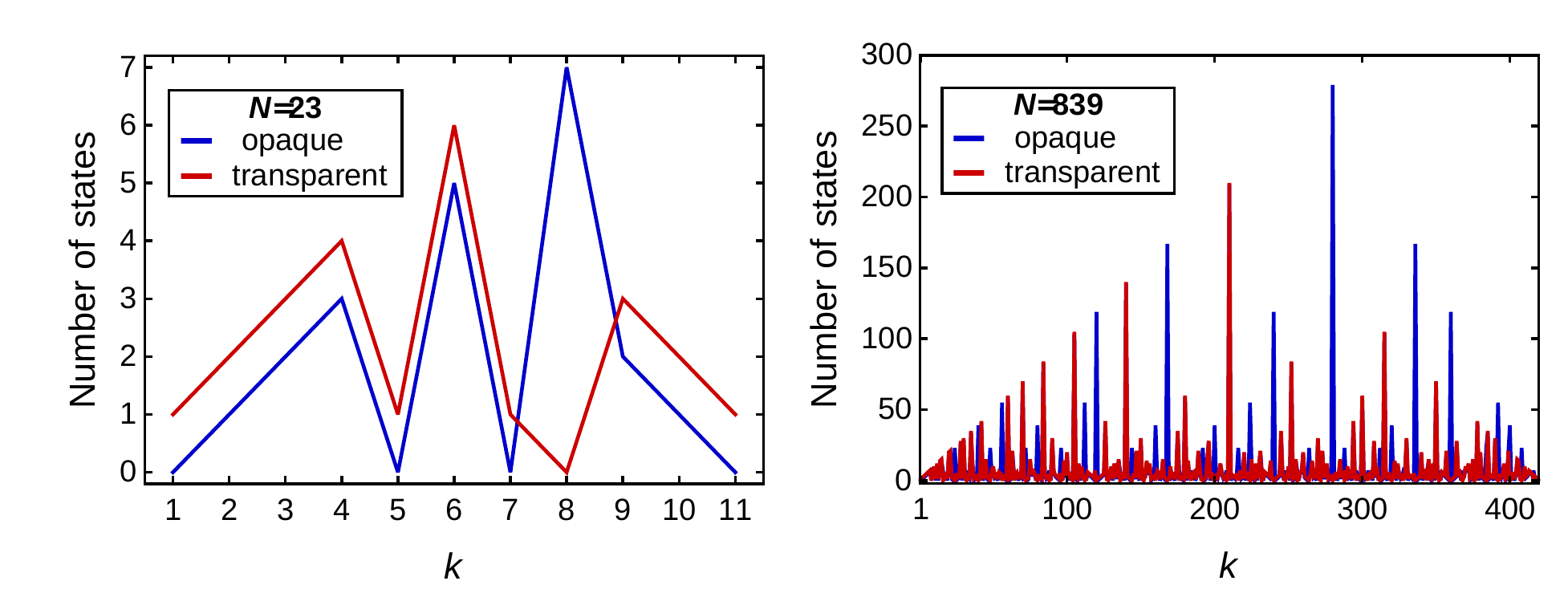}
  \caption{Number of opaque and transparent eigenstates as a function of the contact 
  positions $k$ for two examples of small (left) and large (right) $N$.}
  \label{fig:4}
\end{figure}

We compute the number of opaque and transparent states for $N=23$
for some specific configurations of the leads;
Fig.~\ref{fig:4} shows the complete picture.
For this $N$, the divisors of $N+1$ are $2,3,4,6,8,12$; these are 
all the possible values $M$ and $A$ we may have. As a first example 
we consider $k=8$, and $M=2,4,8$ divide both $N+1$ and $k$; the 
spectrum of this configuration is illustrated in Fig.~\ref{fig:3}(h). 
Since 8 is a multiple of 2 and 4 it suffices to consider $M^*=8$. 
The opaque states are then $\theta^\textrm{Op}_r=r\pi/M^*$ for 
$r=1, \dots M^*-1$, and we have $7$ opaque states for this 
configuration. Likewise, $A=2,4,8$ divide both $N+1$ and $2k=16$, 
but since $A$ also divides $k$, there are no transparent 
states in this configuration.

Consider now the configuration $k=6$ as a second example. In
this case we have that $M=2,3,6$ divide both $N+1$ and $k$,
and, using the same arguments as for $k=8$, we conclude that 
the opaque states are $\theta^\textrm{Op}_r=r\pi/6$,
for $r=1,\dots 5$. With 
regards to the number of transparent states, $A=2,3,4,6,12$ divide 
simultaneously $N+1$ and $2k$. In this case, $A^*=12$ is a multiple
of the remaining values, and it suffices to consider it. From
the 11 states arising from $A^*$ we subtract the 5 opaque states,
finally obtaining that there are 6 transparent states in this
configuration, $\theta^\textrm{T}_r=r\pi/12$ for $r=1,3,5,7,9,11$.

The left panel of Fig.~\ref{fig:4} shows the number of opaque and transparent states in 
terms of $k$ for $N=23$. It is worth stressing that the number of these states 
is a very irregular function of both the system size $N$ and of the position 
of the leads. Indeed, even for large system sizes, this number depends on the 
divisibility properties of $N+1$ and of $k$. For example, a system of length $N=839$ shows 
up to 279 opaque states and no transparent states when $k=280$; there 
are 210 transparent states and 209 opaque states when
$k=210$. The right panel of Fig.~\ref{fig:4} shows the number of opaque 
and transparent states for all values of $k$ for a system of size $N=839$. 
However, a system of size $N=838$ will have neither opaque nor transparent 
states independently of where the leads are placed, because $N+1=839$ is a 
prime number.

\subsection{Perturbation theory around an exceptional point}
\label{sec:perturb}

Before we turn to the discussion of the transport coefficient in this system, we
address the behavior of the eigenstates close to the exceptional
points. In order to simplify the discussion, we shall consider the case of $N=10$ 
with contacts in the end-to-end configuration. In this case, there is only
one exceptional point which occurs at $\eta_0=1$ (see
Fig.~\ref{fig:2}(a)). The eigenvalues are given by
(\ref{eq:2}), where $\theta$ is determined by
(cf. Eq.~(\ref{eq:3}))
\begin{equation}
  \sin(N+1)\theta + \eta^2 \sin(N-1)\theta = 0.
  \label{eq:exconf}
\end{equation}

To calculate analytically the behavior of the eigenvalues
around an exceptional point, we use a simple perturbation
scheme~\cite{Benderbook1978}. We write Eq.~(\ref{eq:exconf})
generically as $F(\theta, \eta^2)=0$, from which we determine 
the values of $\theta$ given the strength of the coupling
constant $\eta$. In the present case, the equation defining 
$\theta$ for $\eta=\eta_0=1$ can be rewritten as
\begin{equation}
 F(\theta,\eta_0^2)= 2\sin N\theta\cos\theta = 0,
\end{equation}
with the solutions $\theta_r=\frac{r\pi}{N}$ ($r=1,2,\dots,N-1$) 
and $\theta_0 = \pi/2$. Since in this case $N$ is even, the 
root $\theta_{N/2}$ is identical to $\theta_0$, hence at
$\eta=\eta_0=1$ these roots coalesce.

If we follow the usual perturbation scheme, we write
$\eta^2=\eta_0^2+\epsilon$, propose a solution of the form
$\theta=\theta_0+\epsilon\theta_{(1)}+\dots$ in powers of 
$\epsilon$, and solve $F(\theta, \eta^2)=0$, which is also 
written as a series expansion in $\epsilon$. Each term 
of that series must be equal to zero, which is used to 
obtain $\theta_{(1)}$. However, this procedure breaks 
down when $\theta_0$ and $\eta_0$ define an exceptional 
point. Indeed, to first order in $\epsilon$ the 
expansion reads
\begin{equation}
  F(\theta, \eta^2) \approx F(\theta_0, \eta_0^2) +
  \epsilon\left(\theta_{(1)} \frac{\partial F}{\partial\theta}(\theta_0,\eta_0^2)
  + \frac{\partial F}{\partial (\eta^2) }(\theta_0,\eta_0^2) \right),
\end{equation}
but at the exceptional point we have
\begin{eqnarray}
  \label{eq:excep_points}
  \frac{\partial F}{\partial \theta}(\theta_0, \eta_0^2) & = &
  0, \\
  \frac{\partial F}{\partial (\eta^2)}(\theta_0, \eta_0^2) & \neq & 0,
\end{eqnarray}
implying that we cannot choose $\theta_{(1)}$ such that 
the first
order term vanishes. We emphasize that the condition given by
Eq.~(\ref{eq:excep_points}) defines the exceptional point.

To overcome the failure of the usual perturbation scheme, 
we write the solution for $\theta$ as 
$\theta=\theta_0+\epsilon^{1/2}\theta_{(1)}+\epsilon\theta_{(2)}\dots$,
which is analogous to the expansion proposed in 
\cite{MoiseyevPRA1980, GarmonRotter2012}.
Then, to first order in $\epsilon$ we have
\begin{equation}
  \label{eq:pert}
  F(\theta, \eta^2) \approx F(\theta_0,\eta_0^2)
  + \epsilon^{1/2} \theta_{(1)}\frac{\partial F}{\partial \theta}(\theta_0, \eta_0^2)
  + \epsilon\left(\frac{1}{2}\theta_{(1)}^2
  \frac{\partial^2 F}{\partial\theta^2}(\theta_0,\eta_0^2) +
  \frac{\partial F}{\partial(\eta^2)}(\theta_0,\eta_0^2) +\theta_{(2)} 
  \frac{\partial F}{\partial \theta}(\theta_0,\eta_0^2)\right).
\end{equation}
The term of order $\epsilon^{1/2}$ and the $\theta_{(2)}$ term vanish identically 
at the exceptional point, due to Eq.~(\ref{eq:excep_points}) above,
and from the rest of the term of order $\epsilon$ we can obtain 
$\theta_{(1)}$. Explicitly, the first order term in $\epsilon$ 
leads to
\begin{equation}
-2N\theta_{(1)}^2\cos{\frac{N\pi}{2}} + \sin{\frac{(N-1)\pi}{2}} = 0,
\end{equation}
and we have $\theta_{(1)} = \pm\frac{\I}{\sqrt{2N}}$.
Consequently, near the exceptional point we obtain
\begin{equation}
    \theta^\pm \approx \frac{\pi}{2} \pm\I\frac{\epsilon^{1/2}}{\sqrt{2N}}
    = \frac{\pi}{2} \pm\I\sqrt{\frac{\eta^2-1}{2N}},
    \label{eq:phipm1}
\end{equation}
in terms of which, the eigenvalues read
\begin{equation}
    \label{eq:Eplusmin}
  E_{\theta^\pm} = 2\cos\theta^\pm \approx \mp 2\I\sinh{\sqrt{\frac{\eta^2-1}{2N}}}.
\end{equation}

Notice that the two solutions of $\theta^\pm$ indicate a coalescence
at the exceptional point, and a ``complexification'' of the 
eigenvalues in the $\mathcal{PT}$-symmetric broken phase 
($\eta>1$). The correction term proportional to the square 
root of $\eta^2 -1$,
describes well the results obtained by direct numerical
diagonalization of the Hamiltonian matrix in the proximity of 
the exceptional point; see Fig.~\ref{fig:2}(a).

The treatment described above and Eq.~(\ref{eq:excep_points}) can 
be generalized to the case when more eigenvalues coalesce at 
the exceptional point. Indeed, if the second (and higher) 
derivatives of $F(\theta,\eta)$ also 
vanish at the exceptional point, the appropriate perturbation
expansion should be proportional to $\epsilon^{1/p}$, where the 
first non-vanishing derivative is 
$\partial^{p} F/ \partial\theta^{p}$, and $p$ points coalesce at
the exceptional point. In this situation $p$ distinct eigenvalues
coalesce to the same value \cite{MoiseyevPRA1980, GarmonRotter2012}.

\section{Transport}
\label{sec:transportPT}
Having now a thorough description of the spectra and
eigenvectors of the $\mathcal{PT}$ symmetric chain, we turn 
to the discussion of the transport properties of the system. 
First of all, for all non-opaque states in the unbroken 
$\mathcal{PT}$-symmetry 
phase we have $\xi_{E_\theta}=1$, as can be checked 
directly using the explicit expression of the coefficients,
Eq.~(\ref{eq:4}). This indicates that transport in the
eigenstates with real eigenvalues 
is efficient. On the other hand, for states in the $\mathcal{PT}$-broken 
symmetry phase, some eigenvalues become complex and for the corresponding 
eigenstates $\xi_{E_\theta}$ is no longer equal to one. 
In view of Eq.~(\ref{eq:sym2}), it is straight forward 
to see that $\xi_{E_\theta}\xi_{E_\theta^*}=1$ as mentioned previously.

The actual values of the transport coefficient for the
eigenstates near the exceptional point can be evaluated 
using the perturbation expansion for $\theta$. From 
Eq.~(\ref{eq:phipm1}) we can write
\begin{equation}
  \xi_{E_\theta,E_\theta^*} \sim 1\pm \sqrt{N(\eta^2-1)/2}\sim e^{\pm \sqrt{N(\eta^2-1)/2}}, 
  \label{eq:xi1}
\end{equation}
for $\eta > 1$, where the exponential form was chosen 
merely to enforce the fact 
that $\xi_{E_\theta,{E_\theta}^*}$ are positive, and 
that their product
$\xi_{E_\theta} \xi_{E_\theta^*}$ must 
be equal to one. 

\begin{figure}[t]
  \centering
  \includegraphics[scale=0.32]{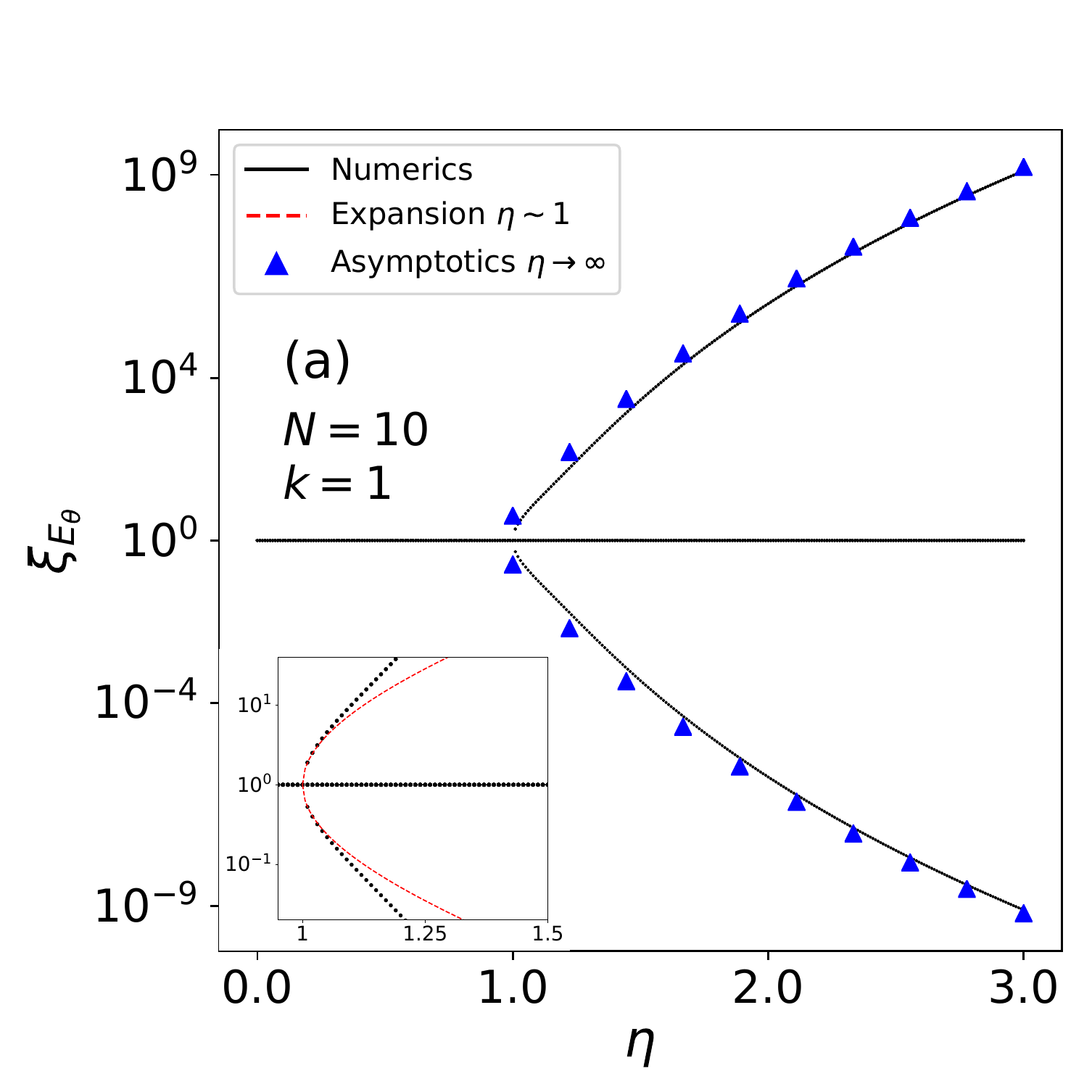}
  \includegraphics[scale=0.32]{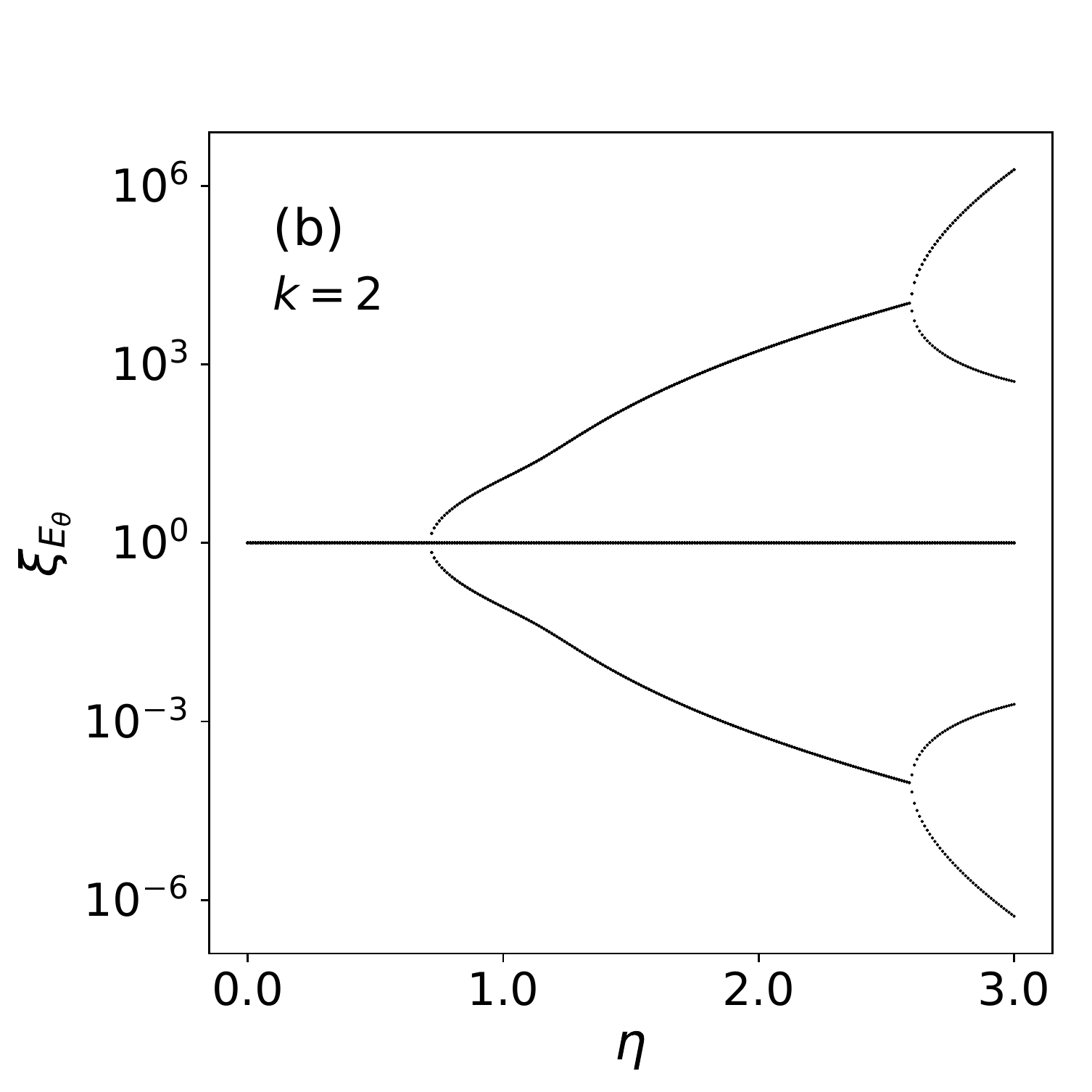}
  \includegraphics[scale=0.32]{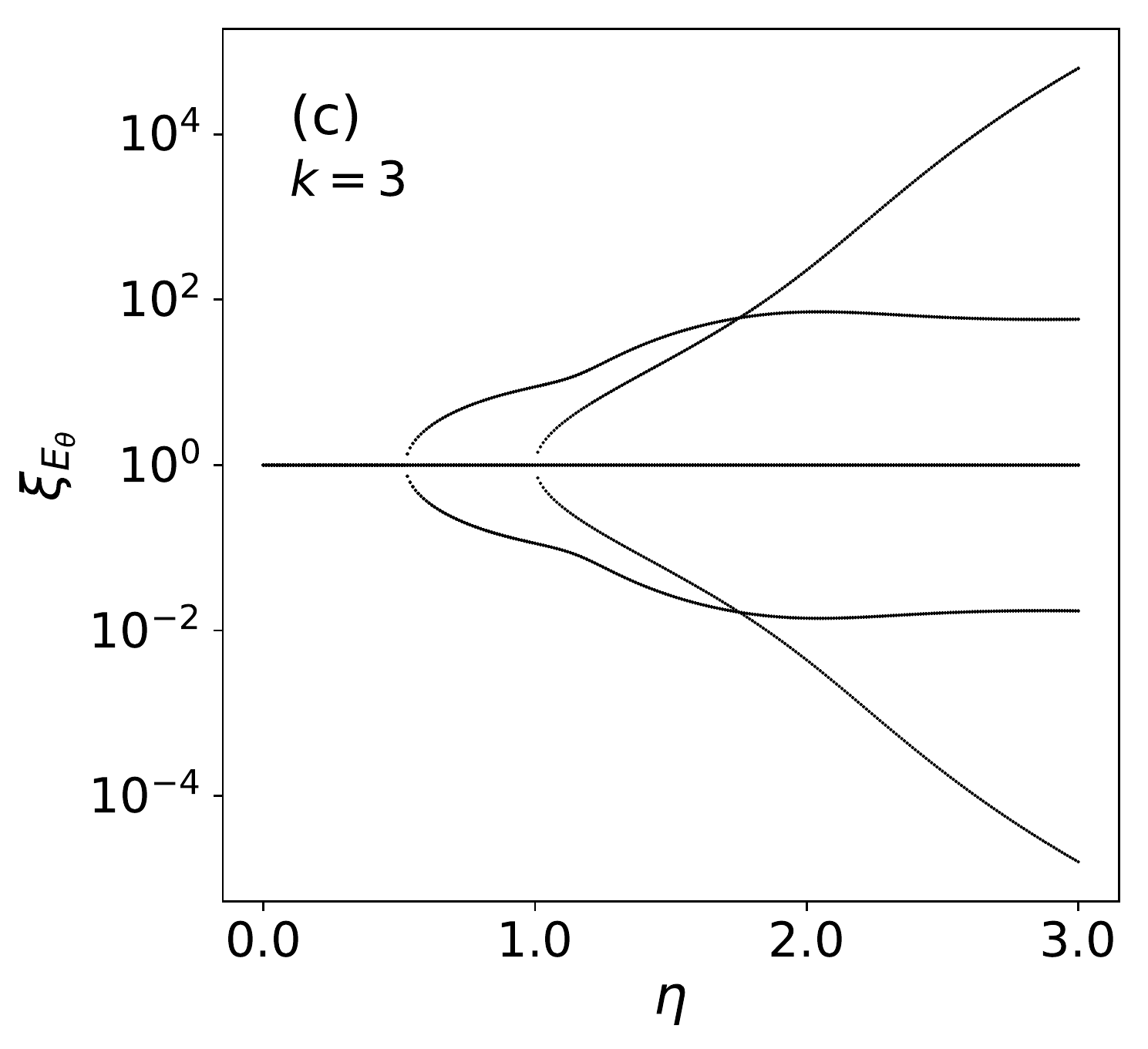}\\
  \includegraphics[scale=0.32]{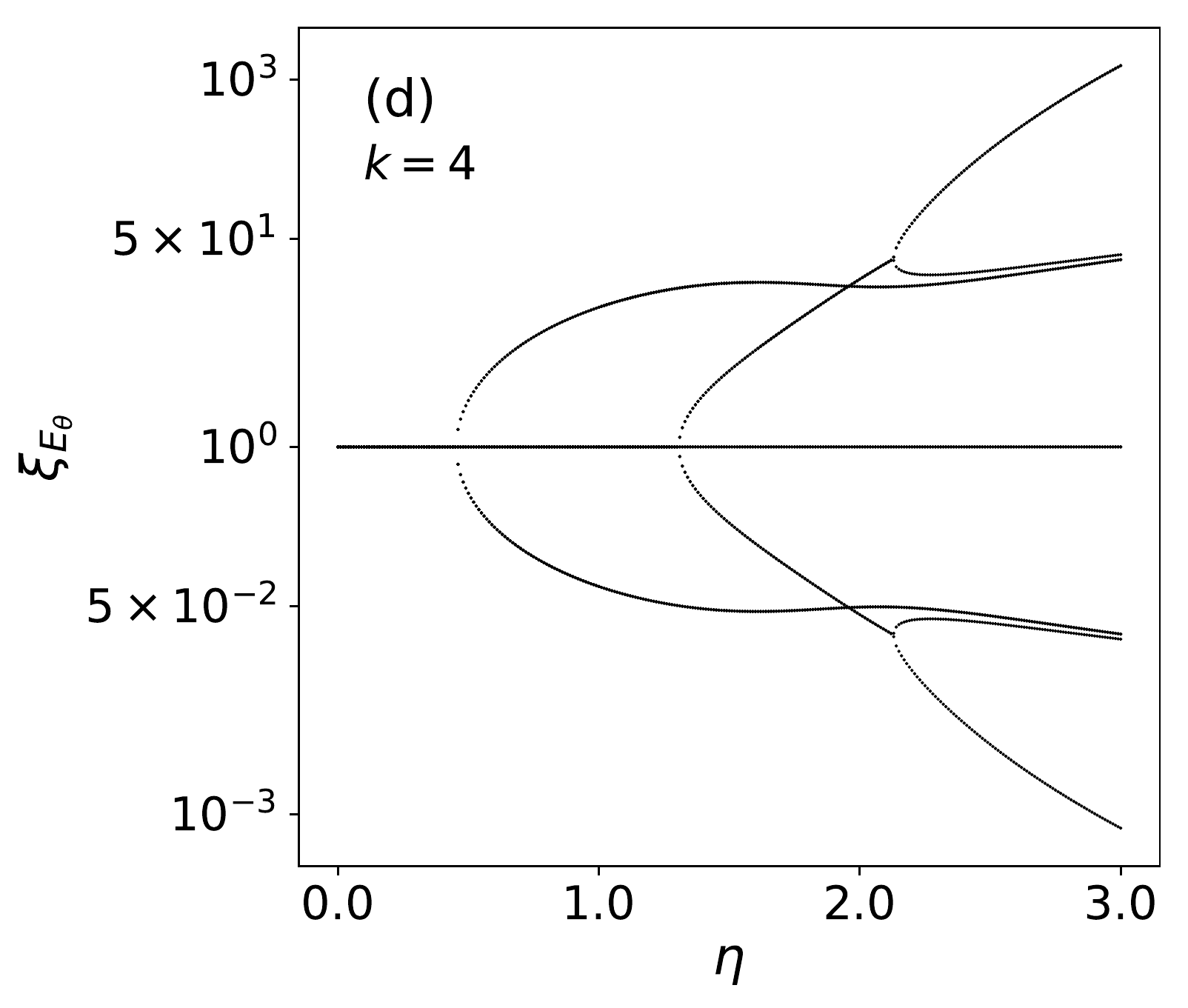}
  \includegraphics[scale=0.32]{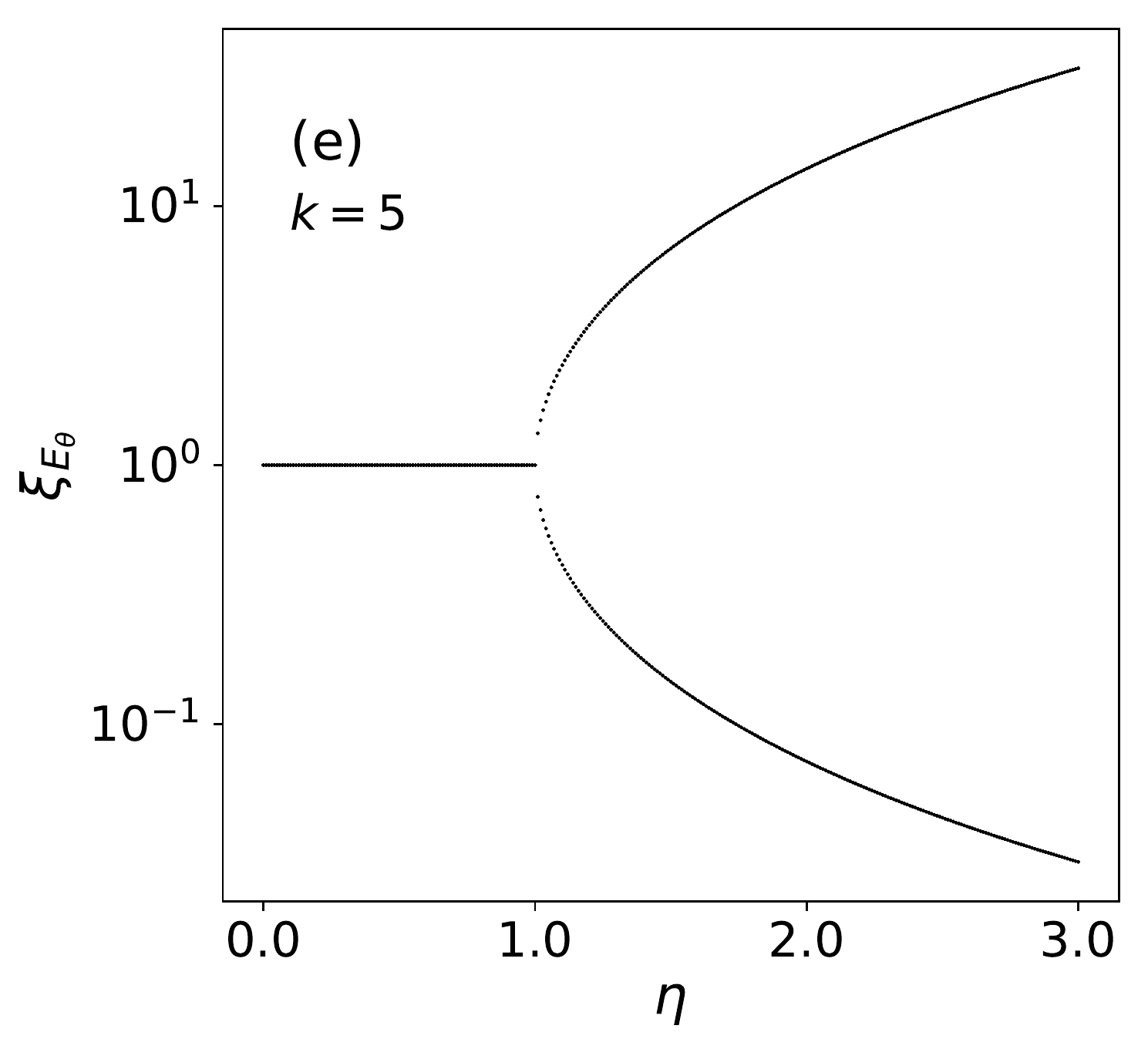}
  \caption{Transport coefficient $\xi_{E_\theta}$ as a 
  function of $\eta$ for all eigenstates and all 
  configurations of the contacts for $N=10$. 
  Sub-figure (a) includes the
  analytical expansion of the transport coefficient (dashed red curve
  in the inset) for $\eta\approx 1$ Eq. \ref{eq:xi1}, as well as the asymptotic expansion (blue triangles) for $\eta\rightarrow \infty$ Eq. \ref{eq:zetaetainf}.}
  \label{fig:5}
\end{figure}

In the limit $\eta\rightarrow\infty$ the
$\xi_{{E_\theta,E_\theta^*}}$ are given by
\begin{equation}
  \xi_{E_\theta,E_\theta^*} \sim [4\eta^{2(N-1)}]^{\pm1}.
  \label{eq:zetaetainf}
\end{equation}

\begin{figure}[t]
  \centering
  \includegraphics[scale=0.32]{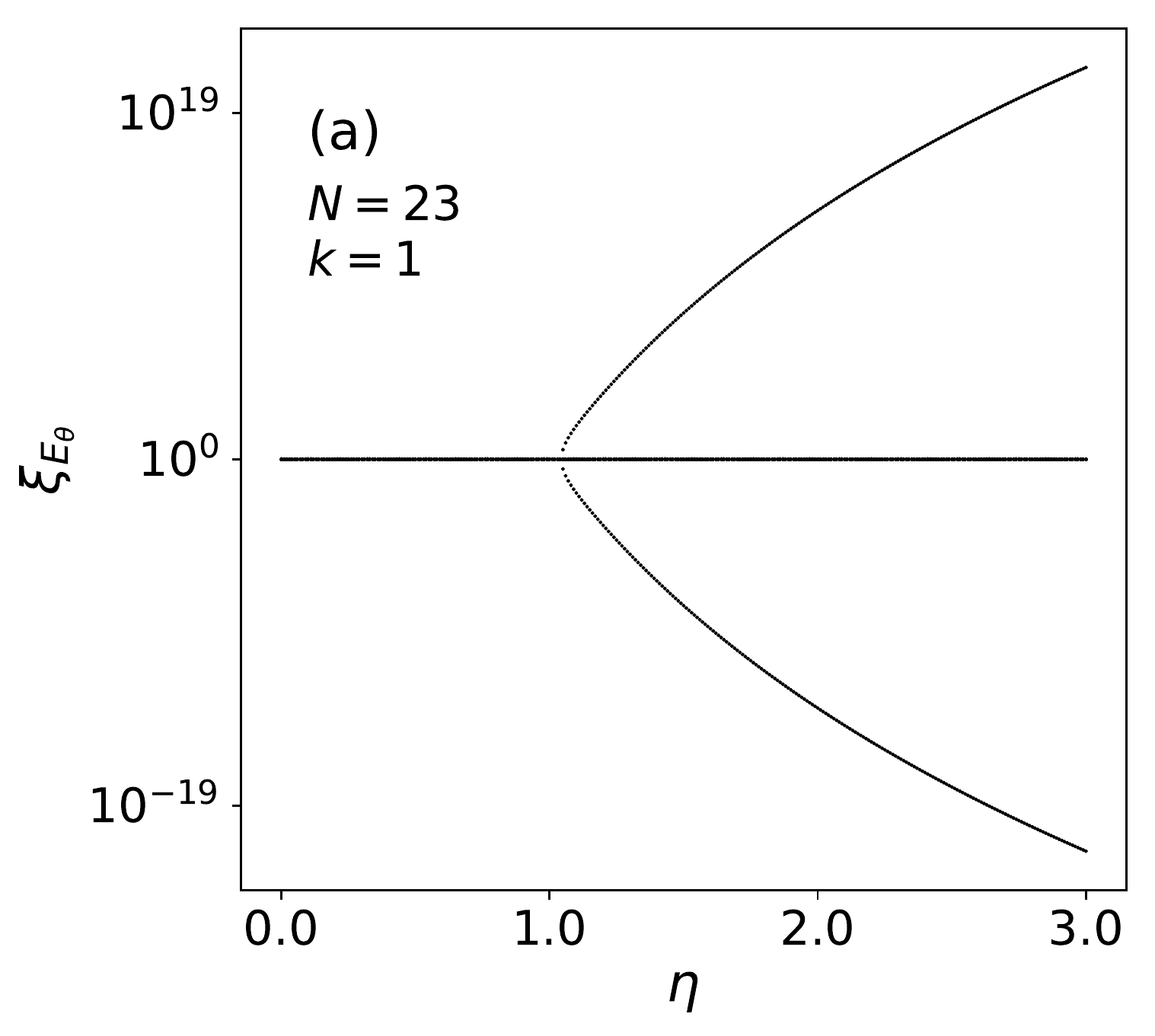}
  \includegraphics[scale=0.32]{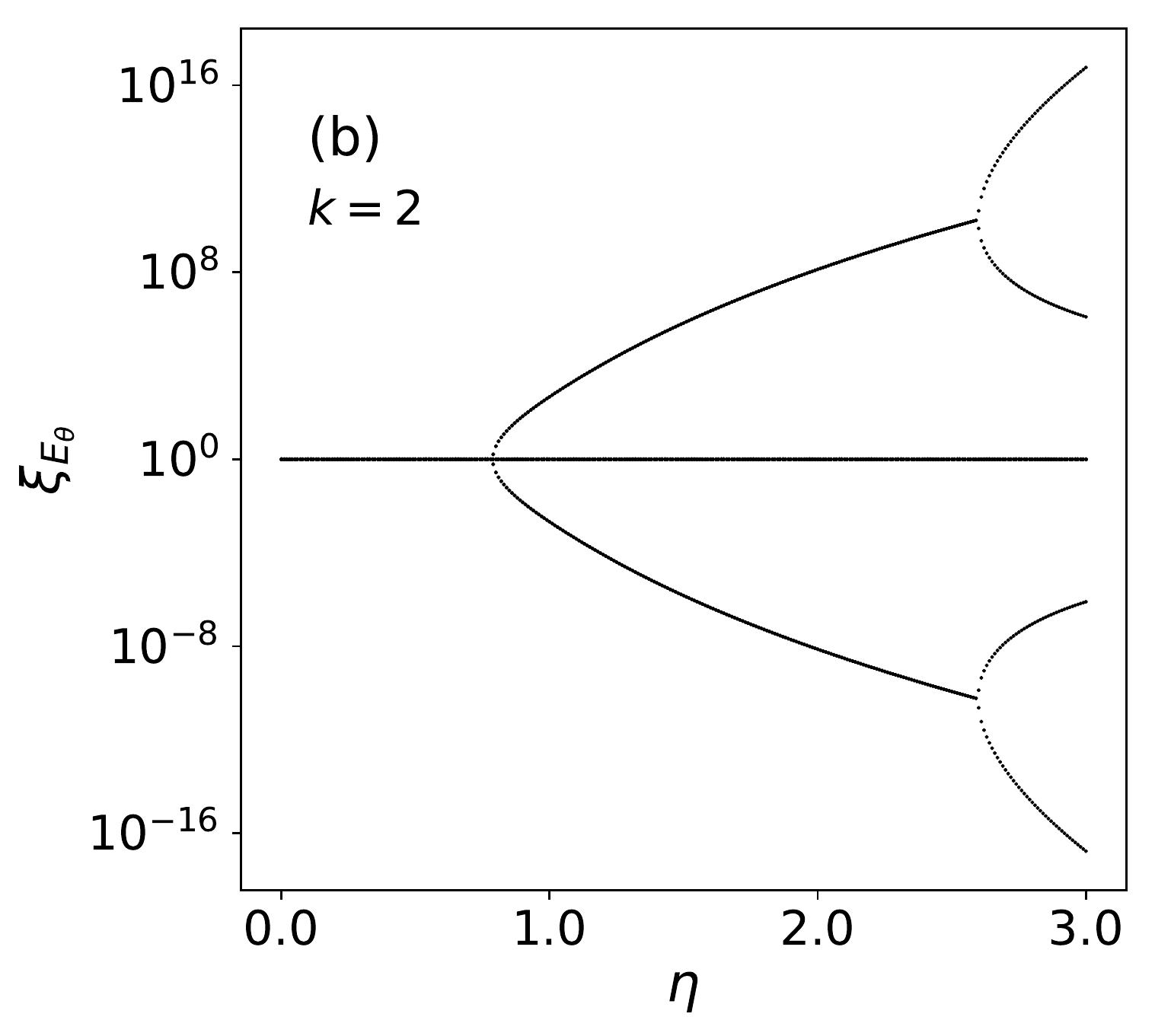}
  \includegraphics[scale=0.32]{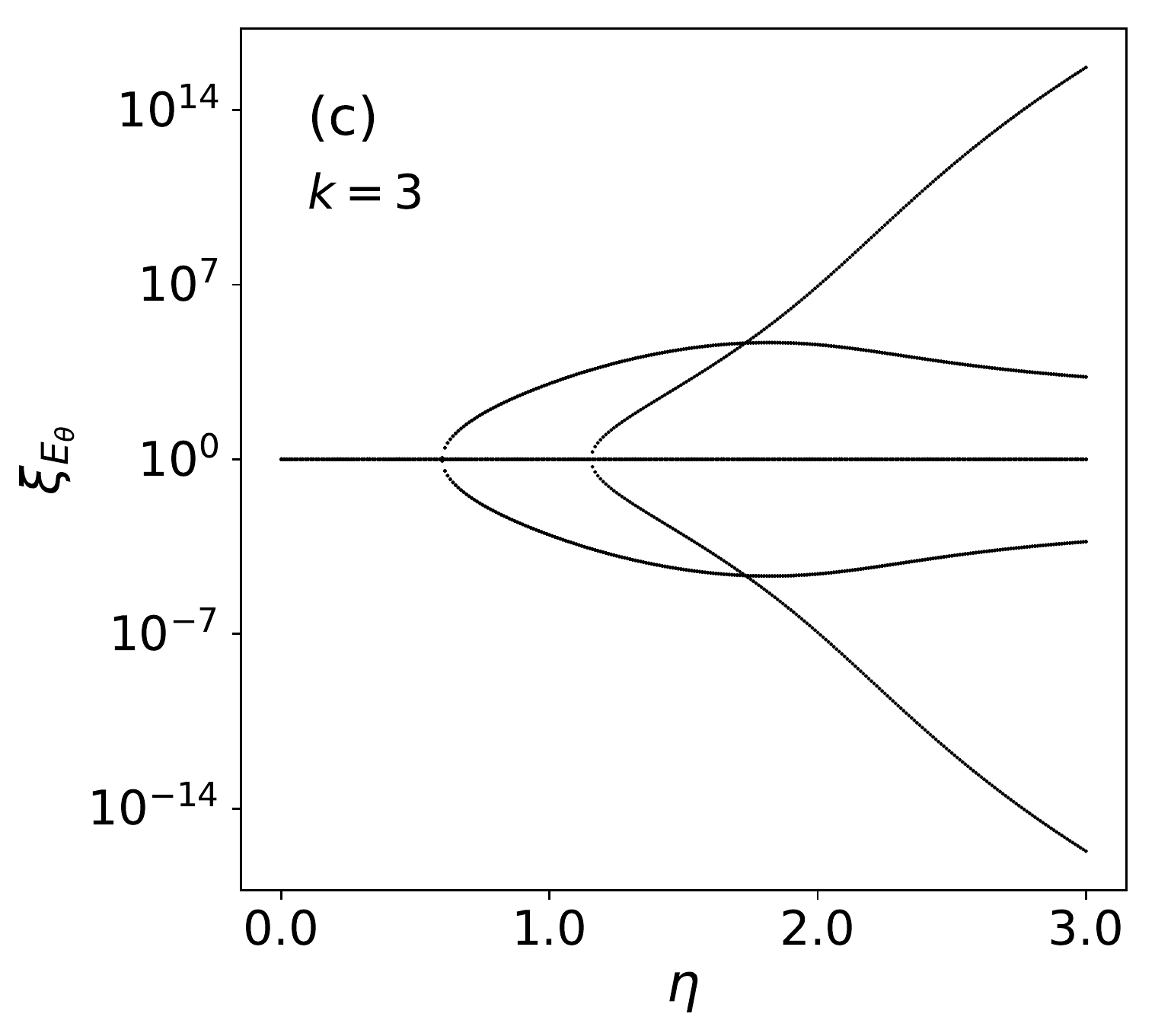}\\
  \includegraphics[scale=0.32]{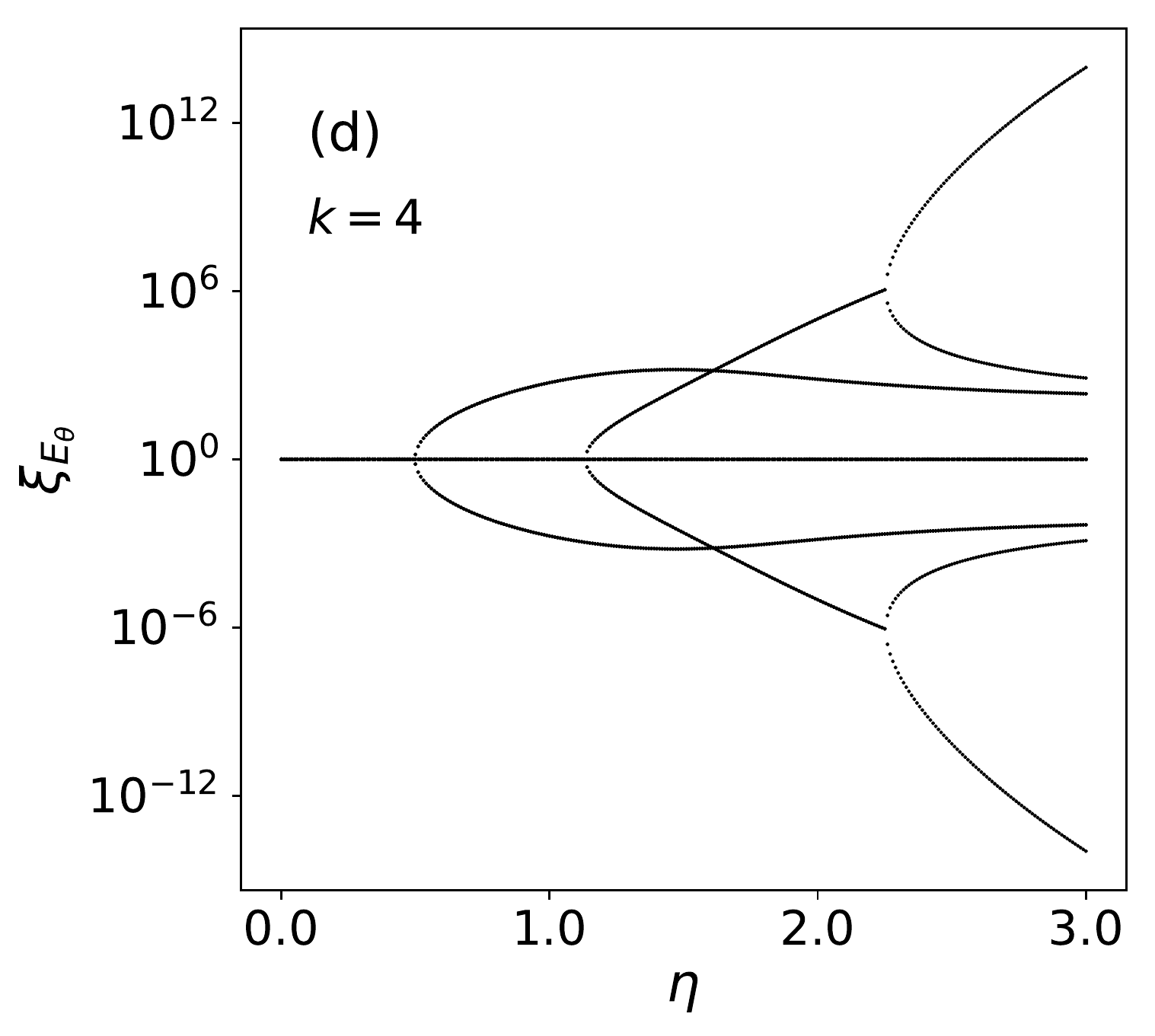}
  \includegraphics[scale=0.32]{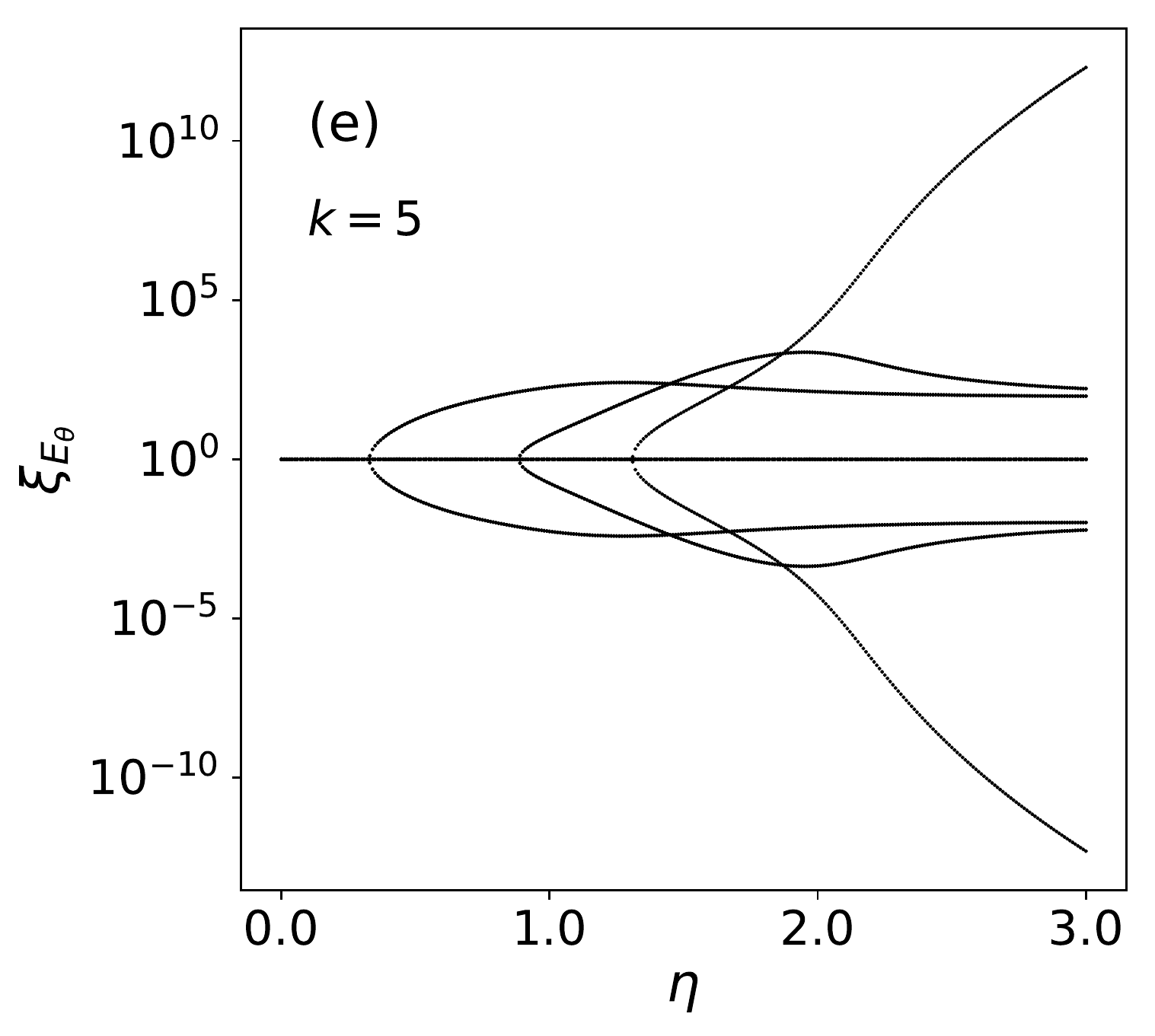}
  \includegraphics[scale=0.32]{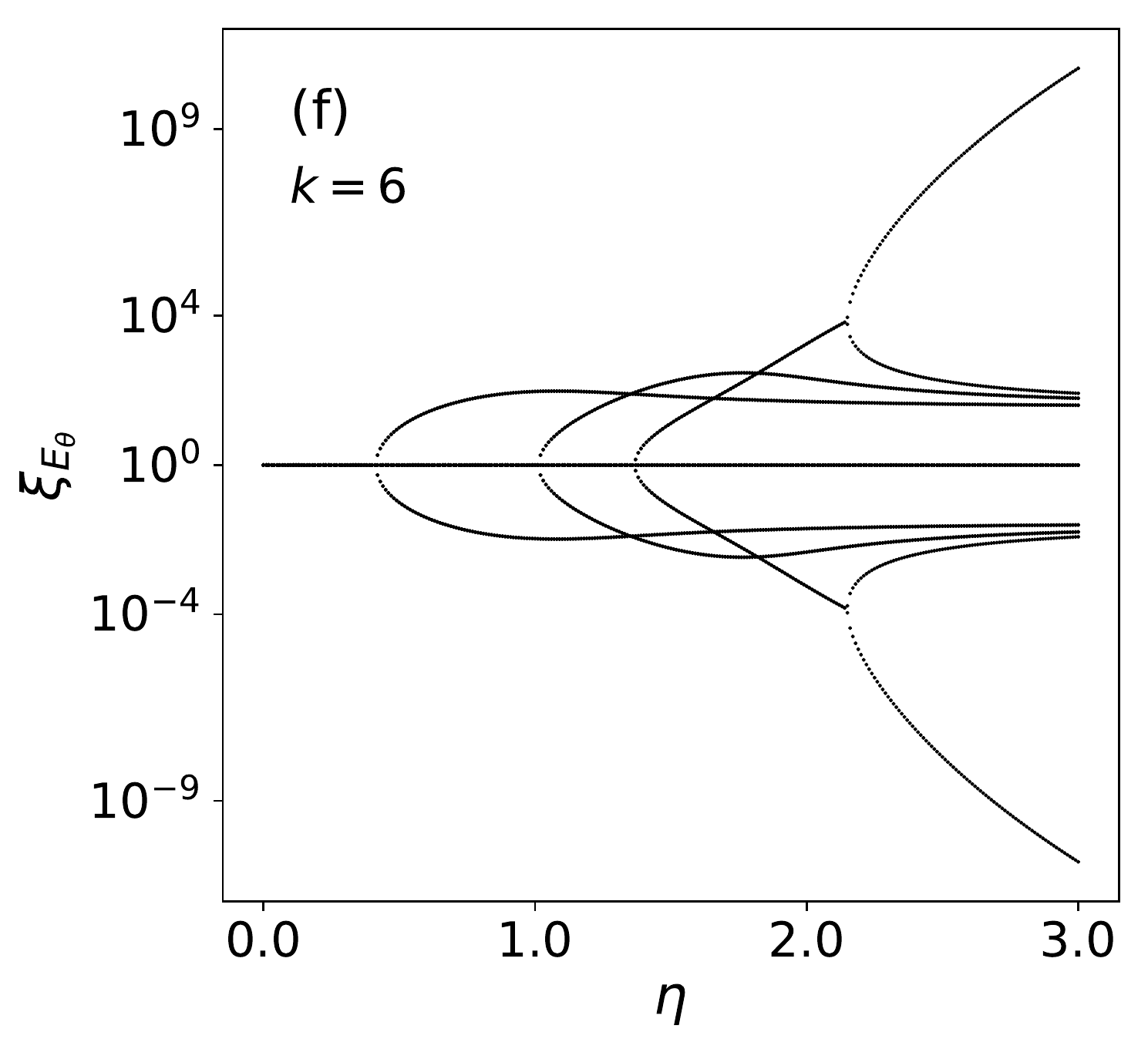}\\
  \includegraphics[scale=0.32]{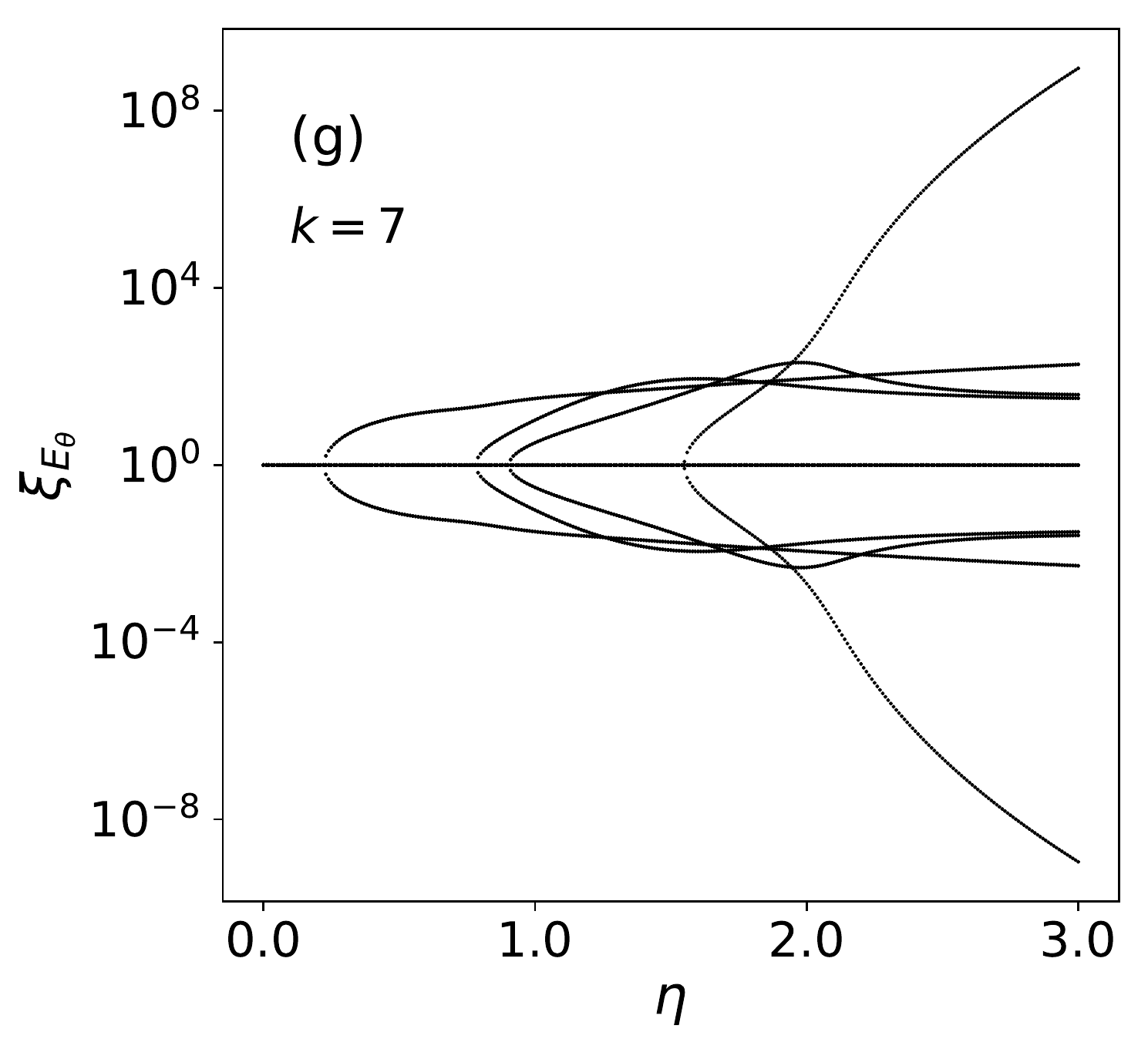}
  \includegraphics[scale=0.32]{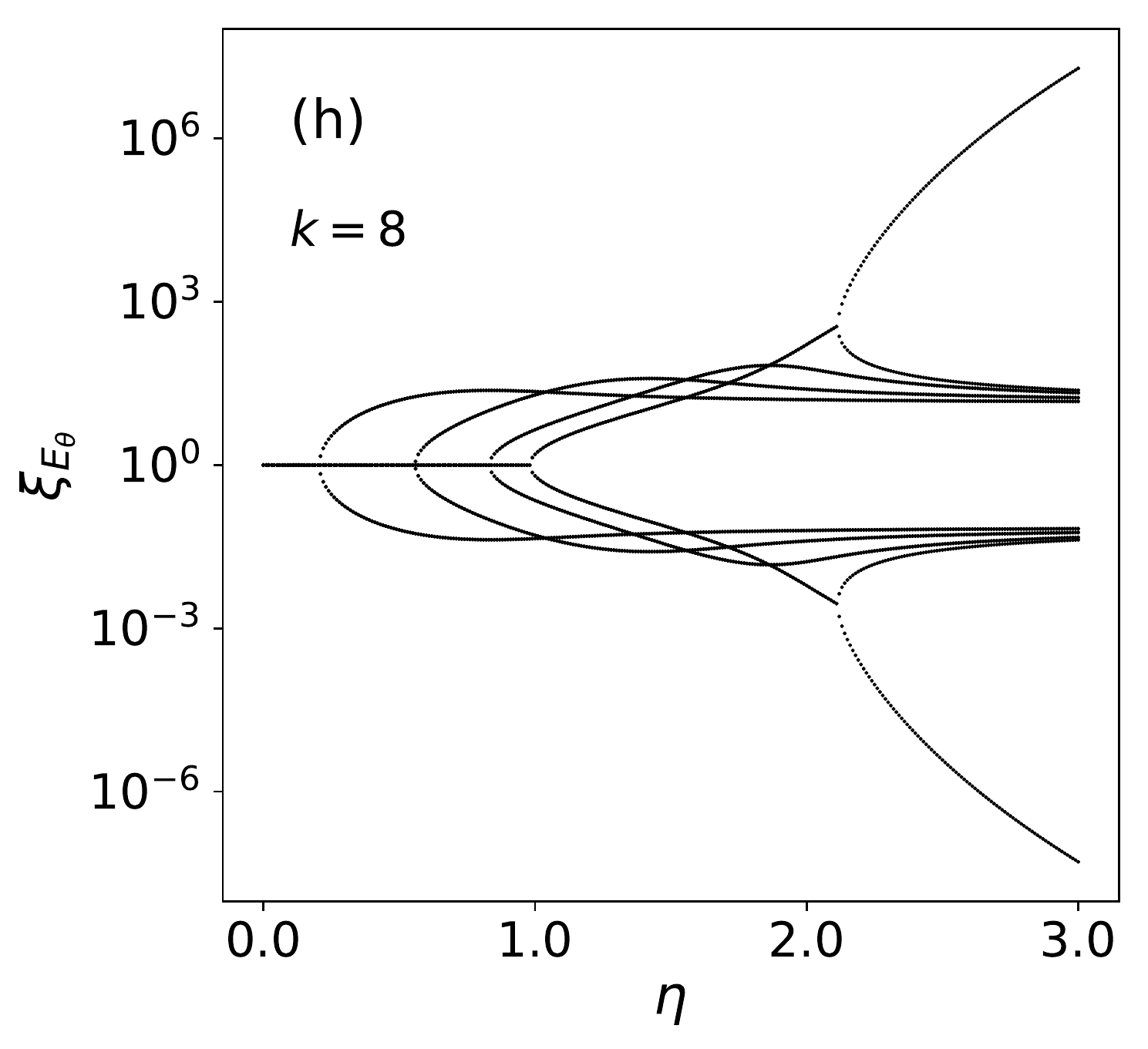}
  \includegraphics[scale=0.32]{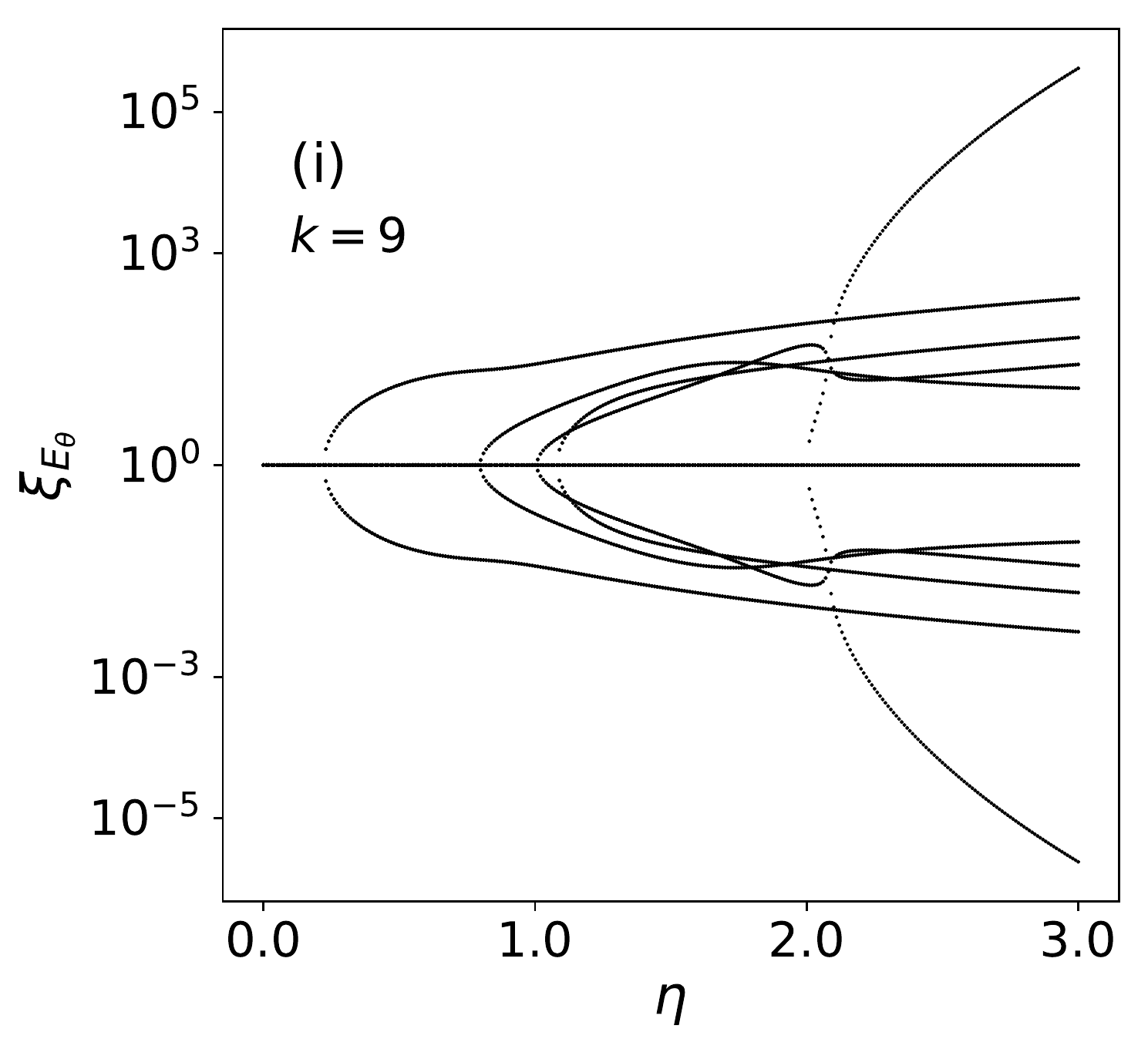}\\
  \includegraphics[scale=0.32]{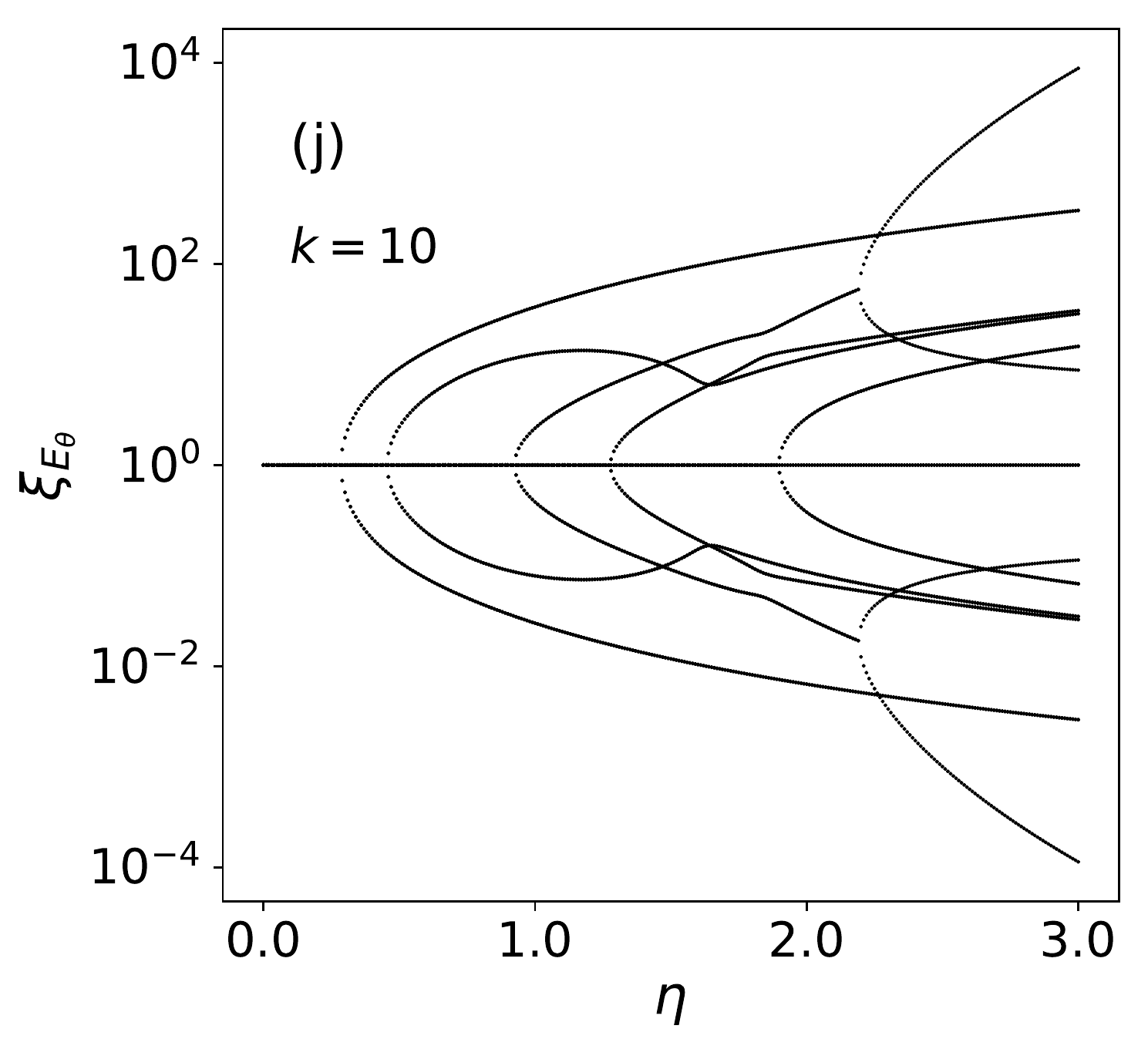}
  \includegraphics[scale=0.32]{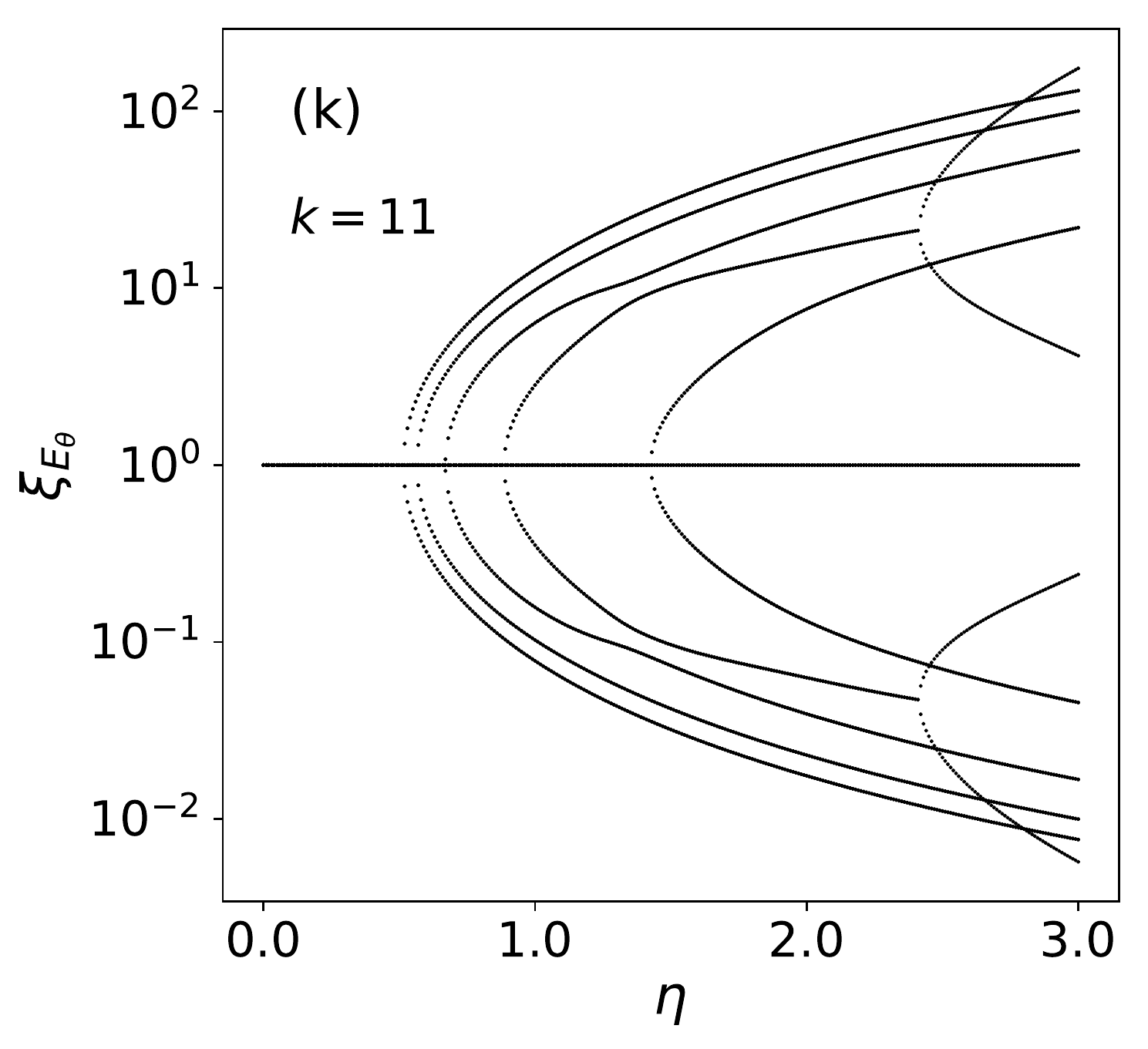}
  \caption{Same as Fig.~\ref{fig:5}, for $N=23$ and all 
  configurations of the contacts. Only the non-opaque states 
  are included.}
  \label{fig:6}
\end{figure}

Figure~\ref{fig:5} shows the transport coefficient 
$\xi_{E_\theta}$  as a function of $\eta$ for all the 
eigenfunctions and all contact configurations for $N=10$. 
As we have seen, since $N+1$
is prime there are no opaque nor transparent states, and all eigenstates
have a well-defined transport coefficient. The perturbative 
approximation, Eq.~(\ref{eq:xi1}), is shown 
in Fig.~\ref{fig:5}(a) by the red dashed curves in the inset;
the asymptotic limit of $\xi_{E_\theta}$,
Eq.~(\ref{eq:zetaetainf}), is illustrated by the 
blue triangles. We note that there are some configurations
that display states with efficient transport; see 
Fig.~\ref{fig:5}(b), (c) and~(d). These states with efficient
transport are not transparent states, but still have real eigenvalues, despite the 
fact that some eigenvalues for other states are complex for the same value of $\eta$.

Similarly, Fig.~\ref{fig:6} shows the transport coefficient 
$\xi_{E_\theta}$ as a function of $\eta$ for all non-opaque 
states and all contact configurations for $N=23$. 
Again, we note that in most configurations
there are some transport coefficients 
which are identical to $1$ for all values of $\eta$, i.e., 
transport is effective for some eigenstates in those
configurations. Some of those states correspond to 
transparent states, but not necessarily all of them.
Notice that for
$k=8$ in Fig.~\ref{fig:6}(h) the opposite is observed: 
beyond certain value of $\eta$, all transport 
coefficients are different from $1$ and transport
is deficient; this configuration ($k=8$, $k'=16$)
corresponds to the maximum number of opaque states
for $N=23$, having no transparent states. In this 
case all states are either 
opaque, or have complex eigenvalues, and transport 
is always deficient beyond $\eta \geq 1$. 

\begin{figure}[t]
  \centering
  \includegraphics[scale=0.3]{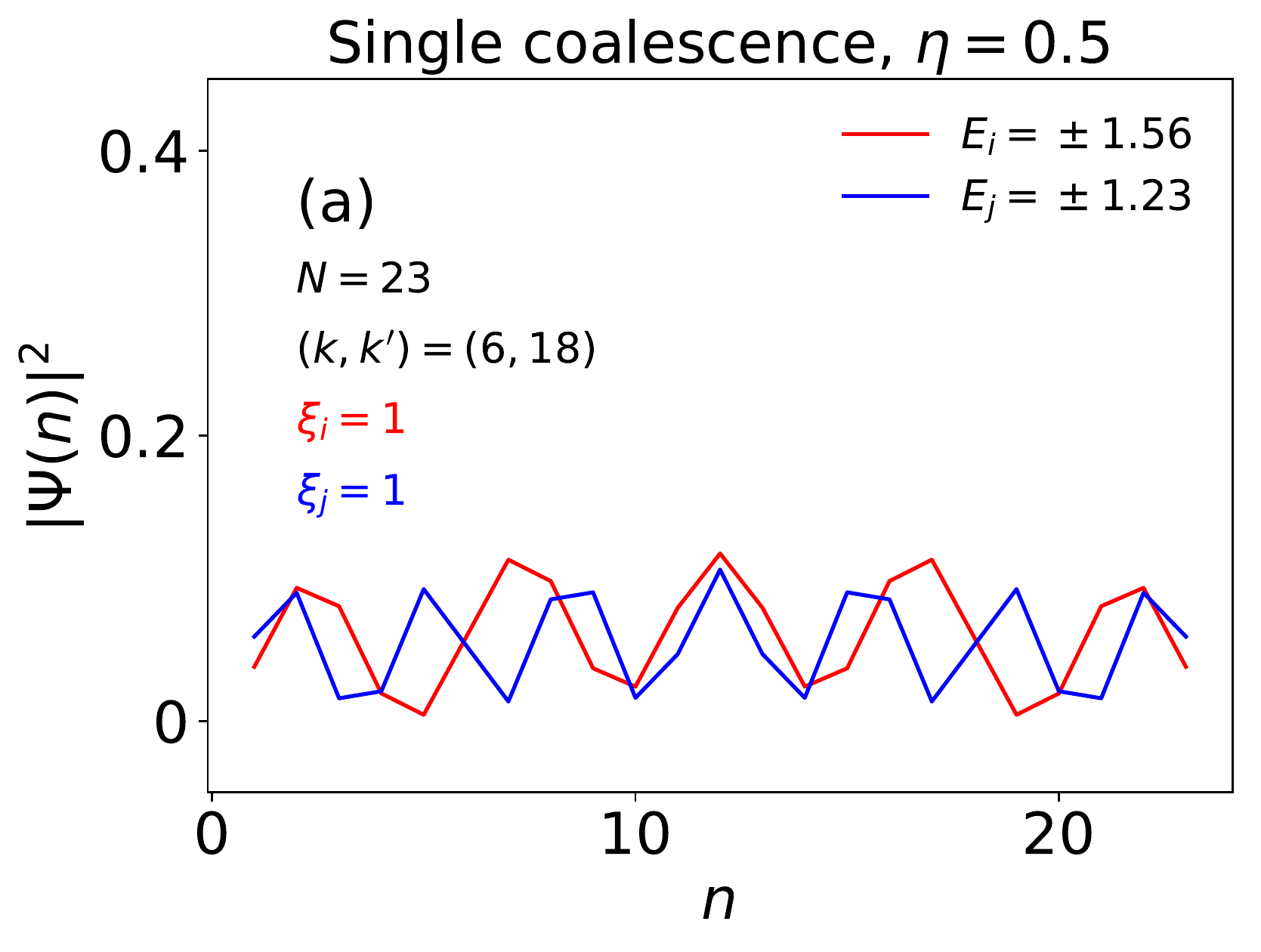}
  \includegraphics[scale=0.3]{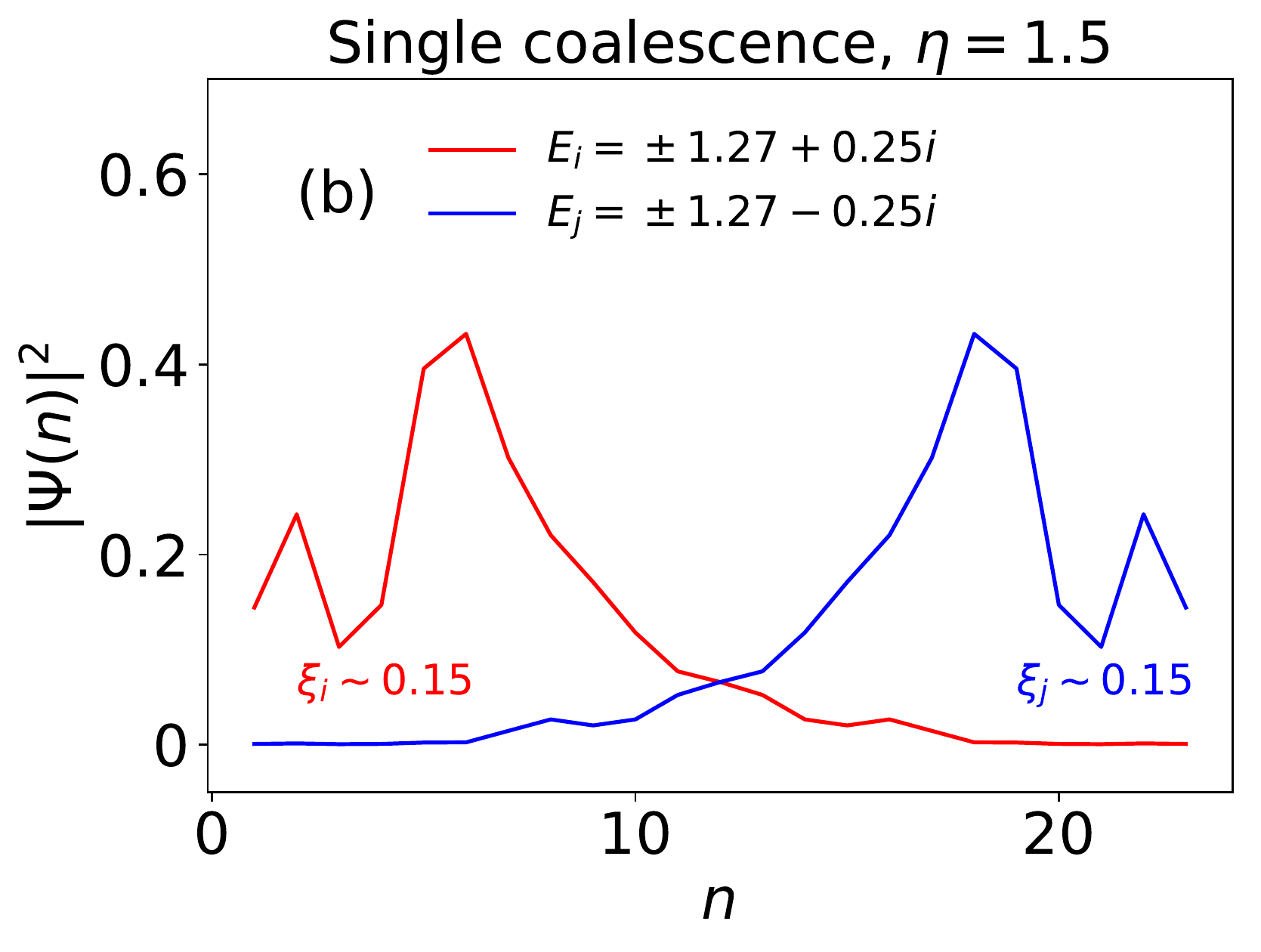}
  \includegraphics[scale=0.3]{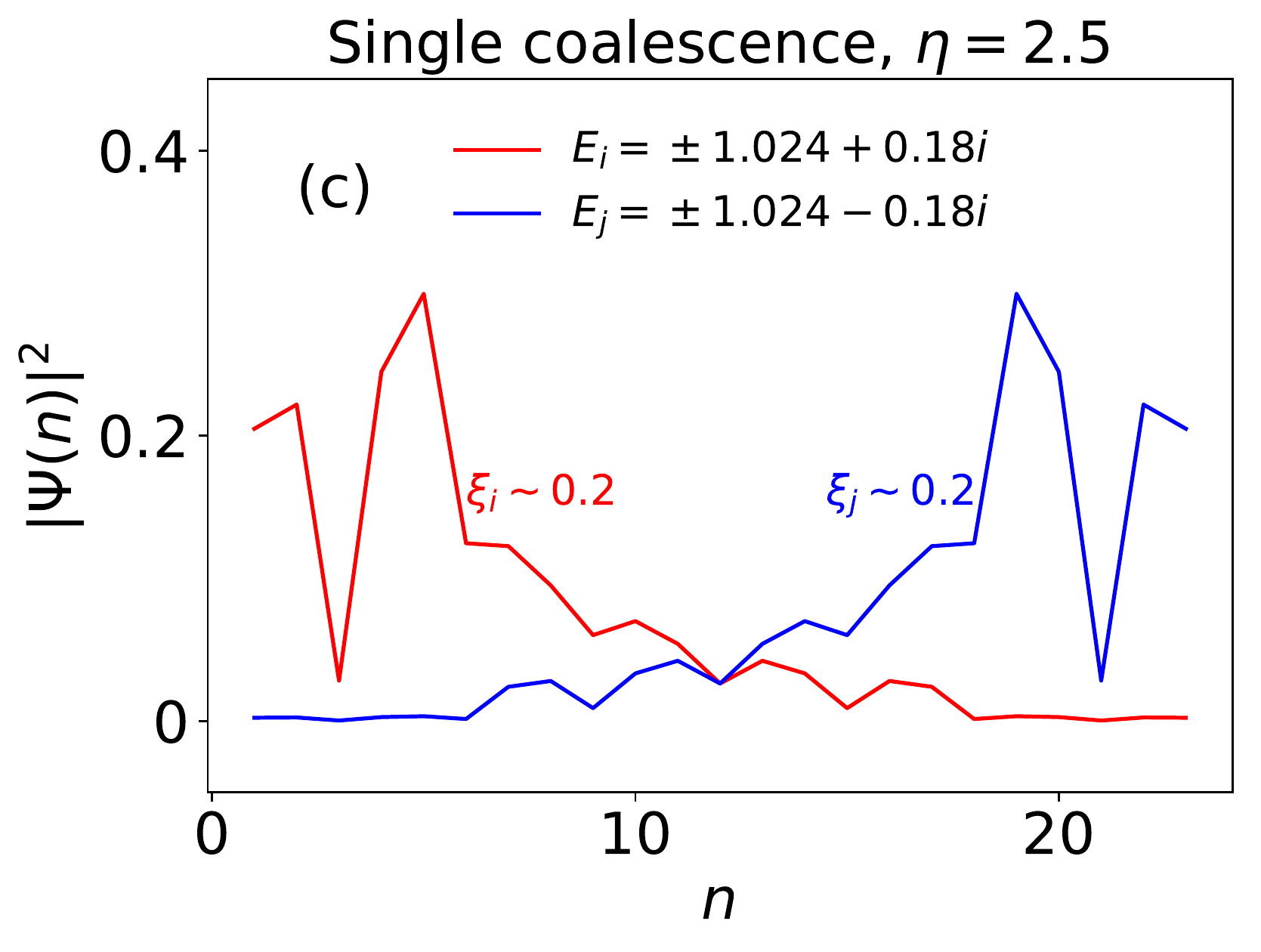}\\
  \includegraphics[scale=0.3]{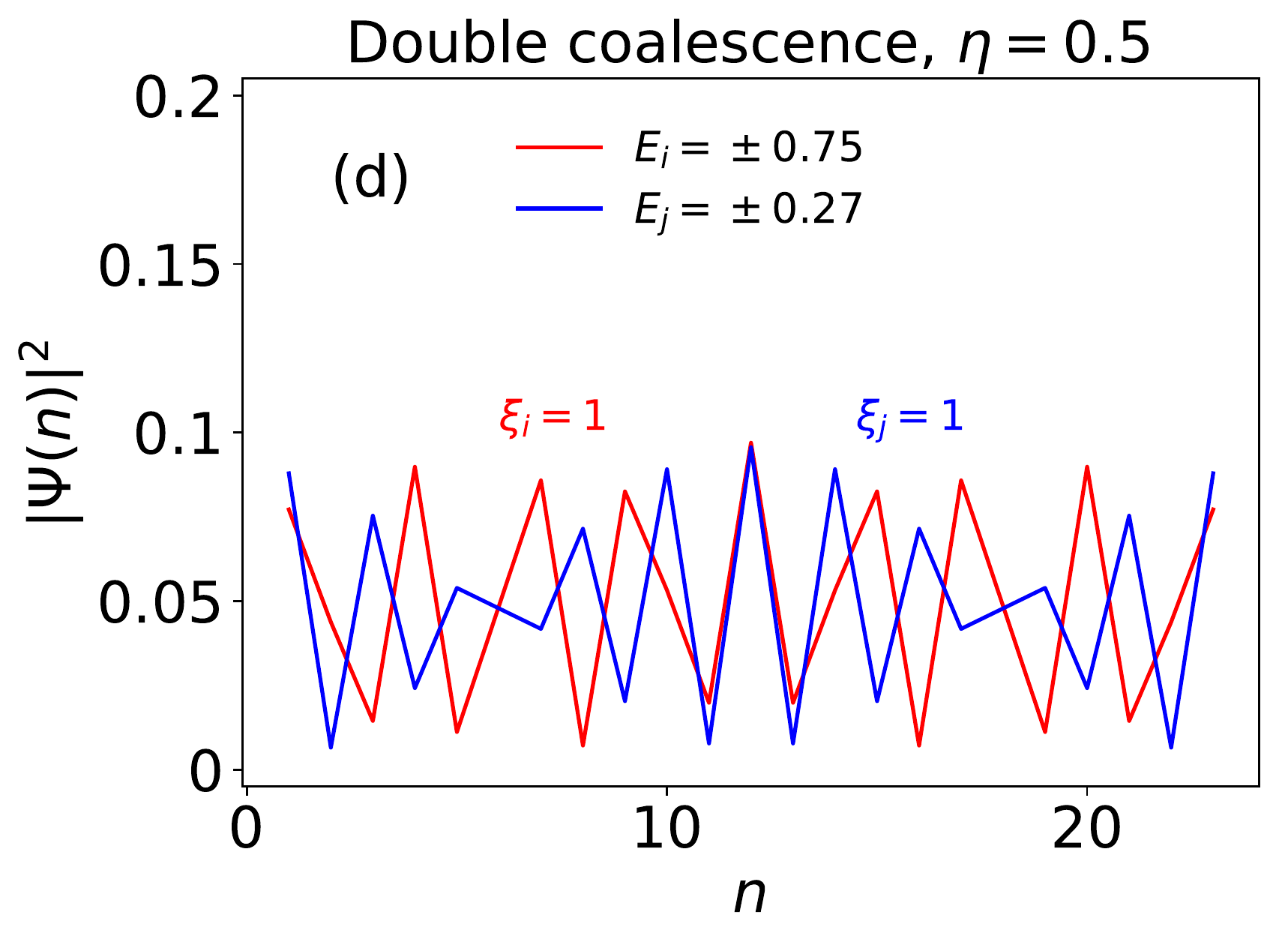}
  \includegraphics[scale=0.3]{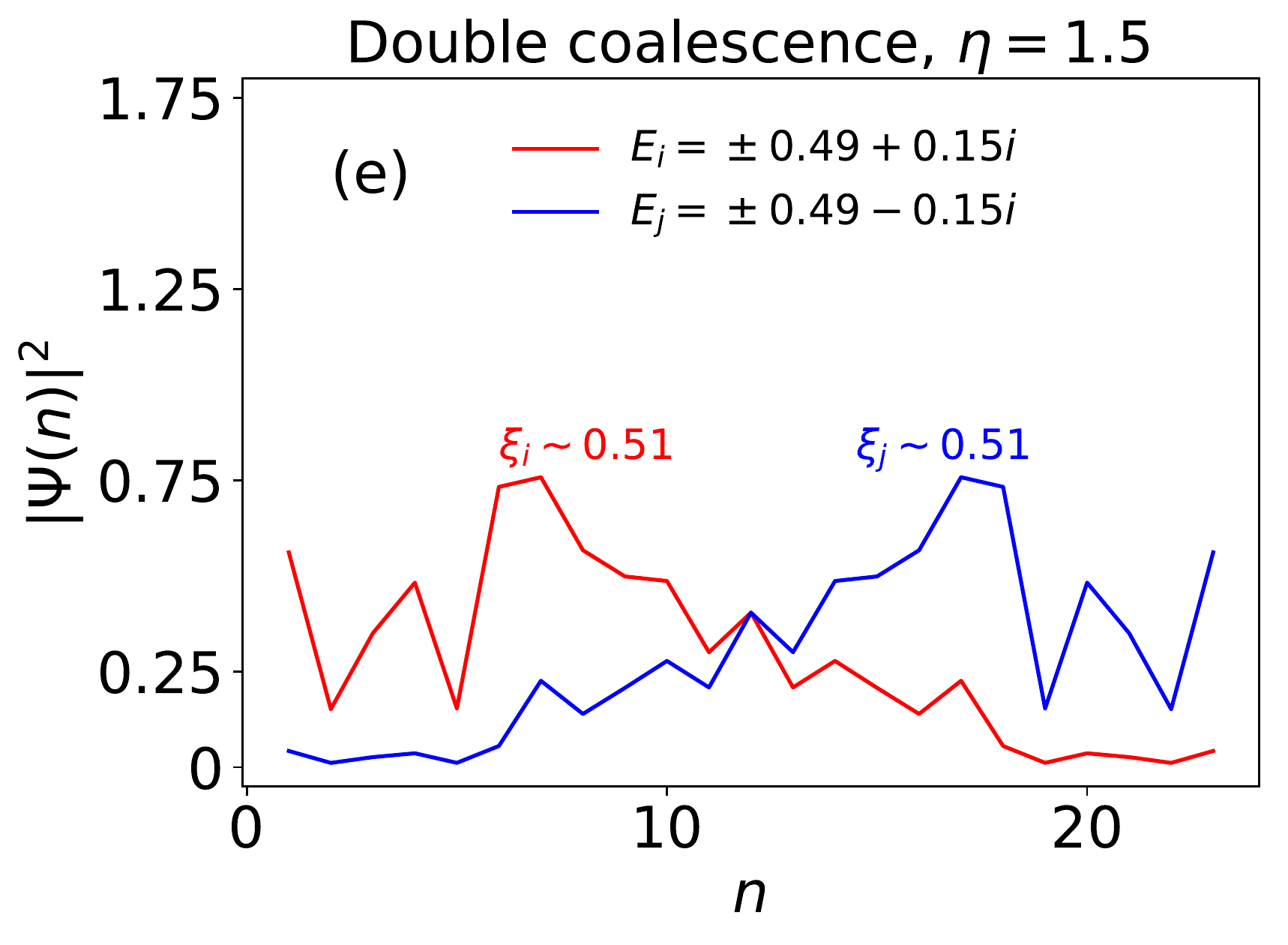}
  \includegraphics[scale=0.3]{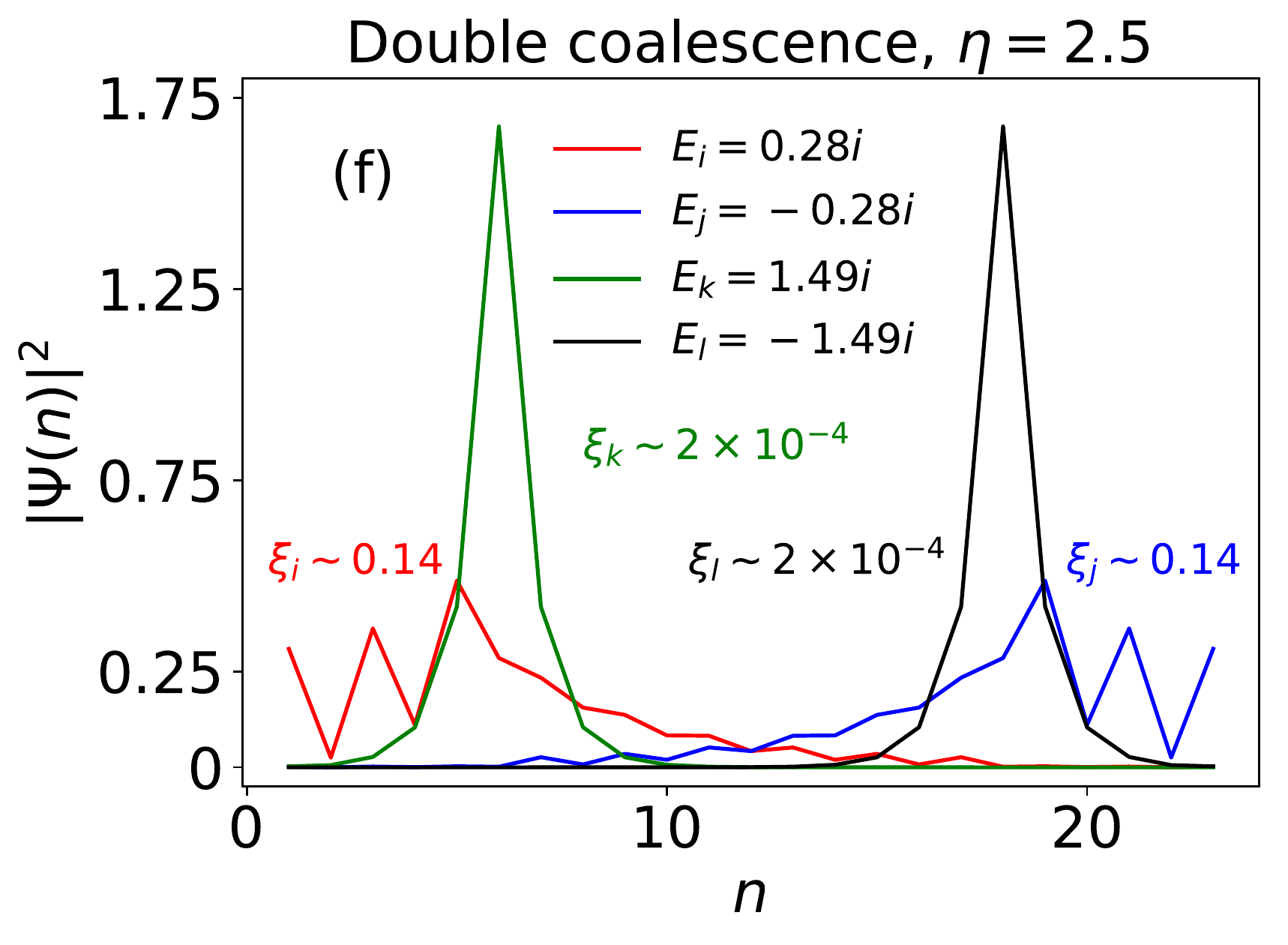}
  \caption{Absolute squared value of eigenstates as function of the site position
  $n$ that undergo a single (upper panels) or a pair (lower panels) of exceptional points, for different values of $\eta$ and $N=23$ in the configuration $k=6$ and $k'=18$. For each eigenstate, the corresponding value of the energy at a certain $\eta$ is given with $E_i$. In a similar way, we have
  included the values of the transport coefficient $\xi_i$. (a) and~(d) correspond to $\eta=0.5$, where all corresponding eigenvalues are real; (b) and~(e) $\eta=1.5$, and all illustrated states have experienced one exceptional point; (c) and~(f) correspond to $\eta=2.5$.}
  \label{fig:7}
\end{figure}

The behavior of the transport coefficients
$\xi_{E_\theta,E_\theta^*}$ is consistent with the strong
localization towards the contacts for the
$\mathcal{PT}$-broken symmetry states. This is 
illustrated in
Fig.~\ref{fig:7} where we show the modulus squared of
two pairs of eigenfunctions whose eigenvalues coalesce 
at different values of $\eta$, in the $k=6$ and $k'=18$
configuration for $N=23$. In Figs.~\ref{fig:7}(a) 
and~(d) for $\eta=0.5$, the eigenvalues 
are real, and their eigenfunctions are extended.
Figures~\ref{fig:7}(b) and~(e) display the states at
$\eta=1.5$ after crossing an exceptional point; 
localization around the contacts is apparent. In
Figs.~\ref{fig:7}(c) and~(f)
we illustrate the case for $\eta=2.5$. The pair of states
in (c) localize in one or the other side of the chain,
between the edge of the chain and the contact. 
A similar situation occurs with the states 
illustrated in (e). The states
in panel (f) are strongly localized in the gain or 
in the loss. These states correspond to energies approximated 
asymptotically by Eq.~(\ref{eq:5}).

\begin{figure}[h!]
  \centering
  \includegraphics[scale=0.31]{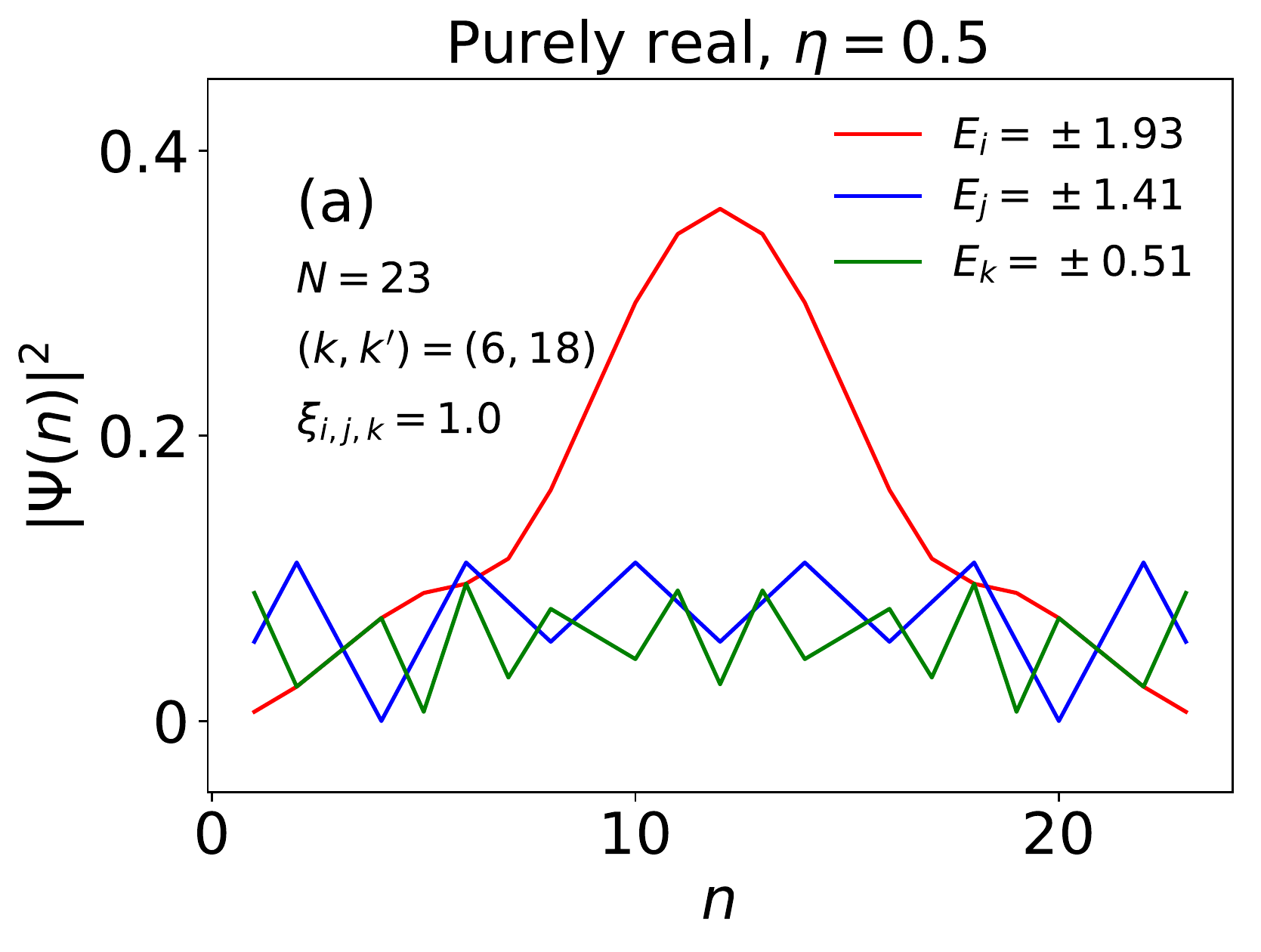}
  \includegraphics[scale=0.31]{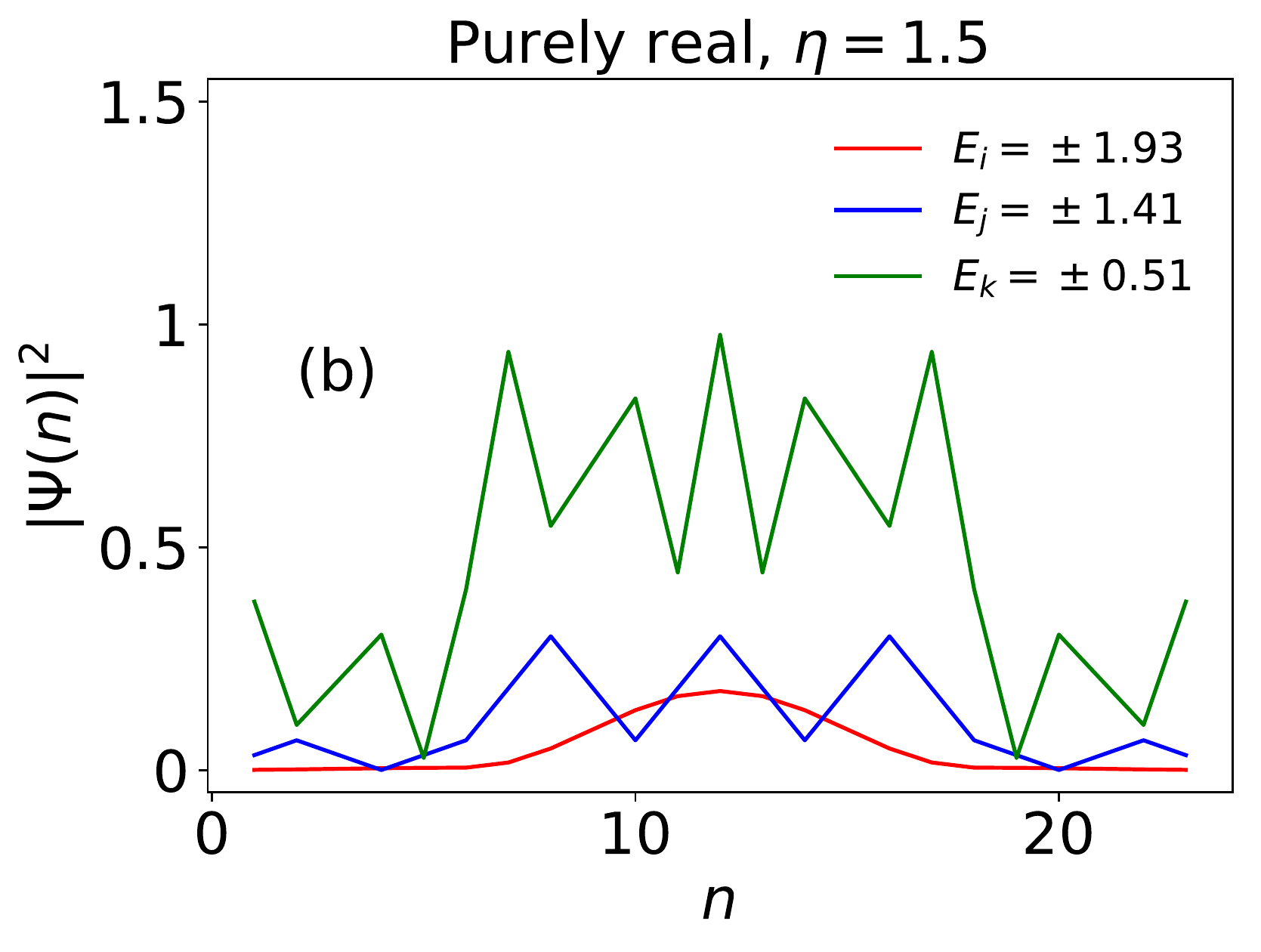}
  \includegraphics[scale=0.31]{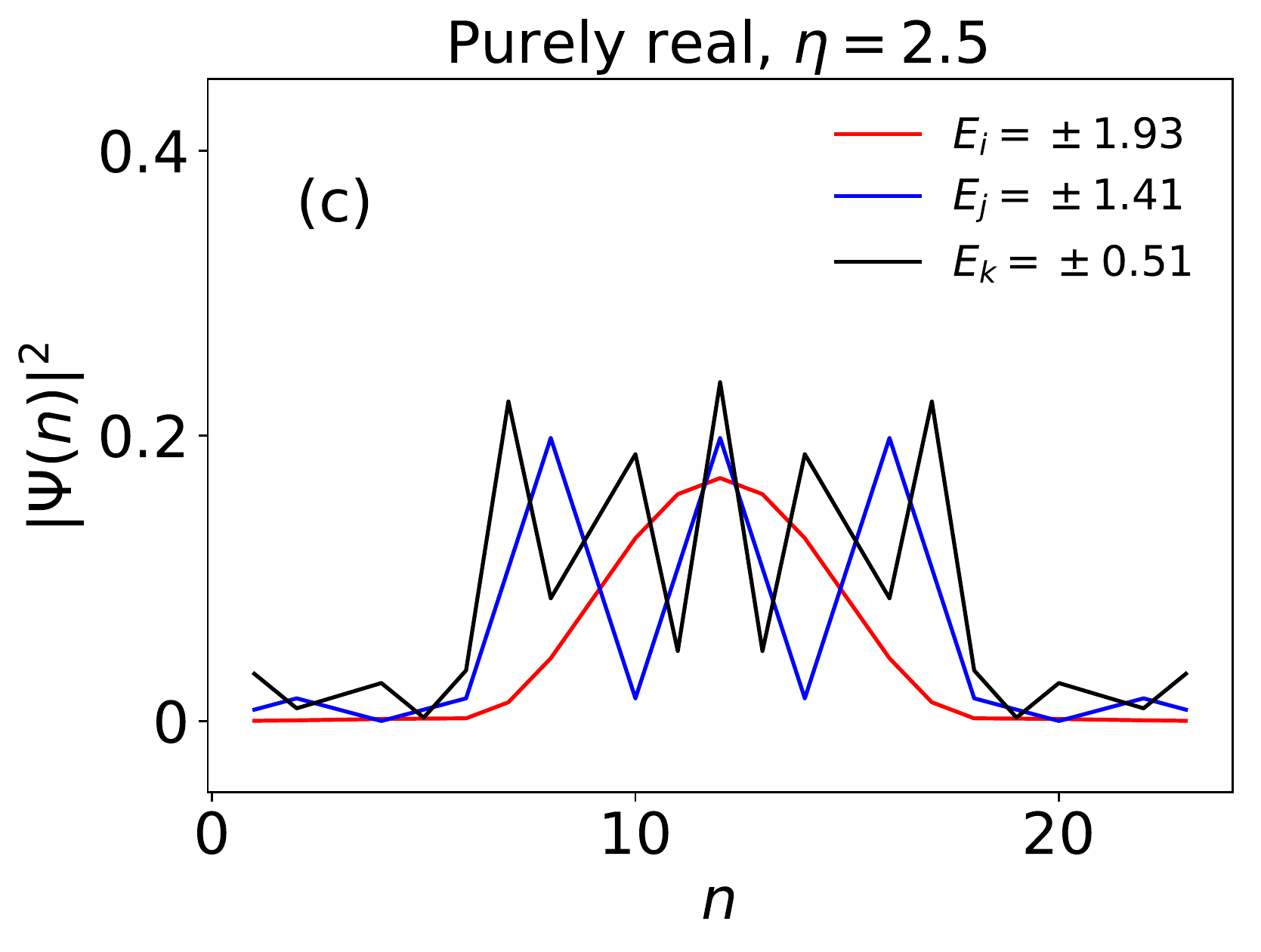}
  \caption{Absolute squared value of three transparent states 
  for $N=23$ in the configuration $k=6$ and $k'=18$.}
  \label{fig:8}
\end{figure}

\begin{figure}[t]
  \centering
  \includegraphics[scale=0.5]{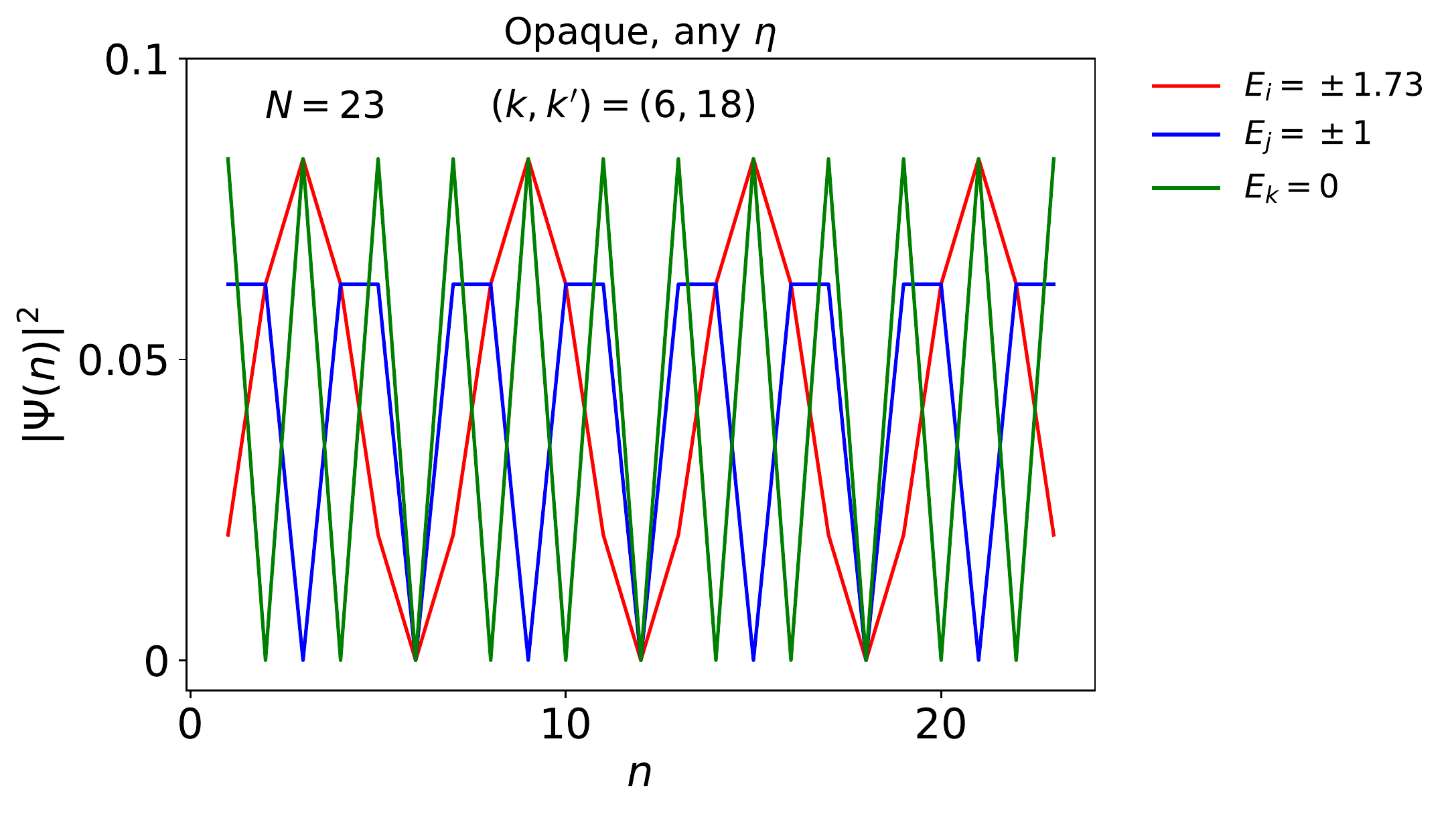}
  \caption{Modulus square of the 5 distinct opaque states 
  for $N=23$ and $k=6$. As shown, all the opaque states 
  have nodes at the gain and loss ($k=6$ and $k'=18$).}
  \label{fig:9}
\end{figure}

As we established above, in several configurations of the
contacts some states have
real eigenvalues that are independent of $\eta$.
The modulus squared of some examples of transparent
states is illustrated
in Fig.~\ref{fig:8} for the configuration $k=6$
and $k'=18$ ($N=23$) and $\eta=0.5$, $\eta=1.5$ and
$\eta=2.5$, respectively. These states 
become increasingly 
concentrated in the center of the chain, between the 
gain and the loss, as the
magnitude of the corresponding eigenvalues increases. In 
Fig.~\ref{fig:9} we show the modulus square 
of the 5 distinct opaque states of this configuration, 
illustrating that all of them have nodes at the 
gain and loss. In addition, we observe that in this
case they also have nodes at the central site $j=12$.

\section{Summary and Conclusions}
\label{sec:Conclusions}

We have presented a detailed analysis of the transport
properties in a one dimensional
$\mathcal{PT}$ tight binding chain. This was achieved by 
first deriving a generalized continuity equation for the 
density. The terms responsible for the non-hermiticity of 
the Hamiltonian appear as gain and loss terms in the 
continuity equation. Transport can be quantified via a single 
number which we called the transport
coefficient. In the $\mathcal{PT}$-unbroken symmetry 
phase this coefficient is
equal to one, which implies that transport is efficient in 
the sense that the inflow of the density equals the 
outflow. For states with broken $\mathcal{PT}$ symmetry
(complex eigenvalues), 
the transport coefficient is different from one, implying 
that density either accumulates or is depleted within the 
system. To study the detailed behavior of the chain, we 
obtained general expressions for the eigenvalues and 
eigenvectors of the system. This analysis led us to note 
that if $N+1$ (where $N$ is the length of the chain) and 
$k$ (the position of the gain) have common divisors, the 
system may have eigenstates that do not couple to the gain 
and loss, and thus do not transport density through 
the chain. We call these states opaque. Similarly,
if $N+1$ and $2k$ have common divisors which do not divide $k$,
the eigenstates have real eigenvalues independently of the
coupling $\eta$ and transport is efficient; we call 
these states transparent. To illustrate these 
phenomena we analyze the eigenvalues for chain
lengths $N=10$ and $23$. In the first case $N+1$ is prime 
and there are no opaque nor transparent states; interestingly, 
for $k=5$ and $\eta>1$ transport is deficient for all states. In the 
second case $N+1$ is a highly composite integer, 
opaque and transparent eigenstates may be present and the behaviour 
is richer in terms of the transport. 
For instance, we
find that in addition to transparent states, there are some 
states with real eigenvalues
for all values of $\eta$, which have a weak dependence on 
$\eta$, and 
their eigenfunctions do not vanish at the contacts.
The existence of such states as well as transparent states allows 
having efficient transport beyond the value of $\eta$ at which $\mathcal{PT}$-symmetry is broken.
Interestingly, for $N=23$ and $k=8$, which corresponds to 
the case with the maximum number of opaque states, beyond
certain $\eta$ transport is deficient in all states. This suggests 
that opaque states play a role inhibiting transport.
For completeness, we have 
presented a simple perturbation scheme to study the
eigenvalues and eigenvectors around the exceptional 
points. The development of the perturbation scheme 
provides a simple rule to obtain the value of 
$\eta$ for which exceptional points appear. 

Our results show that the simple $\mathcal{PT}$-symmetric
tight binding chain displays rather complex spectral 
properties in 
terms of the position of the gain and the loss. These imply rich 
transport behavior depending on the presence of opaque and
transparent states, as well as
possibly other states whose eigenvalues remain real 
independently of the strength of the contacts. This amounts 
to a classification of eigenstates in terms of transport 
which should be observed in experiments using, for example, optical
wave-guides or microwave resonators .


\begin{acknowledgments}
  The authors gratefully acknowledge financial support from 
  CONACyT Proyecto Fronteras 952, CONACyT Posdoctoral Fellowship
  Programme (AO), CONACyT Proyecto A1-S-13469 and the UNAM-PAPIIT 
  research grants IG-100819 and IA-103020.
\end{acknowledgments}

\appendix

\section{Solution of the eigenvalue problem}
\label{sec:egvalsol}

In this section, following Ref.~\cite{Losonczi1992,Yueh2005}, we obtain 
the eigenvalues and eigenvectors of the general complex
tridiagonal matrix defined by
\begin{equation}
  \begin{split}
  A & = b\sum_{j=1}^N \ket{j}\bra{j} + a\sum_{j=2}^N \ket{j-1}\bra{j}
  + c \sum_{j=1}^{N-1} \ket{j}\bra{j+1}\\
  & - \alpha\ket{k}\bra{k} - \beta\ket{N-k+1}\bra{N-k+1}, \\
  \end{split}
  \label{eq:Amatrix}
\end{equation}
where $a,b,c,\alpha,\beta \in \mathbb{C}$, $N$ is the dimension of
the Hilbert space,
and the contacts are in the positions $k$ and $N-k+1$. Since the
location of the contacts is related by parity we may choose
$k\le N/2$; note that in-spite of this, we have not
assumed any specific relation among $\alpha$ and $\beta$.

Let $\lambda$ be an eigenvalue of $A$ and write its associated
(complex) eigenvector in the site basis as 
$u = \sum_i u_i \ket{j}$.
The eigenvalue problem $Au = \lambda u$ can be written as the set
of linear equations
\begin{equation}
  \begin{split}
    u_0 &= 0,\\
    au_0 + bu_1 + cu_2 &= \lambda u_1,\\
    \vdots \\
    au_{k-1} + bu_k + cu_{k+1} &= (\lambda + \alpha) u_k,\\
    \vdots \\
    au_{N-k} + bu_{N-k+1} + cu_{N-k+2} &= (\lambda + \beta) u_{N-k+1},\\
    \vdots \\
    au_{N-1} + bu_N + cu_{N+1} &= \lambda u_N, \\
    u_{N+1} &= 0.
  \end{split}
  \label{eq:Asyseq}
\end{equation}
Note the introduction of two boundary equations, which are used to have a
uniform way of writing the linear system. Further, we assume $ac\neq0$,
otherwise the solution is trivial.

The idea of the approach developed in Refs. \cite{Losonczi1992,Yueh2005} is to rewrite
this system of equations as an algebraic problem on infinite sequences. Then,
one can use the symbolic calculus of Cheng \cite{Chengbook2003} to obtain the
required eigenproblem solution. We shall view the components of the eigenvector,
$u_j$, as the $j$-th term of the complex sequence
$u=\lbrace{u_i \rbrace}_{i=0}^\infty$, with $u_j=0$ for $j=0$ and $j>N$. Notice
that $u_1\neq0$, otherwise if $u_1=0$ we have that $u_2=u_3=\dots=0$.

Similarly, we define the complex sequence $f=\lbrace{f_j\rbrace}_{j=0}^\infty$
with all components identical to zero except for $f_k = \alpha u_k$ and
$f_{N-k+1} = \beta u_{N-k+1}$, which define the location of the contacts.
Then, Eq.~(\ref{eq:Asyseq}) can be expressed as
\begin{equation}
  c\lbrace{ u_{j+2} \rbrace}_{j=0}^\infty + b\lbrace{ u_{j+1} \rbrace}_{j=0}^\infty
  + a\lbrace{ u_{j} \rbrace}_{j=0}^\infty = \lambda \lbrace{ u_{j+1} \rbrace}_{j=0}^\infty + \lbrace{ f_{j+1} \rbrace}_{j=0}^\infty.
\end{equation}

We now introduce the shift sequence $S = \lbrace 0,1,0,\dots\rbrace$ and the
scalar sequence $\overline{z} = \lbrace z,0,0\dots\rbrace$, where
$z\in \mathbb{C}$. We take the convolution $\star$ of the above equation with $S^2$ 
(see \cite{Chengbook2003} for details and definitions), which yields
\begin{equation}
  c(u - \overline{u_0} - u_1S) + (b - \lambda)S(u - \overline{u_0}) + aS^2u
  = S(f - \overline{f_0}).
\end{equation}
Since $u_0=f_0=0$, solving for $u$ we get
\begin{equation}
  (aS^2 + (b - \lambda)S + \overline{c})u = (f+c\overline{u_1})S.
\end{equation}

Since $c\neq 0$, the factor $(aS^2 + (b - \lambda)S + \overline{c})$ has a
multiplicative inverse \cite{Chengbook2003}, i.e.
\begin{equation}
  u = \frac{(f + c\overline{u_1})S}{aS^2 + (b - \lambda)S + \overline{c}}.
  \label{eq:u1}
\end{equation}
The next step is to factorize the denominator, namely
\begin{equation}
  aS^2 + (b - \lambda)S + \overline{c} = a( \overline{\gamma_+} - S)( \overline{\gamma_-} - S),
\end{equation}
where
\begin{equation}
  \gamma_{\pm} = \frac{-(b-\lambda) \pm \sqrt{w}}{2a},
\end{equation}
$w = (b-\lambda)^2 - 4ac$ with $ac\neq0$. Using partial fractions we have
\begin{equation}
  \frac{1}{a(\overline{\gamma_-} - S)(\overline{\gamma_+} - S)} =
    \frac{1}{\sqrt{w}}\Big( \frac{1}{\overline{\gamma_-} - S} -
    \frac{1}{\overline{\gamma_+} - S} \Big).
  \label{eq:partialfractions}
\end{equation}

Since
\begin{equation}
  (\gamma_\pm - S)\star \lbrace \gamma_\pm^{-(j+1)}\rbrace_{j=0}^\infty =
  (\gamma_\pm - S) \lbrace \gamma_\pm^{-(j+1)}\rbrace_{j=0}^{\infty}
  = \lbrace 1, 0, \dots, \rbrace = \overline{1},
\end{equation}
we have $( \overline{\gamma_\pm} - S)^{-1} = \lbrace
\gamma_\pm^{-(j+1)}\rbrace_{j=0}^\infty$. Using this last result in
Eq.~(\ref{eq:partialfractions}) we get
\begin{equation}
  \frac{1}{a(\overline{\gamma_-} - S)(\overline{\gamma_+} - S)} =
    \frac{1}{\sqrt{w}} \lbrace \gamma_-^{-(j+1)} -
      \gamma_+^{-(j+1)}\rbrace_{j=0}^\infty,
\end{equation}
which we insert into Eq.~(\ref{eq:u1}), to obtain
\begin{equation}
  u = \frac{1}{\sqrt{w}}\lbrace \gamma_-^{-(j+1)} -
    \gamma_+^{-(j+1)}\rbrace_{j=0}^{\infty}\,
    (f + c\overline{u_1})S.
  \label{eq:u2}
\end{equation}

At this point, it is convenient to introduce the following notation.
Since $\gamma_\pm$ are complex numbers, we write $\gamma_\pm = p \pm \I q$ and
\begin{equation}
  \gamma_\pm = \sqrt{p^2 + q^2}(\cos\theta \pm\I\sin\theta) =
    \frac{e^{\pm \I\theta}}{\rho},
\end{equation}
where
\begin{equation}
  \rho = \sqrt{a/c},\; \cos\theta = \frac{p}{\sqrt{p^2 + q^2}} = \frac{\lambda - b}{2\sqrt{ac}},
  \label{eq:newvars}
\end{equation}
with $p,q,\rho,\theta\in\mathbb{C}$. With these new definitions, Eq.~(\ref{eq:u2})
now reads
\begin{equation}
\begin{split}
  u &= \frac{1}{\sqrt{w}}\Big\lbrace \Big(\frac{a}{c}\Big)^{j+1} \Big(
  \gamma_+^{j+1} - \gamma_+^{j+1} \Big) \Big\rbrace_{j=0}^{\infty} (f + c\overline{u_1})S \\
  &=  \frac{2\I}{\sqrt{w}}\lbrace{\rho^{j+1}\sin(j+1)\theta \rbrace}_{j=0}^{\infty}
  (f + c \overline{u_1}) S,
  \label{eq:u3}
\end{split}
\end{equation}
where in the last equality we have used De Moivre's theorem.

Up to Eq.~({\ref{eq:u3}}) we have followed the same steps as in
Yueh \cite{Yueh2005}; the remaining of the derivation is a generalization of
Yueh's. Notice that our eigenvalues are determined by the second equality of Eq.~(\ref{eq:newvars})
\begin{equation}
  \label{eq:evalsgen}
  \lambda = b + 2\sqrt{ac}\cos\theta,
\end{equation}
and also our eigenvectors depend on $\theta$. Thus, our first task is
to obtain an equation for $\theta$. To do so, we have to calculate the
convolutions in Eq.~(\ref{eq:u3}). As noted above, the contacts are at
sites $k$ and $N-k+1$, and recall that the $f$ sequence contains
this information.

Then
\begin{equation}
  (f + c \overline{u_1})S = \lbrace 0,cu_1,0,\dots,0,\underbrace{f_k}_{k+1},0,
  \dots,0,\underbrace{f_{N-k+1}}_{N-k+2}, 0, \dots\rbrace,
\end{equation}
where we explicitly stated that due to the action of S over $f$, $f_k$ has been
switched to position $k+1$ and similarly for $f_{N-k+1}$. Next we take the
$j$-th component of Eq.~(\ref{eq:u3})
\begin{eqnarray}
  u_j &=& \frac{2\I}{\sqrt{w}}\Big(\lbrace{\rho^{j+1}\sin(j+1)\theta \rbrace} (f
  +  c \overline{u_1}) S\Big)_j \nonumber \\
  &=& \frac{1}{\sqrt{ac}\sin\theta}\Big( cu_1\rho^j \sin(j\theta)
  + \Theta(j-k-1)\alpha u_k \rho^{j-k}\sin[(j-k)\theta] \nonumber \\
  && + \Theta(j-N+k-2)\beta u_{N-k+1}\rho^{j-N+k-1}\sin[(j-N+k-1)\theta]\Big),
  \label{eq:u4}
\end{eqnarray}
where $\Theta(x)$ is the unit step function defined by $\Theta(x) = 1$
if $x\geq 0$ and $\Theta(x) = 0$ if $x<0$, and in the last identity
we have used $2\I\sqrt{ac}\sin\theta = \sqrt{w}$. In particular, we shall
use below the expressions for $u_k$ and $u_{N-k+1}$, which read
\begin{eqnarray}
  \label{eq:u_source}
  u_k &=& \frac{cu_1\rho^k}{\sqrt{ac}\sin\theta} \sin k\theta , \\
  \label{eq:u_drain}
  u_{N-k+1} &=& \frac{cu_1\rho^{N-k+1}}{\sqrt{ac}\sin\theta}
    \Big(\sin(N-k+1)\theta + 
      \frac{\alpha\sin k\theta \, \sin (N-2k+1)\theta}{\sqrt{ac}\sin\theta}
     \Big).
\end{eqnarray}
In the last two expressions we have eliminated the terms for 
which the argument of the step function is negative.

Using Eq.~(\ref{eq:u4}) for $j = N+1$ and exploiting the explicit
expressions derived for $u_k$ and $u_{N-k+1}$, we obtain
\begin{eqnarray}
  u_{N+1} &=&
  \frac{cu_1\rho^{N+1}}{\sqrt{ac}\sin\theta}
    \Big[ \sin(N+1)\theta + \frac{\alpha + \beta}{\sqrt{ac}\sin\theta}
    \sin(N-k+1)\theta\, \sin k\theta \nonumber\\
  & & + \frac{\alpha\beta}{ac\sin^2\theta} \sin(N-2k+1)\theta\,\sin^2 k\theta \Big],
\end{eqnarray}

and from the boundary condition $u_{N+1}=0$ of the linear
system, Eq.~(\ref{eq:Asyseq}), we get
\begin{equation}
  \sin(N+1)\theta + \frac{\alpha + \beta}{\sqrt{ac}\sin\theta}
  \sin(N-k+1)\theta\,\sin k\theta  + \frac{\alpha\beta}{ac\sin^2\theta}
  \sin(N-2k+1)\theta\,\sin^2 k\theta  = 0.
  \label{eq:thetagen}
\end{equation}
Equation~(\ref{eq:thetagen}) determines $\theta$, with
$\theta \neq m\pi,\; m\in \mathbb{Z}$, which excludes all the
trivial solutions~\cite{Yueh2005}.

To obtain the components
of the eigenvectors $u_j$, $j=1,2,\dots,N$, we proceed similarly
with Eq.~(\ref{eq:u4}), where we substitute Eqs.~(\ref{eq:u_source})
and~(\ref{eq:u_drain}), which leads us finally to
\begin{eqnarray}
    u_j & = & \frac{cu_1\rho^j}{\sqrt{ac}\sin\theta}
      \Big[ \sin j\theta +
        \Theta(j-k-1) \frac{\alpha\sin k \theta \,
          \sin(j-k)\theta}{\sqrt{ac}\sin\theta}\nonumber \\
    & & \quad + \Theta(j-N+k-2) \frac{\beta}{\sqrt{ac}\sin\theta}
      \sin(j-N+k-1)\theta \nonumber\\
    & & \qquad \times \Big( \sin(N-k+1)\theta +
      \frac{\alpha}{\sqrt{ac}\sin\theta} \sin (N-2k+1)\theta\,\sin k\theta\Big)
      \Big].
  \label{eq:ujgen}
\end{eqnarray}

In the specific case that Eq.~(\ref{eq:Amatrix}) is $\mathcal{PT}$-symmetric,
we set $a=c=t$, $\alpha=-\I\eta$, $\beta=\I\eta$. Further, we set $b=0$ since
it amounts to a global shift in the energy.

\bibliographystyle{unsrt}
\bibliography{PTchain2.bib}

\end{document}